\documentclass[fleqn,usenatbib]{mnras}

\usepackage{newtxtext,newtxmath}

\usepackage[T1]{fontenc}
\usepackage{ae,aecompl}
\usepackage{deluxetable}
\usepackage{amsmath}

\DeclareRobustCommand{\VAN}[3]{#2}
\let\VANthebibliography\thebibliography
\def\thebibliography{\DeclareRobustCommand{\VAN}[3]{##3}\VANthebibliography}

\usepackage{graphicx}	
\usepackage{epstopdf}
\usepackage[normal]{threeparttable}
\usepackage[dvipsnames]{xcolor}

\title[M31 cloud property]{The HASHTAG project II. Giant molecular cloud properties across the M31 disc 
}

\author[Y. Deng et al.]{Yikai Deng,$^{1,2,3}$
Zongnan Li,$^{4,1}$\thanks{E-mail: zongnan.li@astro.nao.ac.jp}\thanks{East Asian Core Observatories Association (EACOA) Fellow}
Zhiyuan Li,$^{3,5}$
Lijie Liu,$^{6,7}$
Zhiyuan Ren,$^{1}$
Gayathri Athikkat-Eknath,$^{8}$
\newauthor
Richard de Grijs,$^{9,10,11}$
Stephen A. Eales,$^{8}$
David J. Eden,$^{12,13}$
Daisuke Iono,$^{4,14}$
Sihan Jiao,$^{1}$
Bumhyun Lee,$^{15}$
\newauthor 
Di Li,$^{16,1}$ 
Amelie Saintonge,$^{17,18}$
Matthew W. L. Smith,$^{8}$
Xindi Tang,$^{19}$
Chaowei Tsai,$^{1}$
\newauthor
Stefan A. van der Giessen,$^{20,21}$
Thomas G. Williams,$^{22}$
Jingwen Wu$^{2,1}$
\\
$^{1}$National Astronomical Observatories, Chinese Academy of Sciences, A20 Datun Road, Chaoyang District, Beijing, 100101, China\\
$^{2}$University of Chinese Academy of Sciences, Beijing 100049, China\\
$^{3}$School of Astronomy and Space Science, Nanjing University, Nanjing 210023, China\\
$^{4}$National Astronomical Observatory of Japan, 2-21-1 Osawa, Mitaka, Tokyo, 181-8588, Japan\\
$^{5}$Key Laboratory of Modern Astronomy and Astrophysics, Nanjing University, Nanjing 210023, China\\
$^{6}$Cosmic Dawn centre (DAWN), Technical University of Denmark, DK2800 Kgs. Lyngby, Denmark\\
$^{7}$DTU-Space, Technical University of Denmark, Elektrovej 327, DK2800 Kgs. Lyngby, Denmark\\
$^{8}$School of Physics \& Astronomy, Cardiff University, The Parade, Cardiff, CF24 3AA, UK\\
$^{9}$School of Mathematical and Physical Sciences, Macquarie University, Balaclava Road, Sydney NSW 2109, Australia\\
$^{10}$Astrophysics and Space Technologies Research Centre, Macquarie University, Balaclava Road, Sydney NSW 2109, Australia\\
$^{11}$International Space Science Institute--Beijing, 1 Nanertiao, Zhongguancun, Hai Dian District, Beijing 100190, China\\
$^{12}$Department of Physics, University of Bath, Claverton Down, Bath BA2 7AY, UK\\
$^{13}$Armagh Observatory and Planetarium, College Hill, Armagh, BT61 9DB, UK\\
$^{14}$Astronomical Science Program, The Graduate University for Advanced Studies SOKENDAI, 2-21-1 Osawa, Mitaka, Tokyo 181-8588, Japan\\
$^{15}$Department of Astronomy, Yonsei University, 50 Yonsei-ro, Seodaemun-gu, Seoul 03722, Republic of Korea\\
$^{16}$Department of Astronomy and Tsinghua Centre for Astrophysics, Tsinghua University, Beijing 100084, People's Republic of China\\
$^{17}$Department of Physics \& Astronomy, University College London, Gower Place, London WC1E 6BT, UK\\
$^{18}$Max Planck Institute for Radio Astronomy (MPIfR), Auf dem H$\ddot{u}$gel 69, D-53121 Bonn, Germany\\
$^{19}$Xinjiang Astronomical Observatory, Chinese Academy of Sciences, 830011 Urumqi, PR China\\
$^{20}$Sterrenkundig Observatorium, Ghent University, Krijgslaan 281 - S9, 9000 Gent, Belgium\\
$^{21}$Dept. Fisica Teorica y del Cosmos, E-18071 Granada, Spain\\
$^{22}$Sub-department of Astrophysics, Department of Physics, University of Oxford, Keble Road, Oxford OX1 3RH, UK\\
}

\date{Accepted XXX. Received YYY; in original form ZZZ}

\pubyear{2024}

\begin{document}
\label{firstpage}
\pagerange{\pageref{firstpage}--\pageref{lastpage}}

\maketitle

\begin{abstract}

We present a study of giant molecular cloud (GMC) properties in the Andromeda galaxy (M31) using CO(3-2) data from the James Clerk Maxwell Telescope (JCMT) in selected regions across the disc and in the nuclear ring, and comparing them with CO(1-0) observations from the IRAM 30m telescope in the same regions. We find that GMCs in the centre of M31 generally exhibit larger velocity dispersions ($\sigma$) and sizes ($R$) compared to those in the disc, while their average surface density ($\Sigma$) and turbulent pressure ($P_{\rm turb}$) are lower. This low turbulent pressure in the central region is primarily due to the low density of molecular gas. The estimated GMC properties depend on the choice of CO transitions. Compared to CO(1-0), CO(3-2) exhibits smaller velocity dispersion and equivalent radius but higher surface density. These differences highlight the distinct physical conditions probed by different molecular gas tracers. We estimate the virial parameter $\alpha_{\rm vir}\propto \sigma^2 R/\Sigma$ and find that most molecular clouds exhibit high values ($\alpha_{\rm vir} \sim 4-6$) for both CO transitions, indicating that they are unbound. Furthermore, clouds in the nuclear ring display even larger $\alpha_{\rm vir}$ values of $\lesssim 100$, suggesting that they may be highly dynamic, short-lived structures, although they could potentially achieve equilibrium under the external pressure exerted by the surrounding interstellar medium. 

\end{abstract}



\section{Introduction} 

Star formation plays a crucial role in the evolution of galaxies, as it shapes the galaxy and contributes to the enrichment of heavy elements. Molecular clouds, the birthplaces of stars, are essential for understanding star formation theories, as their properties directly reflect the conditions for star formation and stellar feedback. Studies of resolved molecular clouds in the Milky Way enable the establishment of ``universal'' scaling relations among the properties of the giant molecular clouds (GMCs), including the size, velocity dispersion, and mass, collectively known as Larson's laws \citep{1981MNRAS.194..809L, 2008ApJ...686..948B}. These correlations suggest that the GMCs are in rough virial equilibrium, reflecting a balance state between self-gravity and turbulence in clouds. Supporting this, studies of the fractal distributions of young stars in nearby galaxies, such as the Magellanic Clouds, have found a 2D fractal dimension of about 1.5-1.9, indicating that stars form in regions dominated by supersonic turbulence \citep{2018ApJ...858...31S, 2022MNRAS.512.1196M}. However, subsequent observations show that GMC properties vary with environments \citep[e.g.][]{2013ApJ...779...46H, 2015ApJ...801...25L, 2018ApJ...857...19F, 2021MNRAS.502.1218R, 2018ApJ...860..172S, 2022AJ....164...43S} and deviate from Larson's relations globally, both in the Milky Way \citep{2009ApJ...699.1092H} and other nearby galaxies \citep[e.g.][]{2013ApJ...779...46H, 2021MNRAS.505.4048L, 2022MNRAS.517..632L, 2023MNRAS.522.4078C, 2024MNRAS.531.4045C}.

To understand the cloud-environment correlation, it is essential to conduct spatially resolved observations that span a wide range of environments.
In particular, investigations by \citet{2009ApJ...699.1092H} and \citet{2011MNRAS.416..710F} have manifested the importance of the size-linewidth-surface density relation as a valuable diagnostic for the connection between molecular cloud properties and their surrounding environments in the Milky Way. Subsequently, molecular gas surveys in nearby spiral galaxies at sub-kpc or even GMCs scales have also enabled a close investigation of GMC properties and their scaling relations, such as the HERACLES \citep[HERA CO-Line Extragalactic Survey;][]{2009AJ....137.4670L}, and NGLS \citep[JCMT Nearby Galaxies Legacy Survey;][]{2012MNRAS.424.3050W}; PAWS \citep[PdBI Arcsecond Whirlpool Survey;][]{2013ApJ...779...42S}, CANON \citep[CArma and NObeyama Nearby galaxies;][]{2013ApJ...772..107D}, PHANGS-ALMA \citep[Physics at High Angular resolution in Nearby Galaxies;][]{2021ApJS..257...43L}, WISDOM \citep[mm-Wave Interferometric Survey of Dark Object Masses; e.g.][]{2021MNRAS.505.4048L}. According to the results from these surveys, it is found that clouds close to galactic centres and bars generally have higher surface density, mass and velocity dispersion \citep{2018ApJ...860..172S, 2020ApJ...892..148S, 2021MNRAS.502.1218R, 2023MNRAS.525.4270W}, indicating the impact of galactic environments on GMC properties.

As the closest massive spiral galaxy, M31 provides the most detailed view of the GMCs in a representative spiral galaxy \citep[D$\sim$ 780 kpc, where 1 arcsec $\sim$ 3.8 parsec;][]{2005MNRAS.356..979M, 2014AJ....148...17D}. It is also a useful contrast to our own Galaxy because M31 is the only other massive spiral galaxy in the Local Group, which bears similarity to the Milky Way. Moreover, there exists a wealth of high-resolution multiwavelength observations that capture both the gaseous and stellar components of M31 in great detail, unparalleled by any other external galaxies. 
M31 has a large bulge \citep{2009A&A...505..497Y}, a prominent star-forming ring at a galactocentric distance of 10 kpc, and less prominent spiral arms \citep{2006ApJ...638L..87G}. The origin of this morphology is still debated. Possible explanations include a recent head-on collision between M31 and a satellite galaxy \citep{2006Natur.443..832B} or a bar in M31's stellar disc \citep{2006MNRAS.370.1499A}, which may account for the ring and the less prominent spirals. Additionally, M31's supermassive black hole is extremely quiescent \citep{2011ApJ...728L..10L}, and there is no ongoing star formation in the central region. This raises interesting questions about how the GMC properties vary across these distinct environments.

Previous IRAM 30m CO(1-0) emission line survey data across the M31 disc \citep{2006A&A...453..459N} has revealed the global structure and kinematics of cold gas in M31, enabling investigations of the GMC properties and dynamical conditions across M31. It is found that the linewidth is higher in the spiral arms and lower in the inter-arm region, with an average value of 10 km s$^{-1}$ over the whole disc. This difference could be attributed to higher turbulent energy injected by more active star-forming activities in spiral arms. 
Subsequently, a study combining the CARMA survey of Andromeda observations \citep{2022MNRAS.511.5287A} and IRAM 30m observations found no significant correlation between GMC linewidth and star formation rate at scales of $\sim100$ pc in M31 \citep{2016AJ....151...34C}. Using the same dataset, \citet{2018ApJ...860..172S} reported a much higher virial parameter in M31 (and M33) compared to nearby galaxies. The virial parameter is defined as the ratio between the cloud's kinetic energy and its gravitational potential energy, implying that the molecular gas in M31 may be largely unbound.
More recently, \citet{2019A&A...625A.148D} analysed IRAM PdBI CO(1-0) observations within the central 250 pc of M31 and identified 12 clumps with significantly higher velocity dispersion that deviate from the size-linewidth relation, 
indicating that they may not be in virial equilibrium. 
However, these studies primarily focus on small regions in the disc and neglect the nuclear region due to a deficiency of molecular gas there \citep{2017A&A...607L...7M, 2019MNRAS.484..964L}. The only exception is the PdBI observations \citep{2019A&A...625A.148D}, which have a much higher resolution ($\sim 3''$) and are difficult to compare with other observations. Therefore, a systematic comparison of GMC properties and their scaling relations across the entire galaxy at similar scales is still essential to understand the cloud-environment correlation in M31.

The purpose of this study is to determine the GMC properties and their scaling relations in M31 using high-resolution CO observations in selected regions. These regions span a wide range of physical environments that span the whole disc, including the nuclear region, spiral arms, and interarm regions, which allow the construction of a representative molecular clump sample. The GMCs are identified through IRAM 30m CO(1-0) and JCMT CO(3-2) observations. The CO(3-2) observations in the disc are from the HARP and SCUBA-2 High-Resolution Terahertz Andromeda Galaxy Survey (HASHTAG) project \citep{2020MNRAS.492..195L, 2021ApJS..257...52S}, while the CO data in the nuclear region are from our newly obtained observations covering a gas-rich arm within the central kpc. The footprint of the selected regions is shown in Figure \ref{fig-region} and the basic information is listed in Table \ref{tab-1}. Subsequently, cloud-scale molecular gas properties, including surface density, size, and velocity dispersion, are measured and the turbulent pressure and the virial parameter are estimated based on CO(1-0) and CO(3-2) data, respectively. 
Variations in GMC properties with environments are also analysed and discussed. We describe the data reduction and clump property estimation in Section \ref{sec:data}. The analysis of the properties of the clumps, including the correlations between size, linewidth, and surface brightness, is presented in Section \ref{sec:results}. We discuss our results in Section \ref{sec:discussion} and summarize our findings in Section \ref{sec:sum}.

\section{Data reduction and analysis}\label{sec:data}
\subsection{Observations and data reduction}

We retrieved CO data from the literature: (1) CO(1-0) data from IRAM 30m toward the M31 nuclear ring (Li et al. in prep); (2) CO(3-2) data from JCMT towards M31 nuclear ring (Li et al. in prep); (3) CO(3-2) observations mapping 12 selected regions in the M31 disc as part of HASHTAG \citep{2020MNRAS.492..195L}, a JCMT large program towards M31; (4) IRAM 30m CO(1-0) map fully sampled an area of $2^\circ \times 0.5^\circ$ in the disc \citep{2006A&A...453..459N}.
CO(3-2) and CO(1-0) data have an angular resolution of 15 arcsec and 23 arcsec, and a channel width of 0.42 km s$^{-1}$ and 2.6 km s$^{-1}$, respectively. The regions are selected to span various physical environments in M31, with detailed selection criteria described in \citet{2021ApJS..257...52S}. 

\begin{table*}
\centering
\caption{Basic information of the selected regions\label{tab-1}}
 \begin{threeparttable}
\begin{tabular}{ccccccccc}
\hline
 ID  &   RA(J2000) &  DEC(J2000) & PA($^{\circ}$) &  Coverage($^{\prime}$) \\
\hline
a & 00:46:31.0 & +42:11:51.5 & 160.7 & 2$\times$2\\
b & 00:45:34.8 & +41:58:28.5 & 145.7 & 2$\times$2\\
c & 00:44:37.2 & +41:52:35.6 & 145.0 & 2$\times$2\\
d & 00:44:59.2 & +41:55:10.5 & 141.0 & 2$\times$2\\
e & 00:44:26.5 & +41:37:12.7 & 153.0 & 2$\times$2\\
f & 00:43:03.3 & +41:24:16.2 & 130.0 & 2$\times$2\\
g & 00:42:21.4 & +41:06:21.1 & 130.0 & 2$\times$2\\
h & 00:44:03.1 & +41:42:39.3 & 130.0 & 2$\times$2\\
i & 00:44:13.2 & +41:35:17.1 & 130.0 & 2$\times$2\\
j & 00:45:26.9 & +41:44:54.6 & 37.7 & 2$\times$2\\
k & 00:43:52.2 & +41:33:48.9 & 37.7 & 2$\times$2\\
R & 00:44:40.9 & +41:27:25.2 & 37.7 & 4$\times$4\\
N(CO(3-2)) & 00:42:58.3  & +41:18:18.5 & 335.0 & 2$\times$4$^{(a)}$\\
N(CO(1-0)) & 00:42:58.3  & +41:18:18.5 & 335.0 & 1.5$\times$4.5\\
\hline
\end{tabular}
\begin{tablenotes}
\footnotesize
\item \emph{Notes.} 
$^{(a)}$The coverage of N(CO(3-2)) corresponds to the black solid rectangle in Figure \ref{fig-cl} right panel. 
\end{tablenotes}
\end{threeparttable}
\end{table*}

\begin{figure*}
\begin{minipage}{0.9\linewidth}
\centering\includegraphics[width=\textwidth]{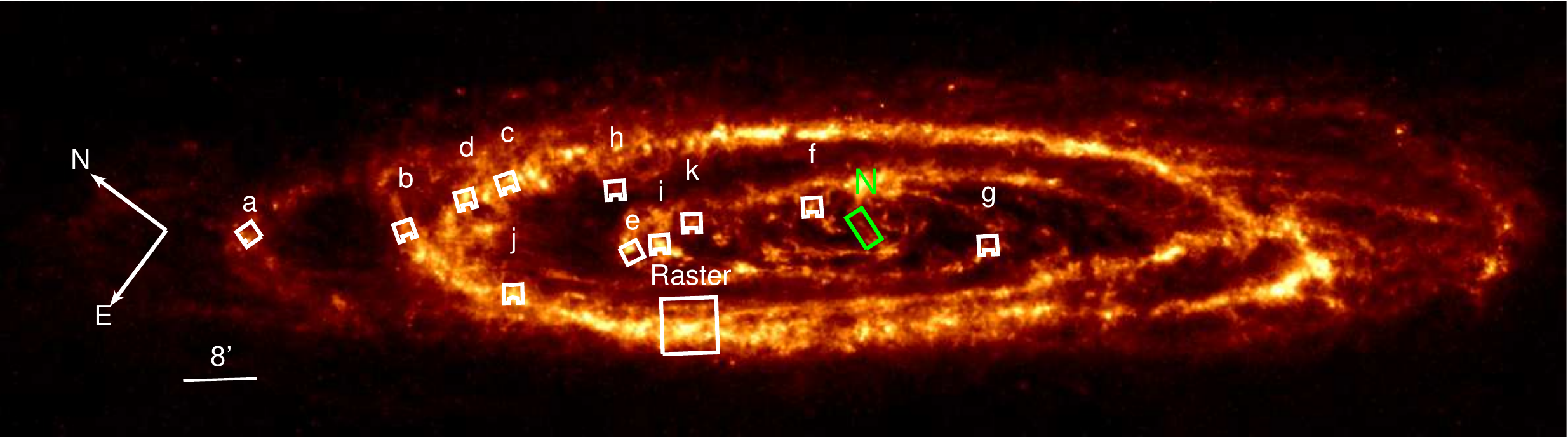}
\end{minipage}
\caption{The footprint of selected regions in M31, similar to that presented in \citet{2020MNRAS.492..195L}, overlaid with \textit{Herschel}/SPIRE 250 $\upmu$m maps \citep{2012ApJ...756...40S}. The 12 HASHTAG fields on the disc are labeled `a-k' and `Raster' (field `R'), while the nuclear ring field is labeled `N'. \label{fig-region}} 
\end{figure*}

For (1) and (2), the data reduction follows Li et al. (2024, in prep), with the main steps described below. For (3) and (4), we use the archival data from the literature \citep{2006A&A...453..459N, 2020MNRAS.492..195L}. 
The IRAM 30m CO(1-0) observations of the nuclear ring are obtained with a total integration time of 20 hours, resulting in a typical RMS of 5.5 mK in 10.4 km s$^{-1}$ channel. The CO(1-0) data reduction follows a standard procedure and is summarized as follows. We employ the CLASS in GILDAS\footnote{http://www.iram.fr/IRAMFR/GILDAS} software package developed by IRAM to examine and process the spectra. First, we checked the quality of individual spectra to ensure there were no spikes or bad channels. Then we performed platform correction on each FTS spectrum using the script FtsPlatformingCorrection5.class\footnote{http://www.iram.es/IRAMES/mainWiki/AstronomerOfDutyChecklist}. The antenna temperature $T\rm_A^*$ is converted to the main beam brightness temperature by $T\rm_{mb} = \it T\rm_A^* \it F\rm_{eff}/\eta_{mb}$, with the main beam efficiency $\eta\rm_{mb}$ = 0.78 and the forward efficiency $F\rm_{eff}$ = 0.94 at 115 GHz.

Data reduction of CO(3-2) observations (2) and (3) follows the standard procedure for JCMT heterodyne observations and is described in detail in \citet{2020MNRAS.492..195L}. The main steps are summarized below. 
We first reduce the data using the ORAC-DR pipeline in the STARLINK software, which can automatically deal with bad spectra. Following \citet{2020MNRAS.492..195L}, we truncate the spectra to a width of 200 km s$^{-1}$ to account for the narrow width of the lines compared to the wide bandwidth. This truncation avoids unnecessary noise in the velocity range without significant signals. For the data in the nuclear ring, we adopted a velocity range between $-600$ and 0 km s$^{-1}$. 
Then we convert the data from Heliocentric to the Local Standard of Rest and correct the temperature with a main beam efficiency of 0.64. For a better comparison with the CO(1-0) data, we smoothed the angular resolution to 23 arcsec and the channel width to 2.6 km s$^{-1}$, respectively, to match those of CO(1-0) data. 

\subsection{Cloud identification}
To obtain the properties of molecular clumps\footnote{Here a clump is used as a technical definition of a GMC identified using the method detailed below.} in these fields, we first need to identify these structures. Given the variations in system temperatures, integration times, and unstable receptors, the signal-to-noise ratio (S/N) differs within and between each field. To facilitate accurate structure identification, the use of S/N cubes is more suitable. We then obtained S/N cubes for both transitions by dividing the reduced cubes with noise maps generated from line-free channels. This approach helps ensure that the resulting cubes provide a reliable representation of the structures while accounting for variations in the S/N ratio across the dataset.
\citet{2020MNRAS.492..195L} used the ClumpFind algorithm \citep{1994ApJ...428..693W} to identify structures. Here we use another widely used algorithm, PYCPROPS \citep{2021MNRAS.502.1218R}, which is an improved version of CPROPS \citep{2006PASP..118..590R}, as a Python package. \citet{2021MNRAS.502.1218R} point out that ClumpFind makes it easier to produce false structures owing to noise. Compared with ClumpFind, CPROPS can identify structures at low S/N, avoid false structures caused by noise, and better identify non-Gaussian structures. 
CPROPS also provides error estimates using bootstrapping, which samples the data with replacement and calculates the required properties for each new sample. This approach provides a robust measure of uncertainty for the derived properties.
Here we sample 1000 times, with the standard deviation of the set taken as the uncertainty of the needed property. In addition, ClumpFind is designed for decomposing data into smaller substructures like clumps, while CPROPS is sensitive to the size of giant molecular clouds. The beam size of our data reaches 23 arcsec ($\sim$90 pc), so using CPROPS is more appropriate. PYCPROPS is faster than CPROPS and has some improvements in the algorithm, providing a better structure analysis. 

Now we summarize the PYCPROPS algorithm. First, we need a Boolean mask, marking regions that are most likely to have signals and omitting low S/N data. 
To make full use of the data, the S/N threshold is selected accordingly. In regions where the S/N is lower, a lower threshold is adopted to ensure that the available data are effectively incorporated. Therefore, for both CO(3-2) and CO(1-0) emission lines, S/N > 2 is adopted in regions g, h, k (Figure \ref{fig-region}), and the nuclear ring (N); and for other regions, S/N > 3 is adopted. The algorithm then uses the dendrogram method \citep{2008ApJ...679.1338R} in the ASTRODENDRO package to identify local maxima and catalog them. Identification is mainly regulated by four criteria. (1) The maxima need to be significant enough, that is, a maximum separates from nearby maxima in the same catalogue above an interval $\delta$. The recommended $\delta$ is twice the noise, which can better balance noise filtering and structure identification \citep{2008ApJ...679.1338R}. (2) The minimum pixels N containing the maxima, to eliminate too small and poorly defined structures. (3) The minimum spatial and spectral separation distance $d_{\rm min}$ and $v_{\rm min}$ between maxima, to better resolve each structure. (4) The difference of the measured properties such as flux and size above a threshold $s$, to test the uniqueness of the structures. Unsatisfactory maxima will be discarded. The remaining maxima will be "seeds" to be grown to the defined edge by the watershed algorithm \citep{2014PeerJ...2..453V} in the SCIKIT-IMAGE package and finally become the "structures".

For our data, a default $\delta$ is adopted. The beam size is close to 90 pc, similar to that of \citet{2021MNRAS.502.1218R}. 
To search for structures close to the resolution limit, which is also the GMC size in crowded conditions, we set parameters the same as \citet{2021MNRAS.502.1218R}, that is, $d_{\rm min}=0, v_{\rm min}=0, s=0$. We chose N = 16 for minimum pixels, with a pixel size of 7.5 arcsec. 
We note that the specific choice of the parameters has an insignificant effect on our conclusions, as our primary objective is to consistently identify GMC-sized clumps for robust statistical analysis and comparison of cloud properties across M31. This approach is inherently insensitive to the properties of individual clouds. We also tested S/N thresholds of 2 and 3 in all regions and found that only a small number of clumps were added or removed, with the results varying by less than 20\%. Importantly, our main conclusions remain unaffected.
It should be noted that we do not match the identification results between CO(3-2) and CO(1-0). Differences between the two sets of data may result in part from the initial resolution, sensitivity, and data quality. If we consider matching the results, some true structures may be discarded due to a mismatch with poor-quality data. Nevertheless, we believe that the results remain reasonably comparable, considering the accuracy of the algorithm. 
The differences between the two transitions will be discussed in Section \ref{sec: CO transition}.

\subsection{Property estimation}

After identifying the structures, we can estimate their properties. The RMS of the size of the structures on three axes was calculated as \citep{2021MNRAS.502.1218R}
\begin{equation}\label{eq-1}
\sigma^2_{l,\rm_{obs}}=\frac{\sum_{{\rm i}\in C}T_{\rm i}(l_{\rm i}-\bar{l})^2}{\sum_{{\rm i}\in C}T_{\rm i}}
\end{equation}
where $T_{\rm i}$ is the measured brightness temperature, $l_{\rm i}$ is the coordinate on each axis, $x, y, v$, $\bar{l}$ is the intensity (temperature) weighted mean coordinate value, and summation range $i\in C$ is all pixels of each structure. 
Due to sensitivity limitations, signals below the limit cannot be detected, resulting in a smaller estimated size. We thus extrapolate the size to 0 K sensitivity and correct for resolution effects. The final velocity dispersion is \citep{2021MNRAS.502.1218R}
\begin{equation}\label{eq-2}
\sigma_v=\sqrt{\sigma^2_{v,{\rm extrap}}-\sigma^2_{v,{\rm chan}}}
\end{equation}
where $\sigma_{v,{\rm extrap}}$ is the extrapolating size and $\sigma_{v,{\rm chan}}$ is the equivalent Gaussian width of a velocity channel \citep{2006PASP..118..590R}
\begin{equation}\label{eq-3}
\sigma_{v,{\rm chan}}=\frac{\Delta_v}{\sqrt{2\pi}}
\end{equation}
$\Delta_v$ is the velocity channel width (velocity resolution).
The equivalent radius is \citep{2006PASP..118..590R}
\begin{equation}\label{eq-4}
R=1.91\sqrt{\sigma_{\rm maj,d}\sigma_{\rm min,d}}
\end{equation}
where $\sigma_{\rm maj,d}$ and $\sigma_{\rm min,d}$ are sizes along the major and minor axes of the structure deduced from $\sigma_{x,{\rm obs}}$ and $\sigma_{y,{\rm obs}}$, corrected for effects from resolution \citep{2006PASP..118..590R}
\begin{equation}\label{eq-5}
\sigma_{\rm j,d}=\sqrt{\sigma^2_{\rm j,extrap}-\sigma^2_{\rm beam}}  
 \rm{(j=maj,min)}
\end{equation}
where $\sigma_{\rm beam}$ is the rms beam size, $1.91$ is a widely-used parameter \citep{2006PASP..118..590R}.
CO luminosity is estimated as \citep{2021MNRAS.502.1218R}
\begin{equation}\label{eq-6}
L_{\rm CO}=A_{\rm pix}\sum_{{\rm i}\in C}T_{\rm i}\Delta_v
\end{equation}
where $A_{\rm pix}$ is the projected area of the pixels of the structure.
The mean surface density is \citep{2021MNRAS.502.1218R}
\begin{equation}\label{eq-7}
\Sigma_{\rm mol}=\frac{M_{\rm gas}}{\pi R^2}=\frac{\alpha_{\rm CO}L_{\rm CO(1-0)}}{\pi R^2}
\end{equation}
where $M_{\rm gas}$ is the gas (luminosity) mass, $R$ is the GMC radius, $\alpha_{\rm CO}$ is the CO-to-H$_2$ conversion factor converting CO luminosity or intensity to molecular mass or surface density. Conventionally, the conversion is performed on the basis of CO(1-0) emission. Therefore, when dealing with CO(3-2) data, it is necessary to convert the CO(3-2) luminosity to the CO(1-0) luminosity using the intensity ratio, denoted as $R_{31}$:
\begin{equation}\label{eq-8}
L_{\rm CO(1-0)}=\frac{L_{\rm CO(3-2)}}{R_{31}}
\end{equation}
We adopted the mean $R\rm_{31}$ value of each region on the disc based on the measurements from \citet{2020MNRAS.492..195L}. For region a, we use a mean value of $R_{\rm 31}$ over the whole disc.
Assuming turbulence is isotropic, molecular clouds are spherical, and the velocity dispersion is distributed all over turbulence, the internal turbulent pressure is estimated as \citep{2020ApJ...892..148S}
\begin{equation}\label{eq-9}
P_{\rm turb}=\rho\sigma_v^2=\frac{3M_{\rm gas}\sigma_v^2}{4\pi R^3}
\end{equation}
The virial parameter can intuitively reflect the internal motion of clouds. For spherical clouds, the virial parameter is estimated as \citep{2018ApJ...860..172S}
\begin{equation}\label{eq-10}
\alpha_{\rm vir}=\frac{2K}{U_{\rm g}}=\frac{M_{\rm vir}}{M_{\rm gas}}=\frac{5\sigma_v^2R}{fGM_{\rm gas}}
\end{equation}
where $M_{\rm vir} = \frac{5\sigma_v^2R}{fG}$ is the virial mass, i.e. cloud mass in virial equilibrium while gravitational potential energy is equivalent to twice the kinetic energy. $f$ is the geometric factor. For clouds density distribution following $\rho\propto r^{-k}$, $f=\frac{1-\frac{k}{3}}{1-\frac{2k}{5}}$. If expressed in surface density $\Sigma\rm_{mol}$, this relation could be rewritten as \citep{2018ApJ...860..172S}
\begin{equation}\label{eq-11}
\frac{\sigma_v^2}{R}=\frac{f\alpha_{\rm vir}G\pi}{5}\Sigma_{\rm mol}
\end{equation}
which is the size-linewidth relation with virial parameter and surface density.
The CO-to-H$_2$ conversion factor can also be estimated from the virial mass and luminosity \citep{2013ARA&A..51..207B}
\begin{equation}\label{eq-12}
\alpha_{\rm CO}=\frac{M_{\rm vir}}{L_{\rm CO(1-0)}}
\end{equation}
which assumes clouds are in virial equilibrium.
If considering external pressure exerted by the surrounding environment (e.g., atomic gas and stars), the relation among velocity dispersion, radius, and surface density is \citep{2011MNRAS.416..710F}
\begin{equation}\label{eq-13}
\frac{\sigma_v^2}{R}=\frac{1}{3}(\pi\Gamma\Sigma_{\rm mol}+\frac{4P_{\rm e}}{\Sigma_{\rm mol}})
\end{equation}
which is the size-linewidth relation with surface density and external pressure. Here $\Gamma$ is a form factor, which is $\frac{3}{5}$ for a constant density sphere and $0.73$ for an isothermal sphere of critical mass, which is defined as the maximum mass of a cloud that is stable for a given pressure and kinetic energy. $P_{\rm e}$ is the external pressure.

\section{Results}\label{sec:results}
\subsection{Cloud morphology}

\begin{figure*}
\begin{minipage}{0.32\linewidth}
\centering\includegraphics[width=\textwidth]{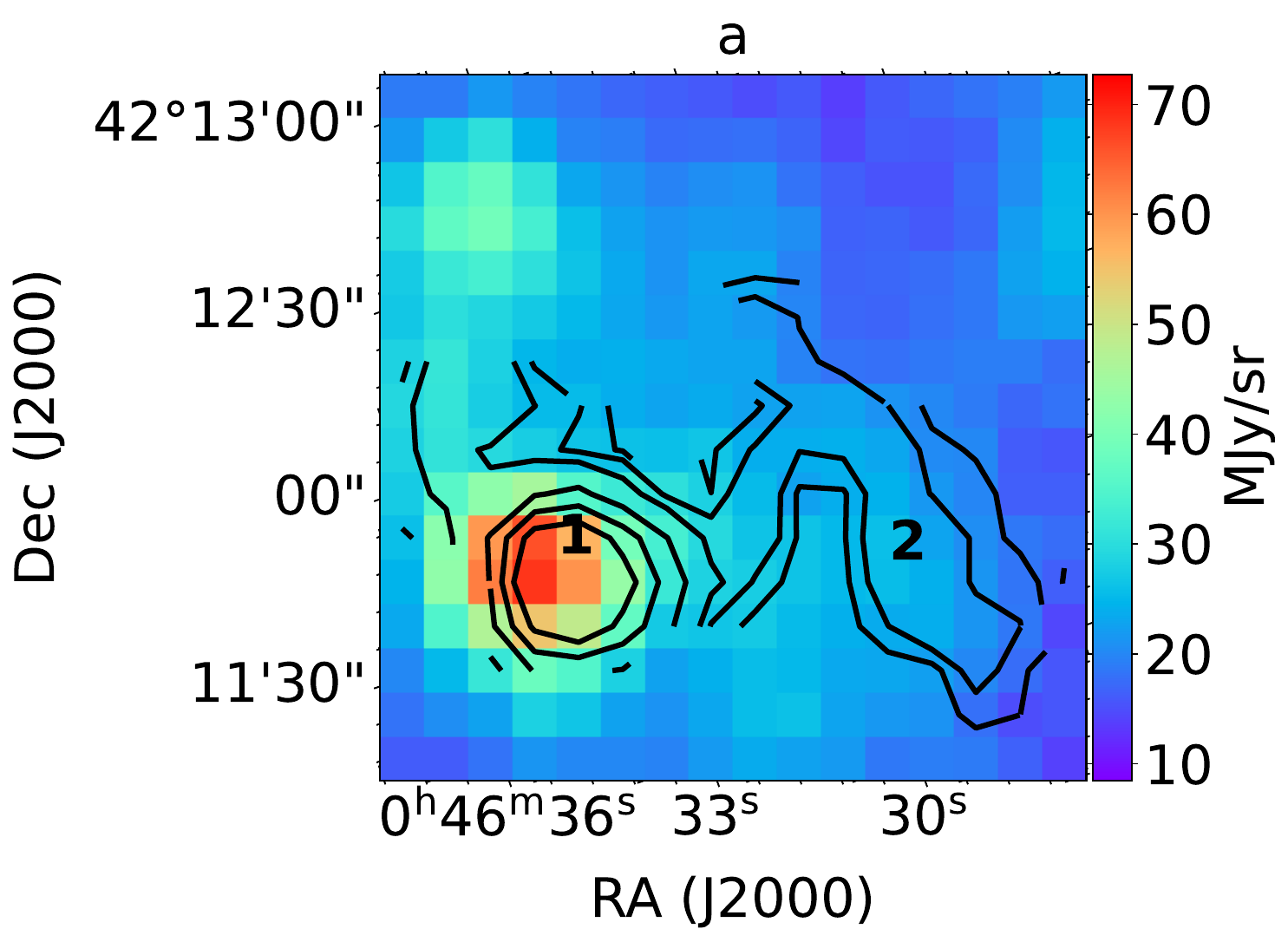}
\end{minipage}
\hfill
\begin{minipage}{0.32\linewidth}
\centering\includegraphics[width=\textwidth]{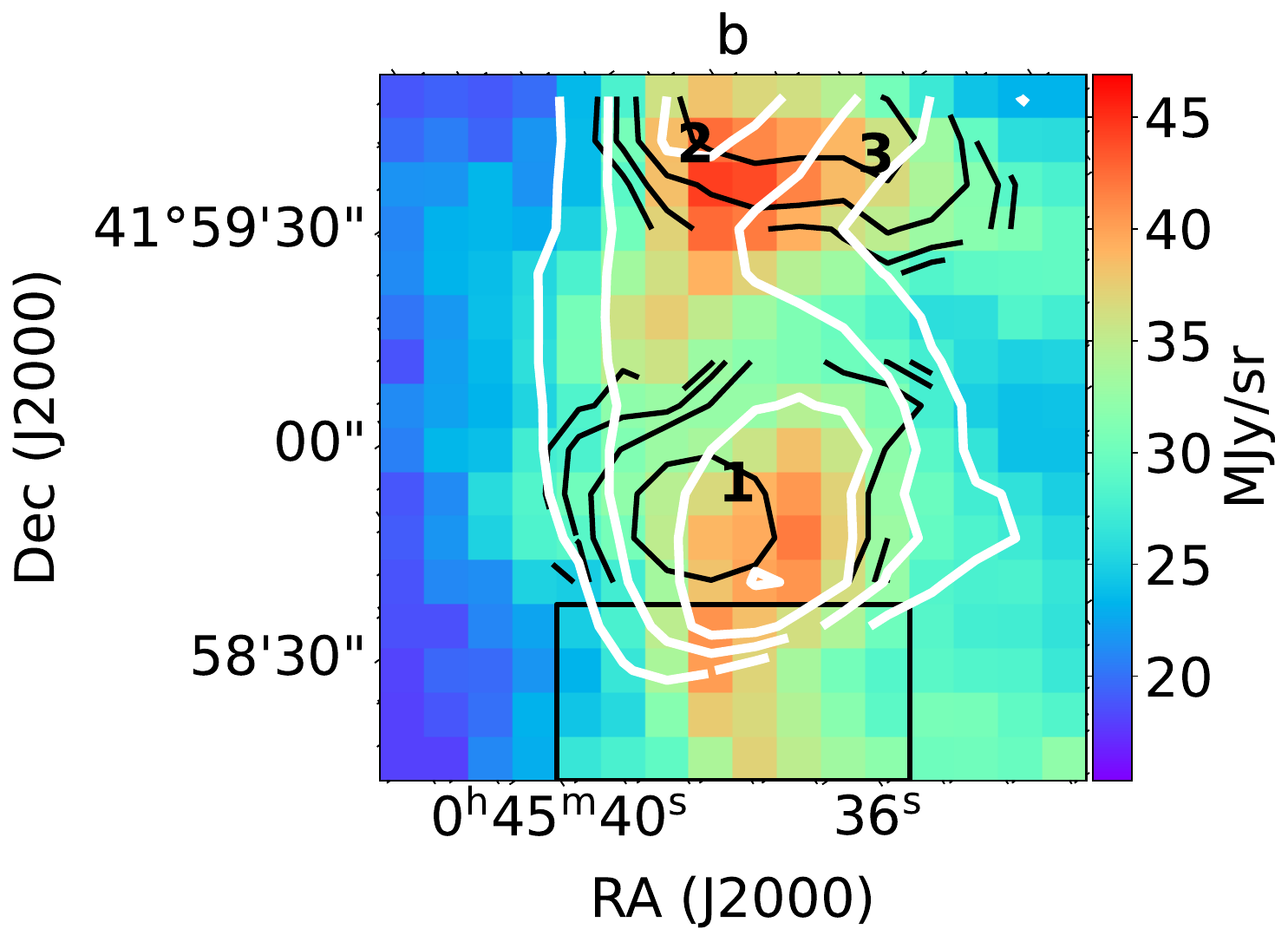}
\end{minipage}
\hfill
\begin{minipage}{0.32\linewidth}
\centering\includegraphics[width=\textwidth]{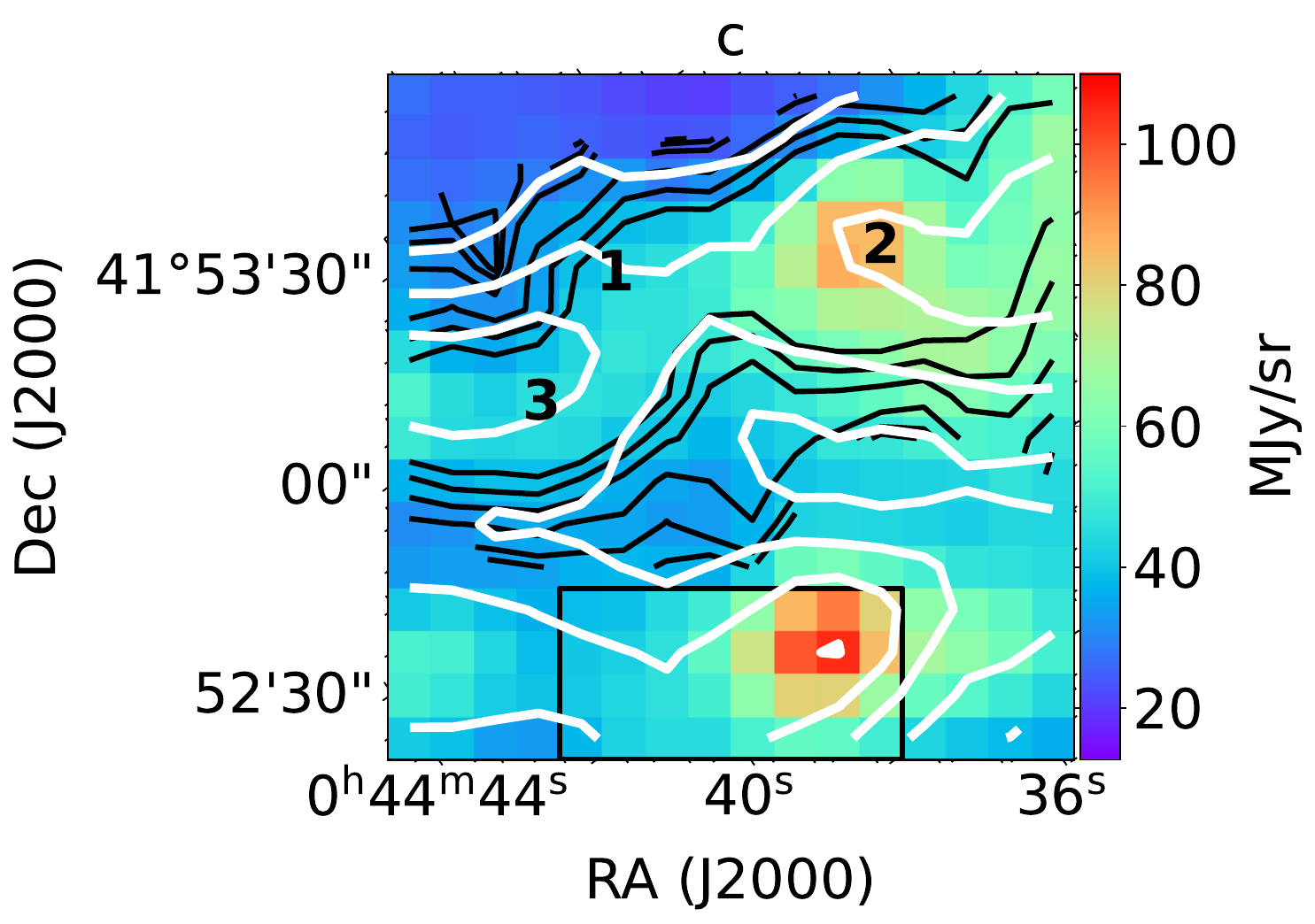}
\end{minipage}
\vfill
\begin{minipage}{0.32\linewidth}
\centering\includegraphics[width=\textwidth]{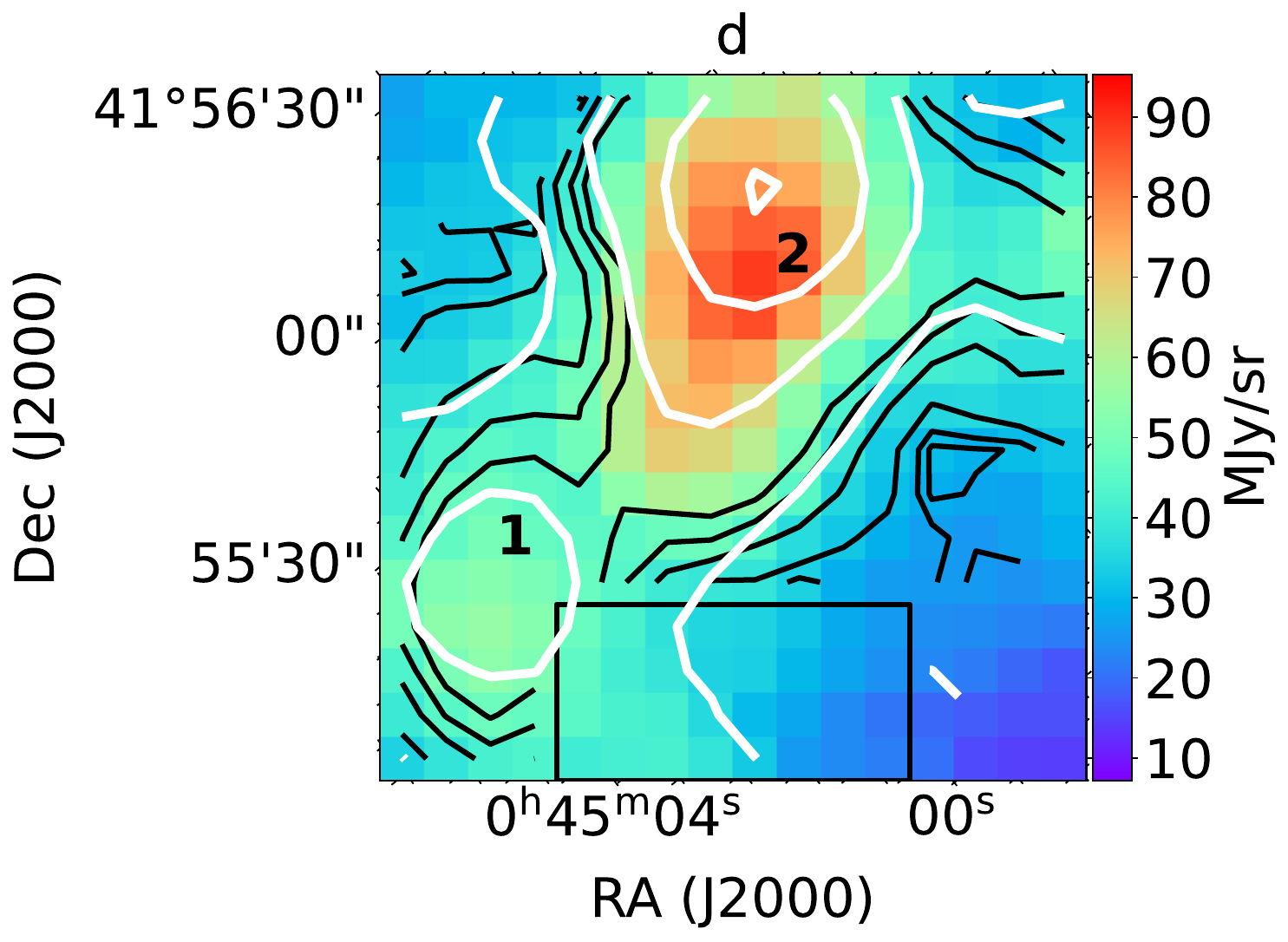}
\end{minipage}
\hfill
\begin{minipage}{0.32\linewidth}
\centering\includegraphics[width=\textwidth]{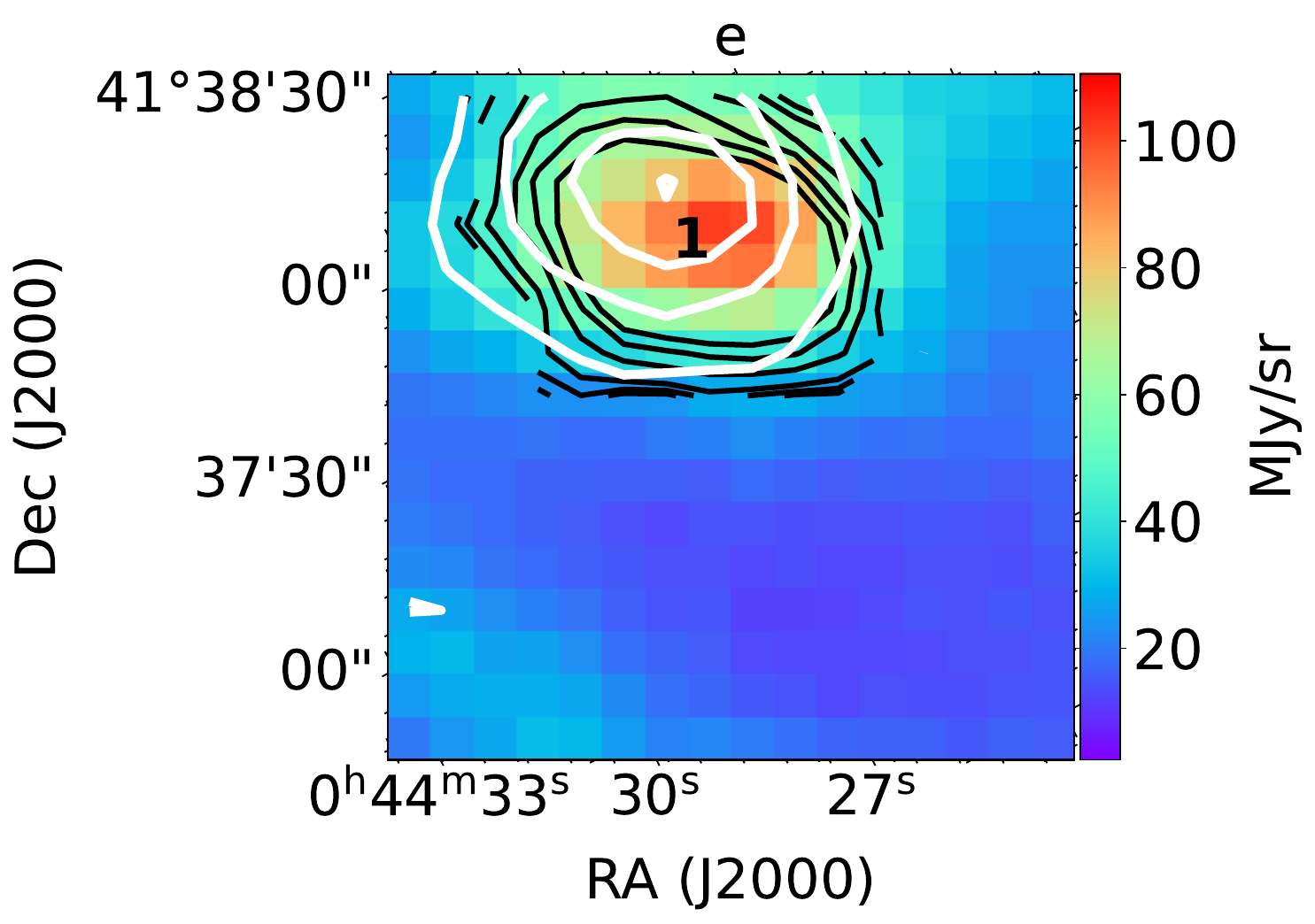}
\end{minipage}
\hfill
\begin{minipage}{0.32\linewidth}
\centering\includegraphics[width=\textwidth]{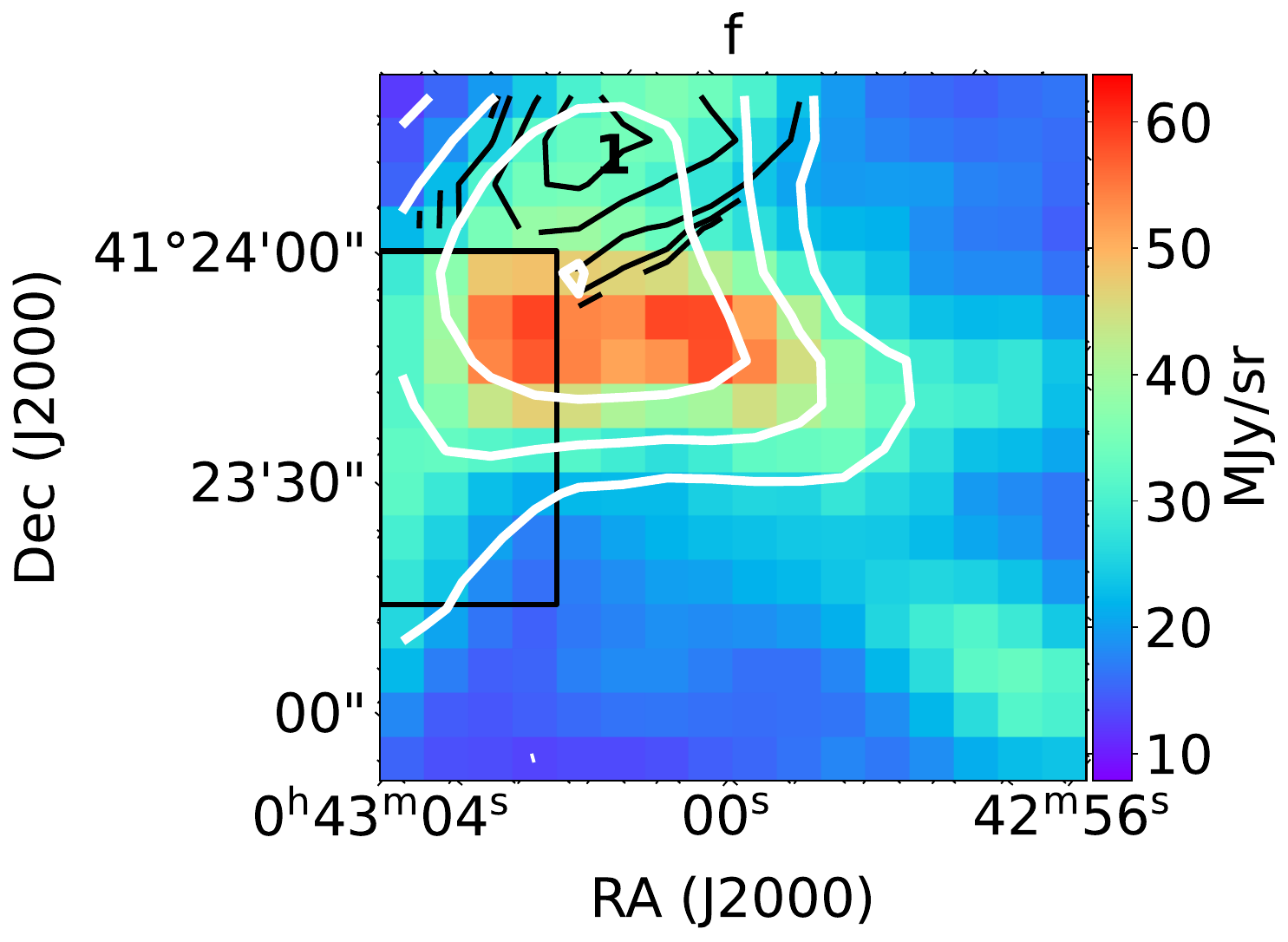}
\end{minipage}
\vfill
\begin{minipage}{0.32\linewidth}
\centering\includegraphics[width=\textwidth]{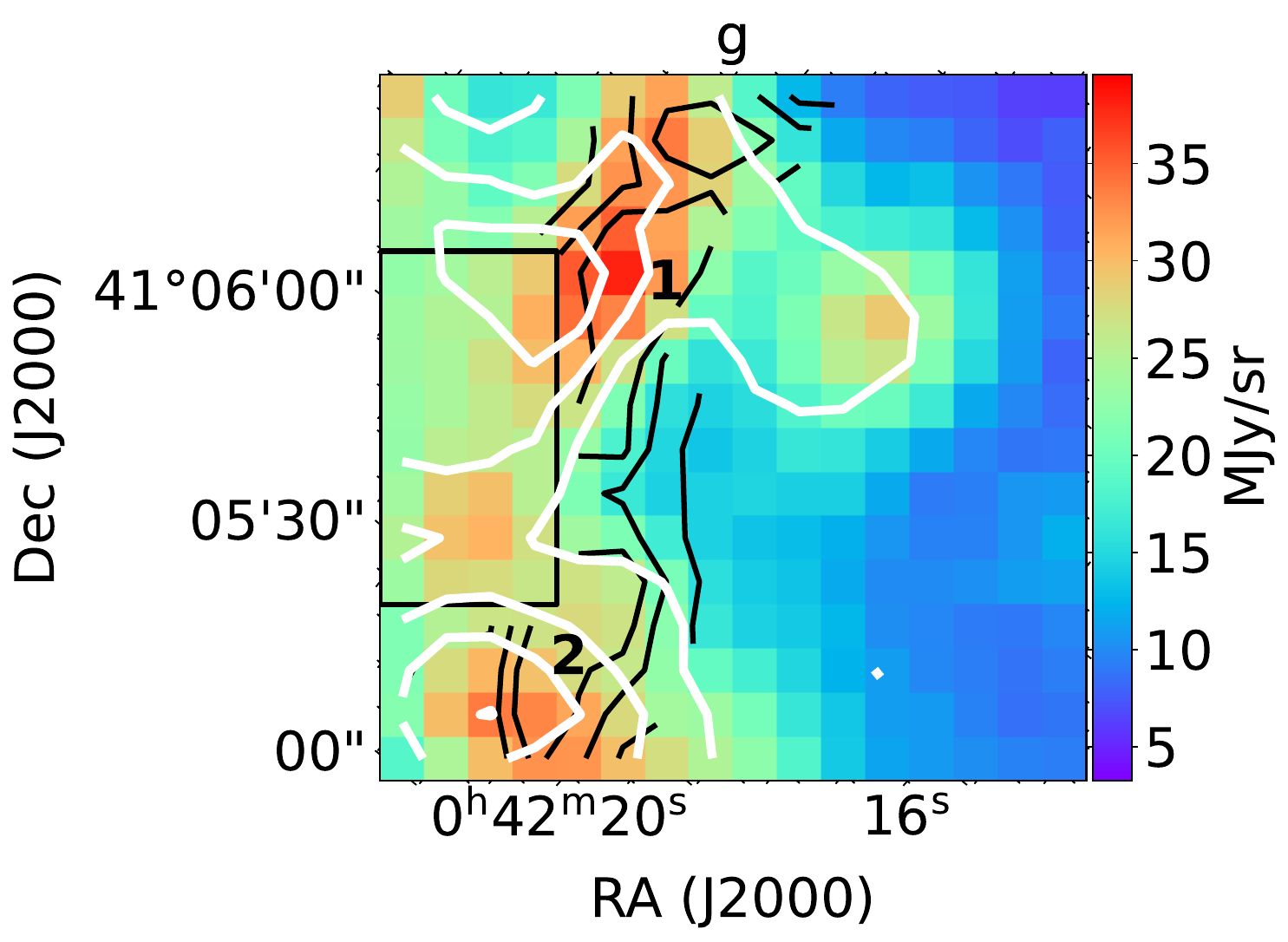}
\end{minipage}
\hfill
\begin{minipage}{0.32\linewidth}
\centering\includegraphics[width=\textwidth]{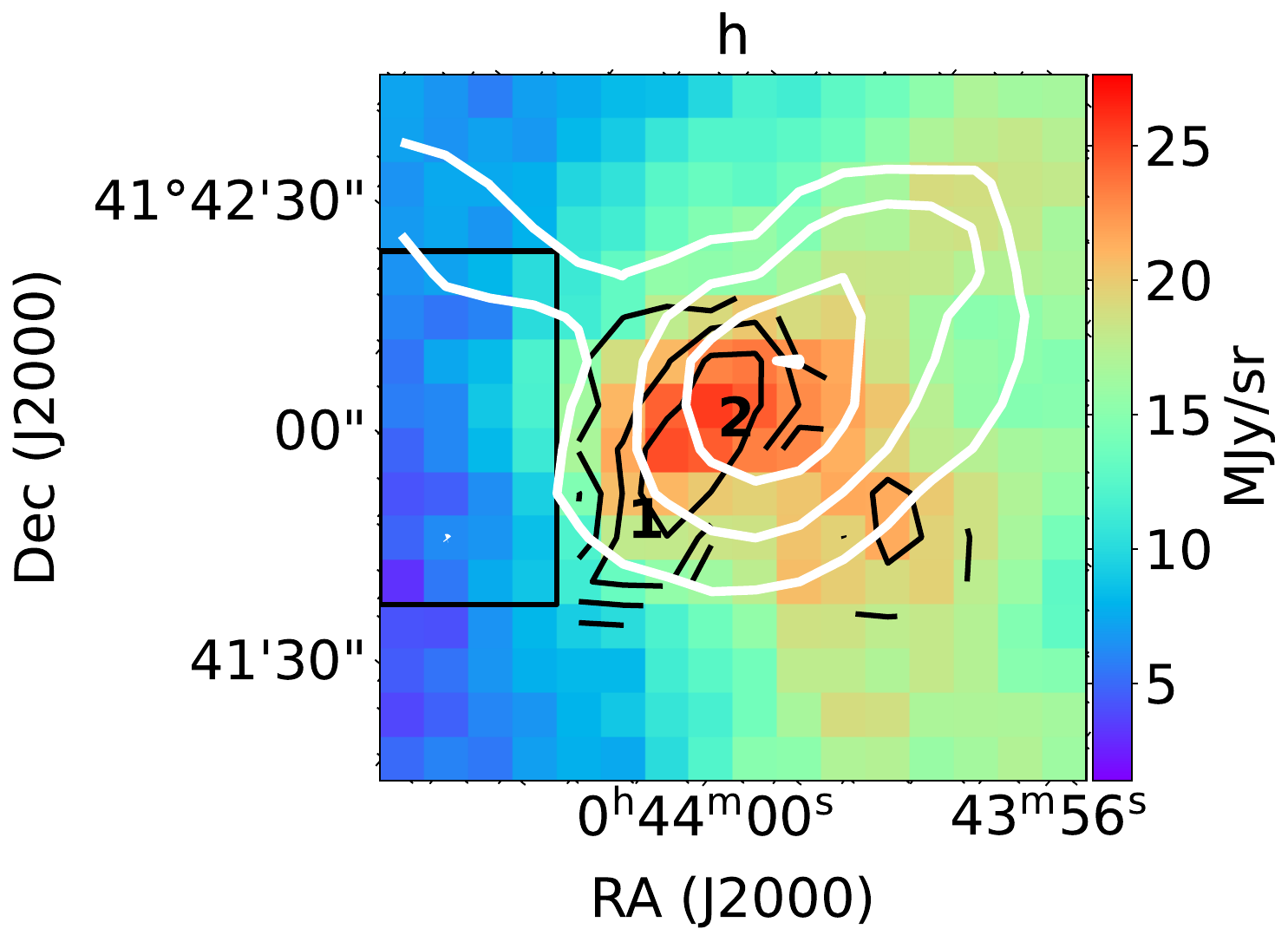}
\end{minipage}
\hfill
\begin{minipage}{0.32\linewidth}
\centering\includegraphics[width=\textwidth]{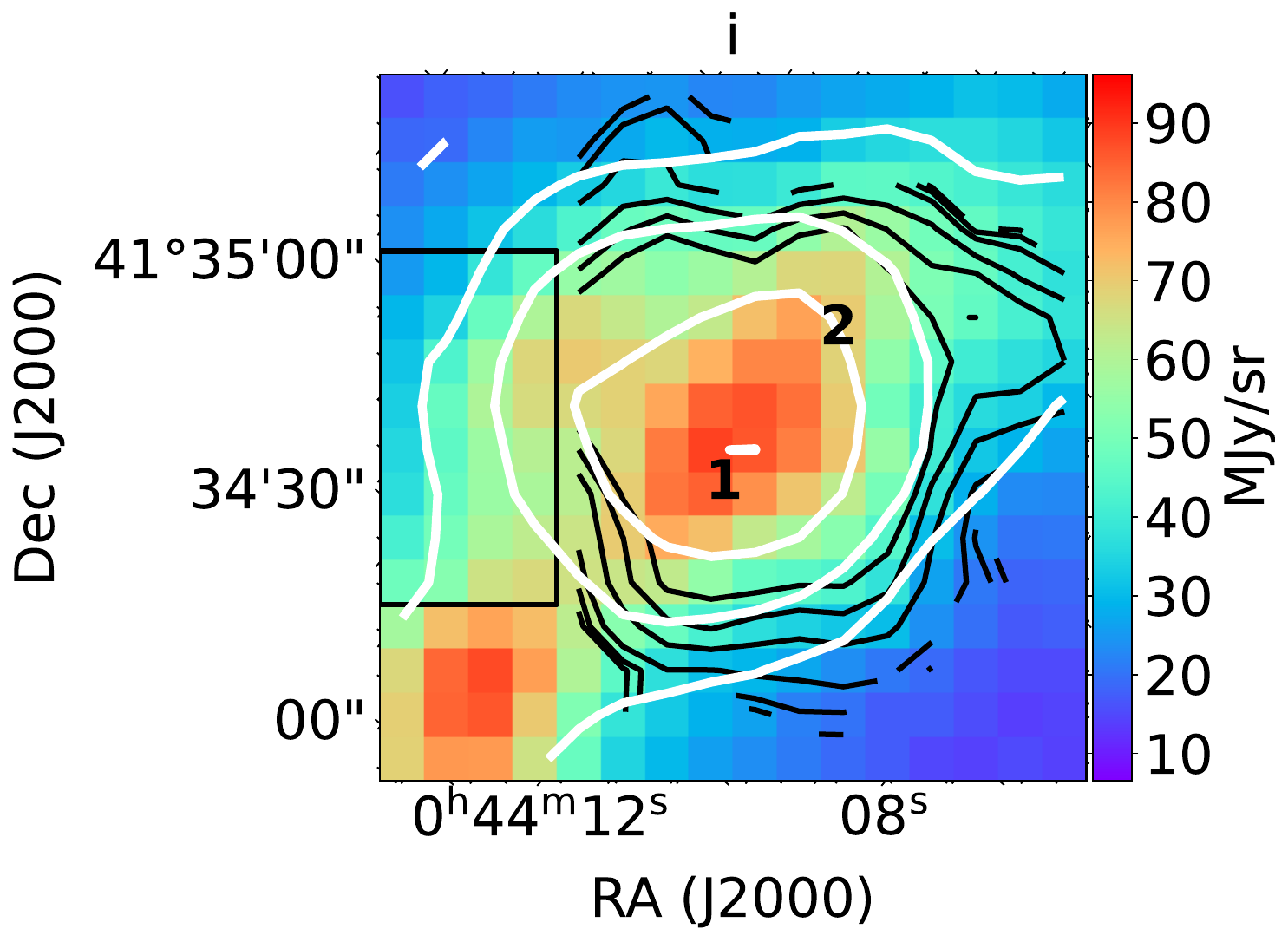}
\end{minipage}
\vfill
\begin{minipage}{0.32\linewidth}
\centering\includegraphics[width=\textwidth]{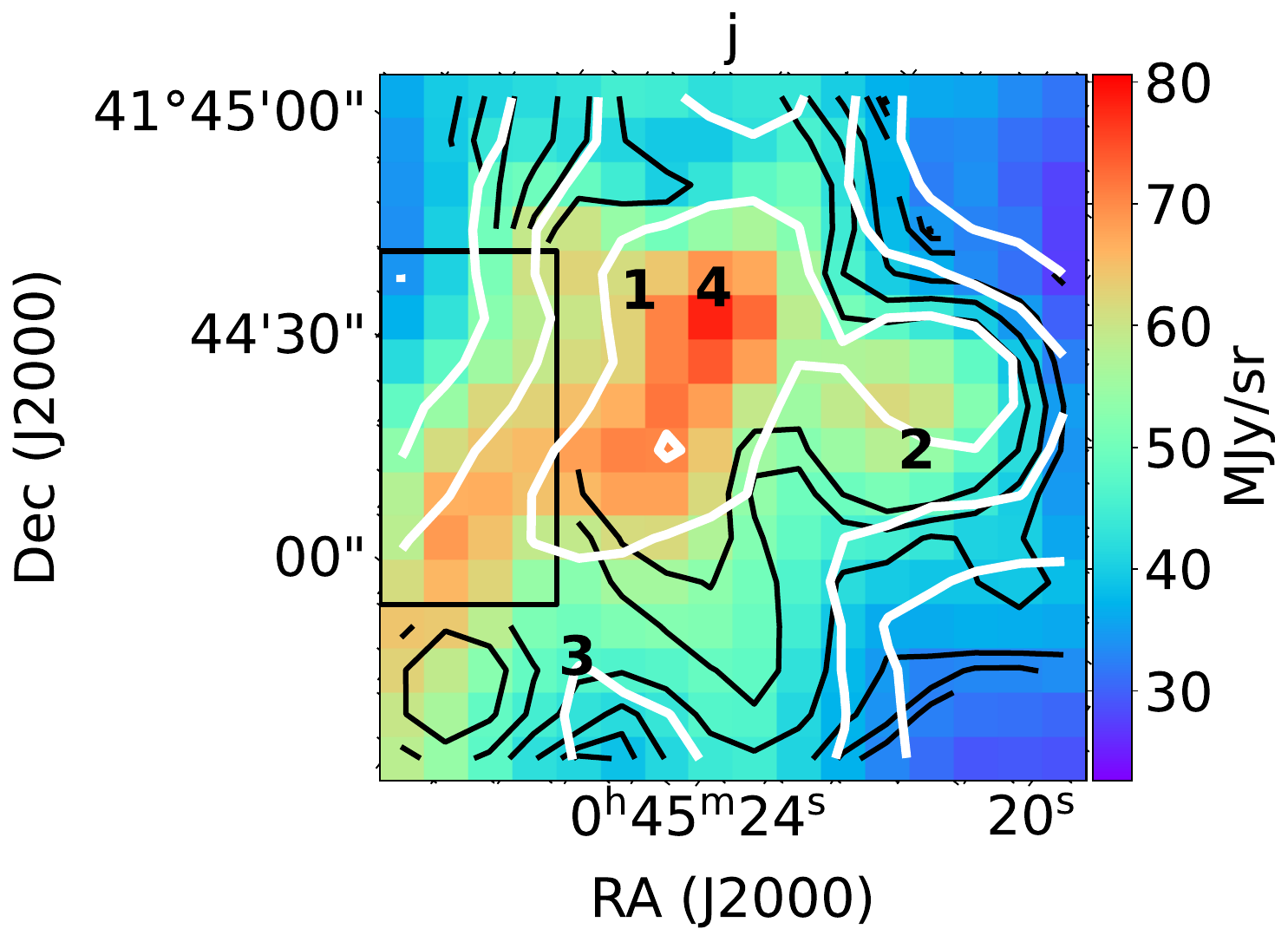}
\end{minipage}
\hfill
\begin{minipage}{0.32\linewidth}
\centering\includegraphics[width=\textwidth]{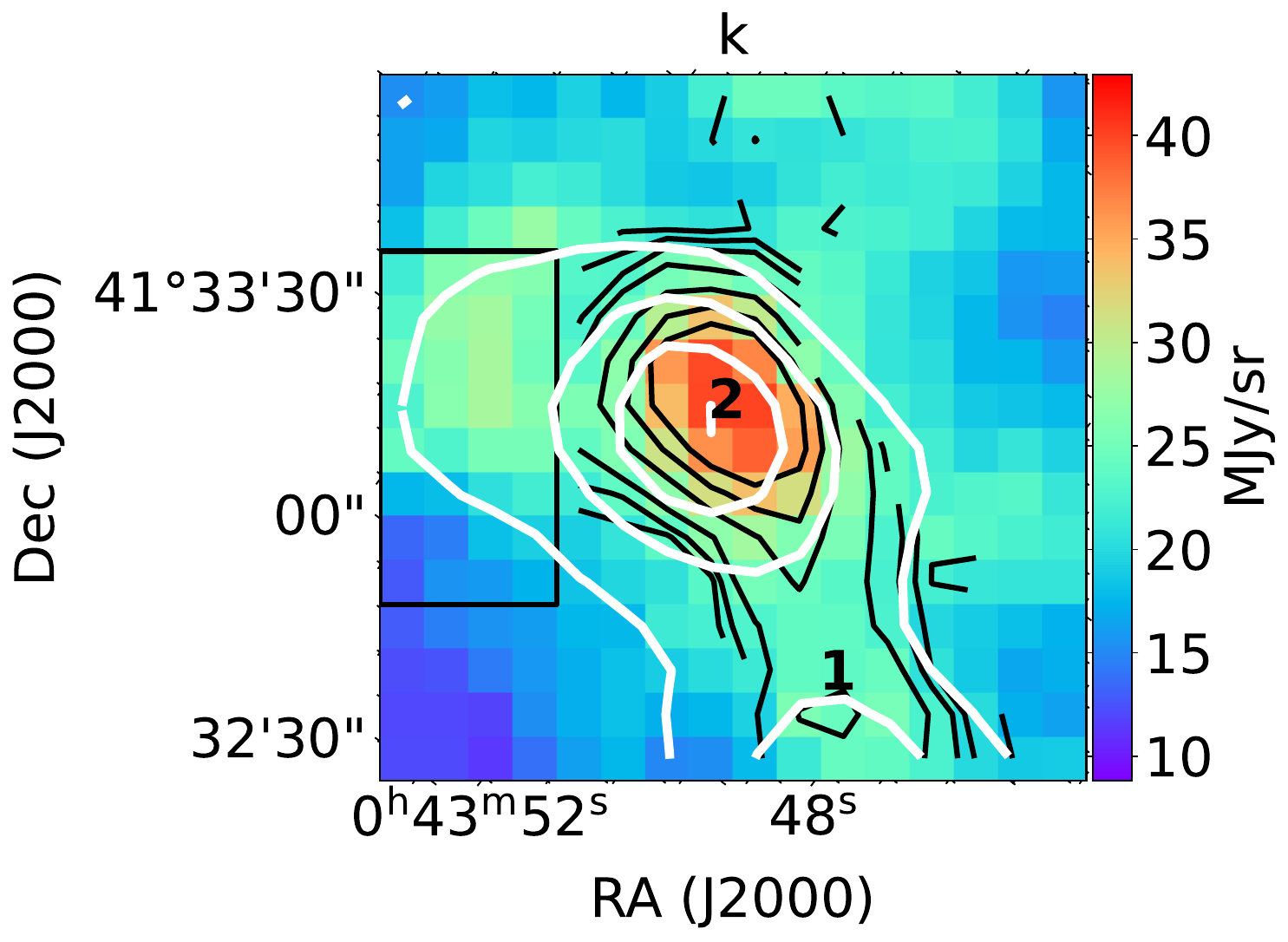}
\end{minipage}
\caption{Contours of the identified clumps of CO(3-2) (black) and CO(1-0) (white) of region a-k. The clumps identified with the CO(3-2) emission lines are labeled with black numbers. The contour levels are 1, 2, 3, 5, 10, 15, and 20 $\sigma$, with 1$\sigma$ corresponding to 0.05 K km s$^{-1}$. The black rectangles represent regions without CO(3-2) observations. The background images are \textit{Herschel}/SPIRE 250 $\upmu$m maps \citep{2012ApJ...756...40S}. Region a has no CO(1-0) data.  \label{fig-cl0}}
\end{figure*}

\begin{figure*}
\vfill
\begin{minipage}{0.51\linewidth}
\centering\includegraphics[width=\textwidth]{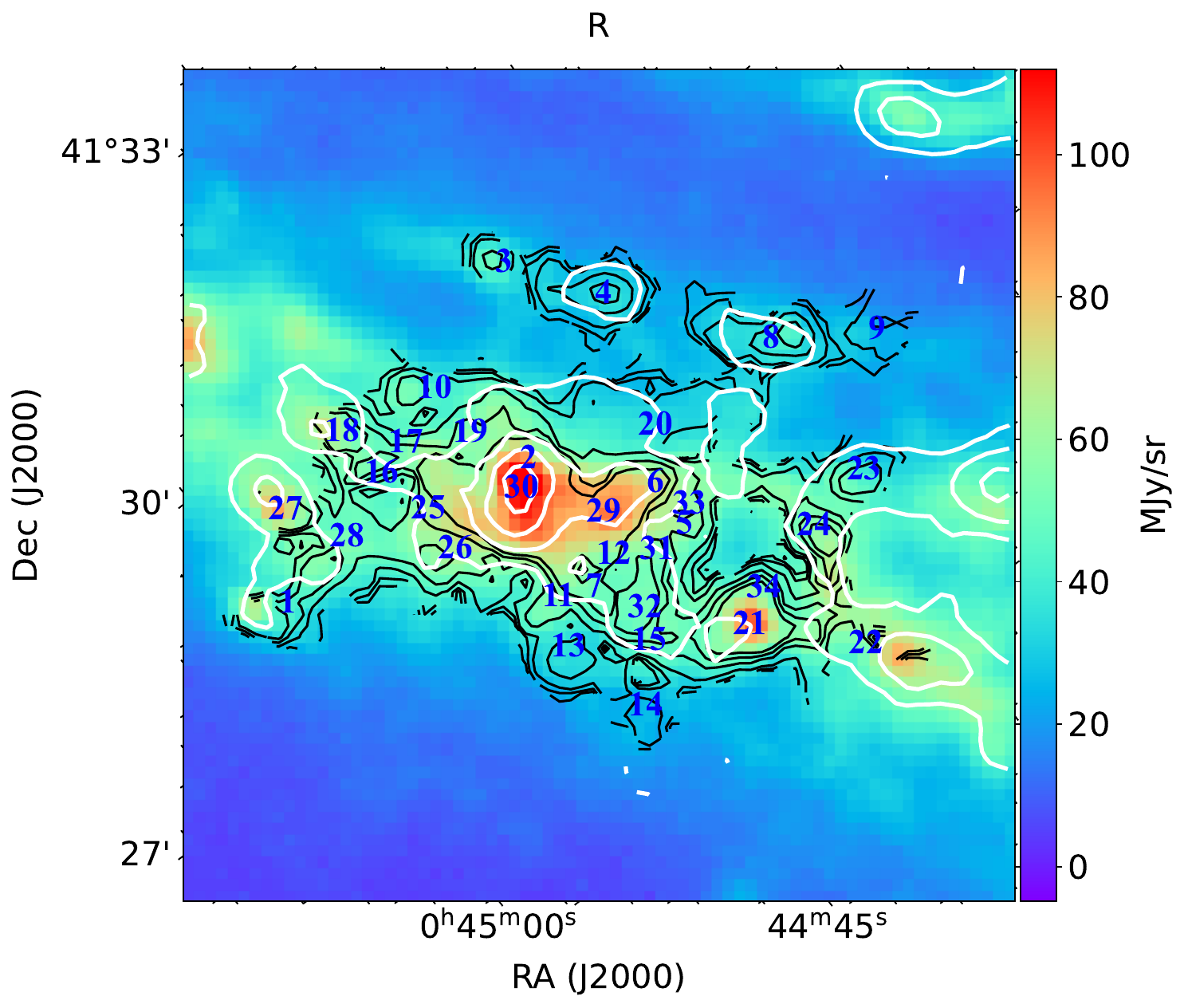}
\end{minipage}
\hfill
\begin{minipage}{0.48\linewidth}
\centering\includegraphics[width=\textwidth]{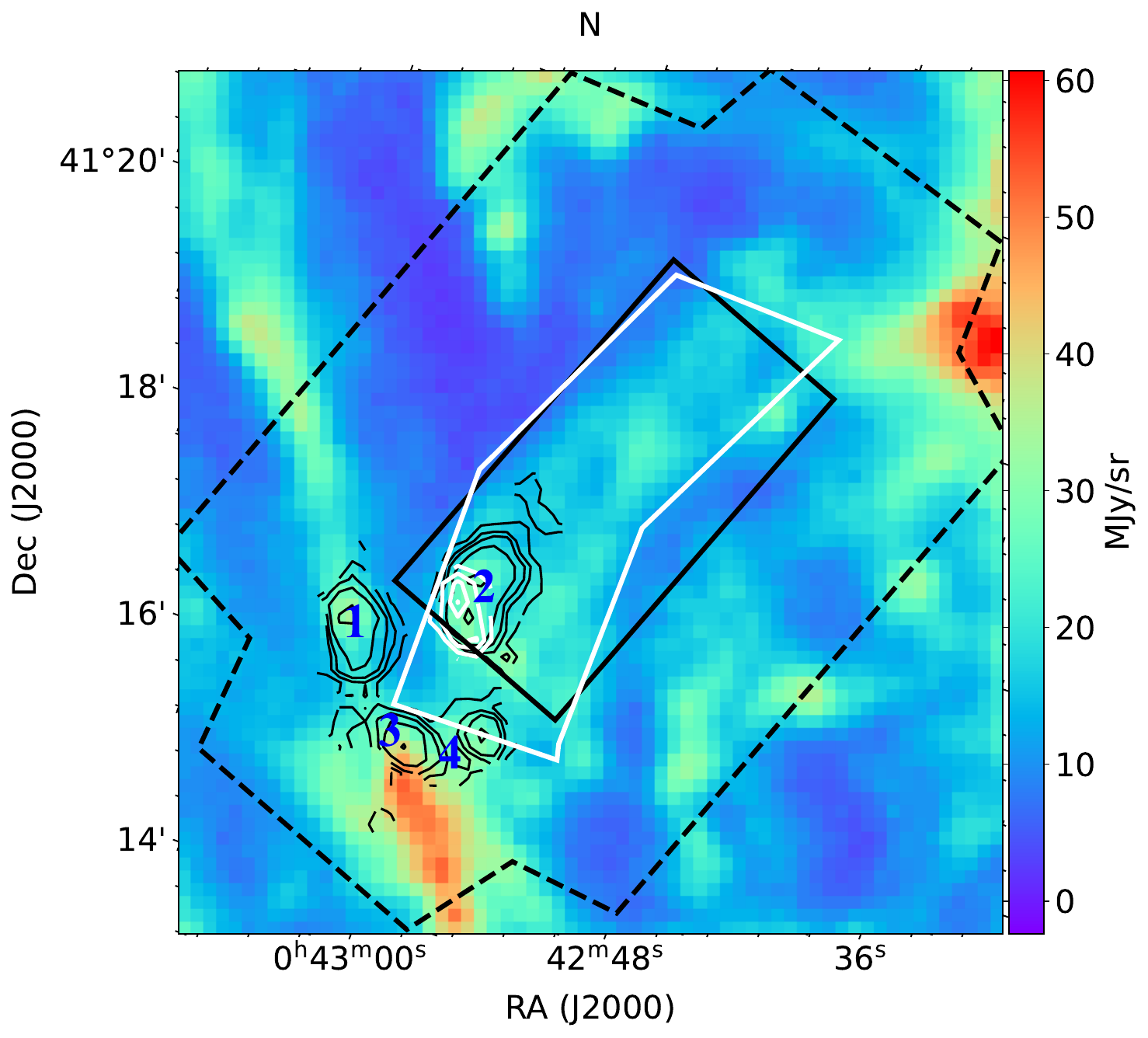}
\end{minipage}
\caption{Contours of the identified clumps of CO(3-2) (black) and CO(1-0) (white) of region R and nuclear ring `N'. The clumps identified with the CO(3-2) emission lines are labeled with blue numbers. In the nuclear ring region, the black dashed polygon represents the field of view (FoV) of CO(3-2) observations, while the black rectangle outlines the region with the highest sensitivity. The white polygon represents the FoV of the CO(1-0) observations. The FoV of CO(3-2) observations in region N is larger than that of CO(1-0). The background images are \textit{Herschel}/SPIRE 250 $\upmu$m maps \citep{2012ApJ...756...40S}.
\label{fig-cl}}
\end{figure*}

In Figures \ref{fig-cl0} and \ref{fig-cl}, we show the contours of the identified clouds in each region for CO(3-2) (black) and CO(1-0) (white). The catalogue of the identified clouds is given in Tables \ref{tab-clumps} and \ref{tab-clumps2}. Given that we use the corrected value from Equations \eqref{eq-2} and \eqref{eq-5}, structures with extrapolated sizes smaller than the rms width of resolution under Gaussian assumption ($\sigma_{v,{\rm chan}}, \sigma_{\rm beam}$) are discarded. As a result, correction for finite spatial beam and velocity channel is made and only robust structures will be retained. However, this selection process can increase discrepancies between CO(3-2) and CO(1-0) due to variations in data quality. In certain regions, the outcomes from the two methods may not align. Nevertheless, the PYCPROPS algorithm offers improved accuracy in avoiding false structures and selecting reliable ones. Consequently, the number of structures identified using PYCPROPS will be lower compared to those identified using ClumpFind. However, the overall distribution of these structures remains consistent with the distribution reported in \citet{2020MNRAS.492..195L}. It is important to note that different transitions reflect distinct physical conditions, which can lead to variations in the identification results. However, we believe that the identified structures are still sufficiently reliable within the context of each transition.

\subsection{GMC Properties}

In this section, we show the distribution of the properties of the clouds with galaxy radius and a comparison between CO(3-2) and CO(1-0).

\subsubsection{Velocity dispersion}

\begin{figure*}
\begin{minipage}{0.32\linewidth}
\centering\includegraphics[width=\textwidth]{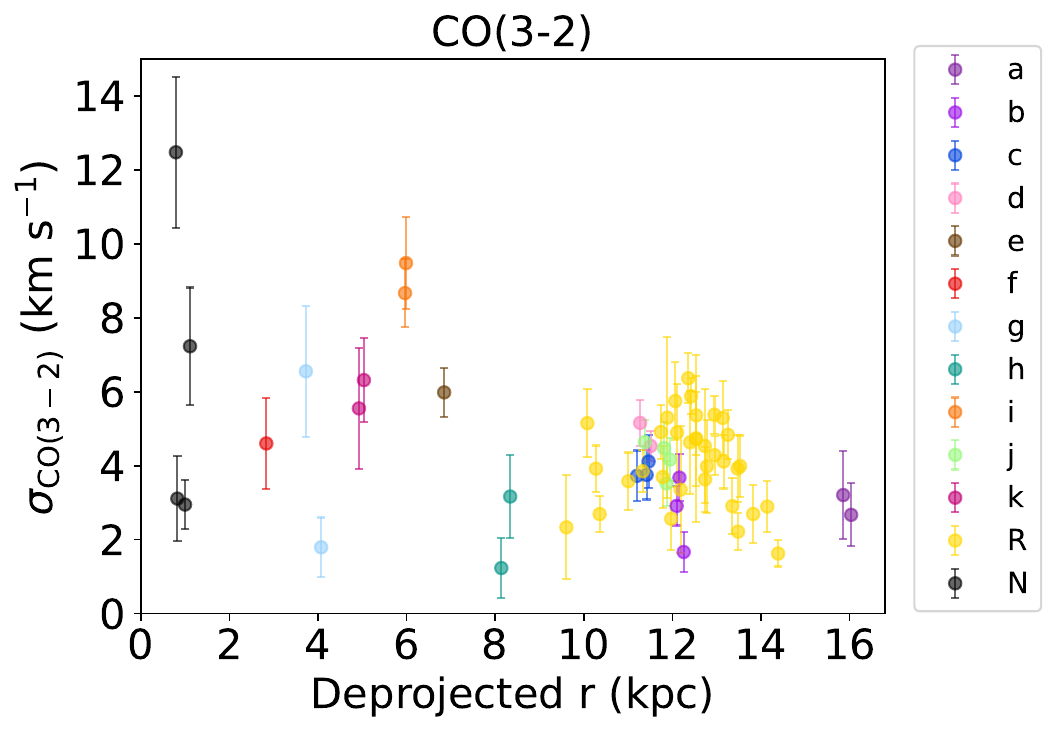}
\end{minipage}
\begin{minipage}{0.32\linewidth}
\centering\includegraphics[width=\textwidth]{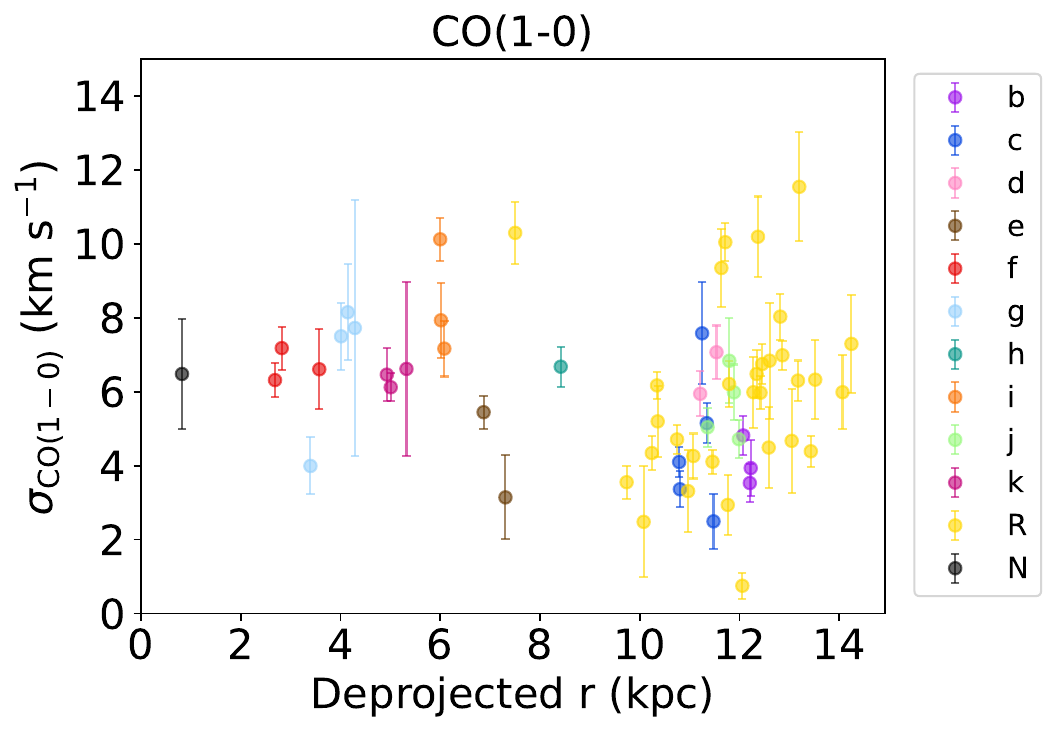}
\end{minipage}
\begin{minipage}{0.32\linewidth}
\centering\includegraphics[width=\textwidth]{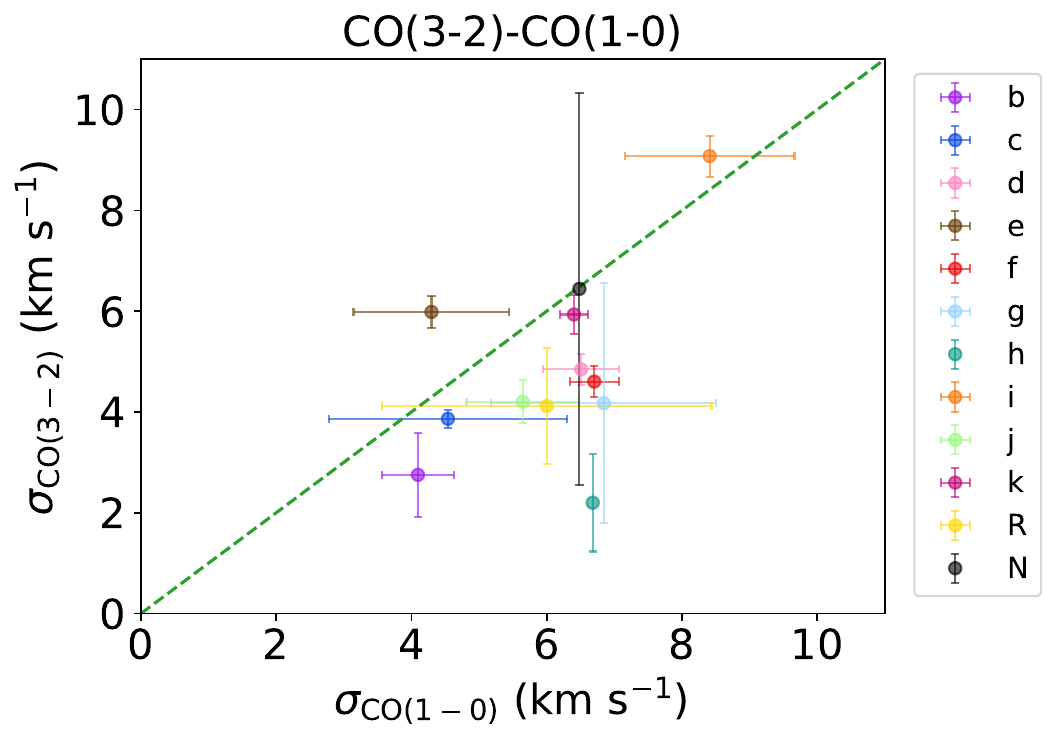}
\end{minipage}
\caption{\textit{Left}: Radial distribution of the velocity dispersion of identified structures from CO(3-2). The error bar represents the estimated error of the structure identified with PYCPROPS. \textit{Middle}: Radial distribution of CO(1-0). \textit{Right}: Comparison of the mean velocity dispersion in each region between CO(3-2) and CO(1-0). The error bar represents the standard deviation of each region. There is no error bar if only one structure exists in the region. The green dashed line represents the locus where the velocity dispersion from two kinds of transitions is equal. \label{fig-velo}}
\end{figure*}

Figure \ref{fig-velo} shows the radial distribution of extrapolated and corrected velocity dispersion and mean value in each region between CO(3-2) and CO(1-0). In the case of CO(3-2), the velocity dispersion ranges from $1.2$ to 12.5 km s$^{-1}$, with the highest value exceeding 10 km s$^{-1}$ within the nuclear ring. A larger scatter is also exhibited in the nuclear ring relative to the disc. 
For CO(1-0), the velocity dispersion spans a range of 0.8 to 11.5 km s$^{-1}$. Due to the low S/N in the nuclear ring, only one structure is identified that shows little difference from the disc (Figure \ref{fig-cl}). Region R shows a large scatter of velocity dispersion due to a relatively large spatial extent. Some structures in regions i, R show large velocity dispersion surpassing 10 km s$^{-1}$. The right panel of Figure \ref{fig-velo} provides a clearer illustration of the trend by comparing the mean values between the two transitions. 
The mean velocity dispersion in region i is the largest, presumably due to the averaging effect in this crowded environment, where component blending along the line of sight is more likely to occur. 
In addition, most of the regions are located below the dashed one-to-one line, 
indicating the velocity dispersion estimated from CO(1-0) is, in general, larger than from CO(3-2).

\subsubsection{Equivalent radius}

\begin{figure*}
\begin{minipage}{0.32\linewidth}
\centering
\includegraphics[width=\textwidth]{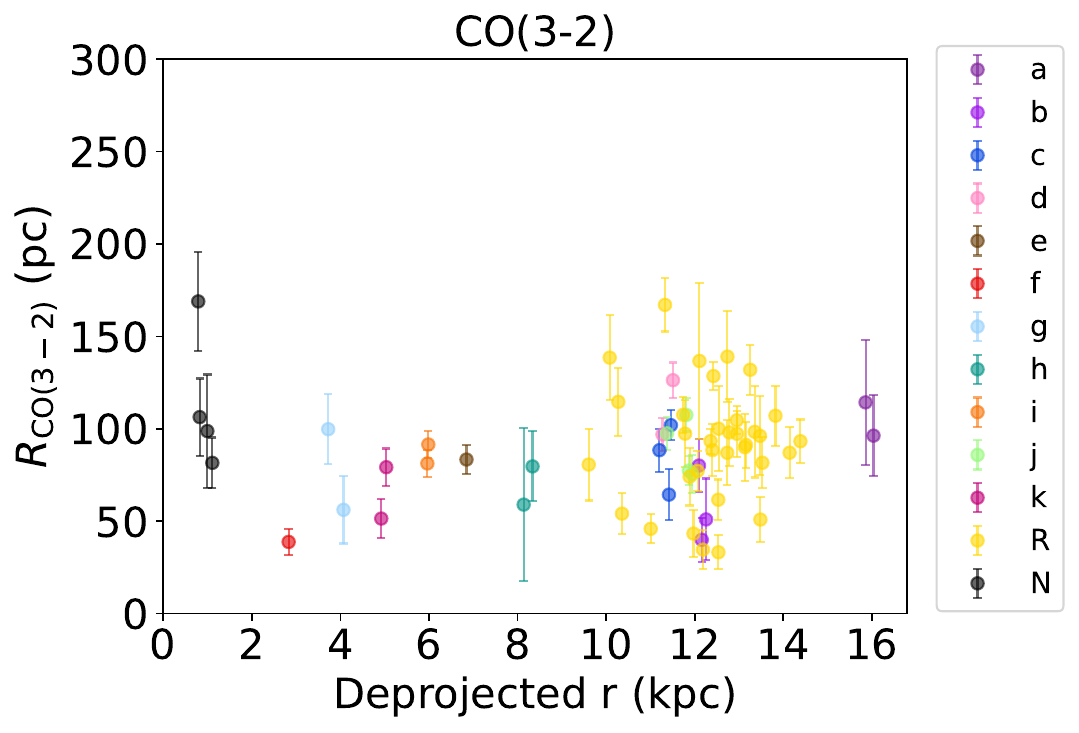}
\end{minipage}
\begin{minipage}{0.32\linewidth}
\centering
\includegraphics[width=\textwidth]{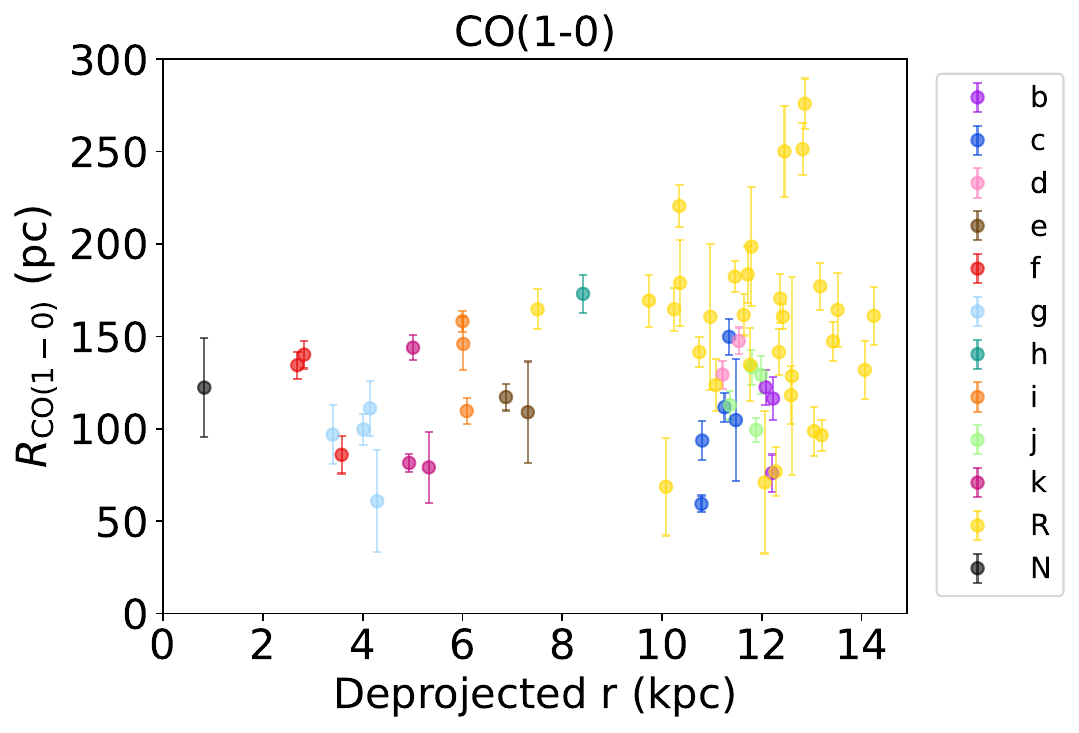}
\end{minipage}
\begin{minipage}{0.32\linewidth}
\centering
\includegraphics[width=\textwidth]{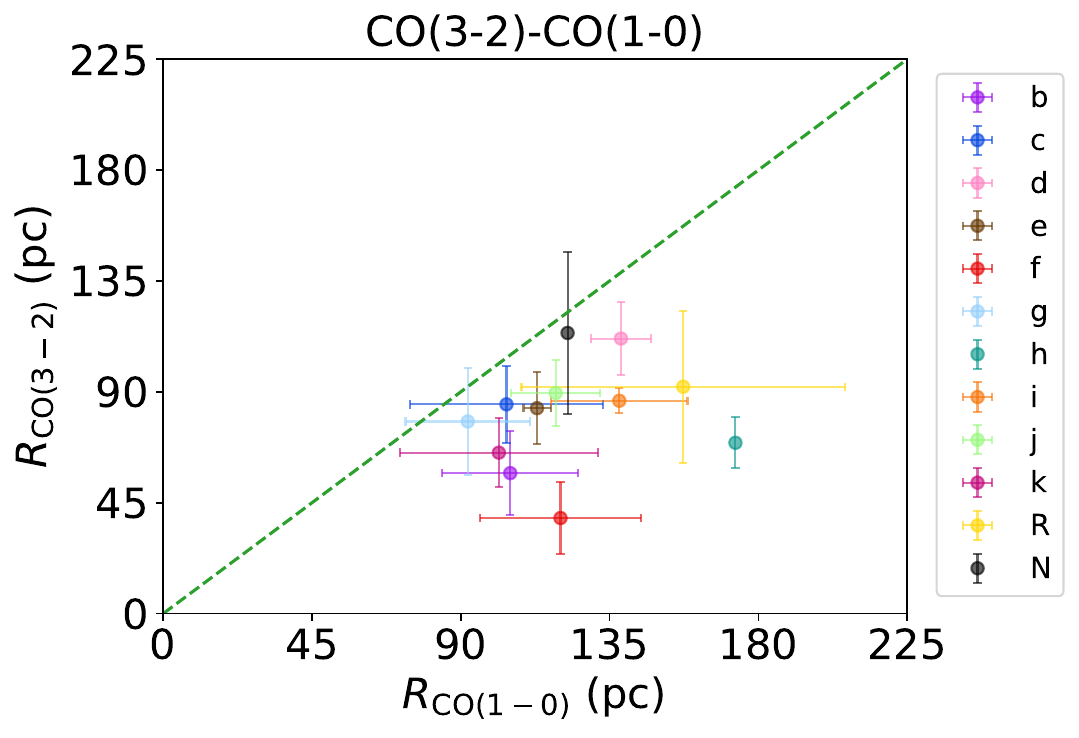}
\end{minipage}
\caption{Radial distribution of equivalent radius of identified structures and average comparison, the same as Figure \ref{fig-velo}. \label{fig-R}}
\end{figure*}

A spherical geometry is assumed since our resolution is smaller than typical molecular gas disc thickness ($\sim 100$ pc). The equivalent radius is thus calculated using equation \eqref{eq-4}, and the results are shown in Figure \ref{fig-R}. The radii show a wide range for both CO(3-2) and CO(1-0), from tens of pc to $\sim300$ pc. One structure in the nuclear ring has a large radius, while others show little difference from the disc. Large radii are mainly in region R for CO(1-0). Mean radii are relatively concentrated in the comparison figure in the right panel, indicating a relatively regular situation across these regions. 
Furthermore, the data points are concentrated below the one-to-one line in the figure, indicating that the estimated radii based on CO(1-0) are generally larger than those based on CO(3-2).

\subsubsection{Mean surface density}

\begin{figure*}
\begin{minipage}{0.32\linewidth}
\centering\includegraphics[width=\textwidth]{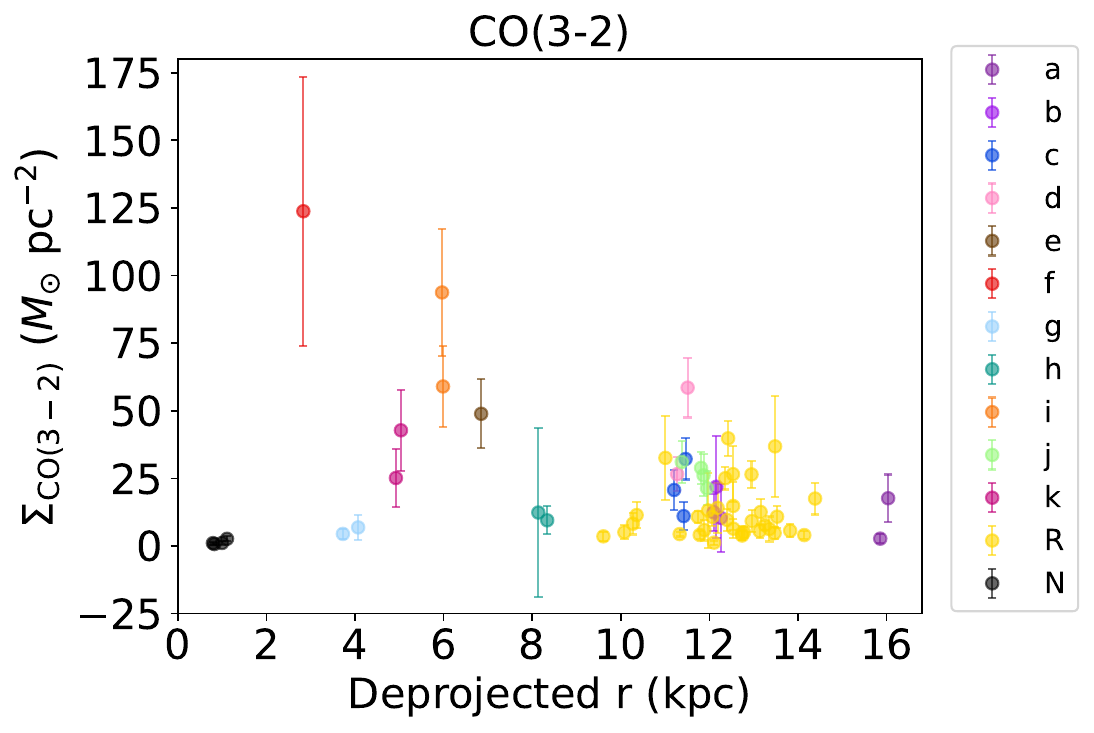}
\end{minipage}
\begin{minipage}{0.32\linewidth}
\centering\includegraphics[width=\textwidth]{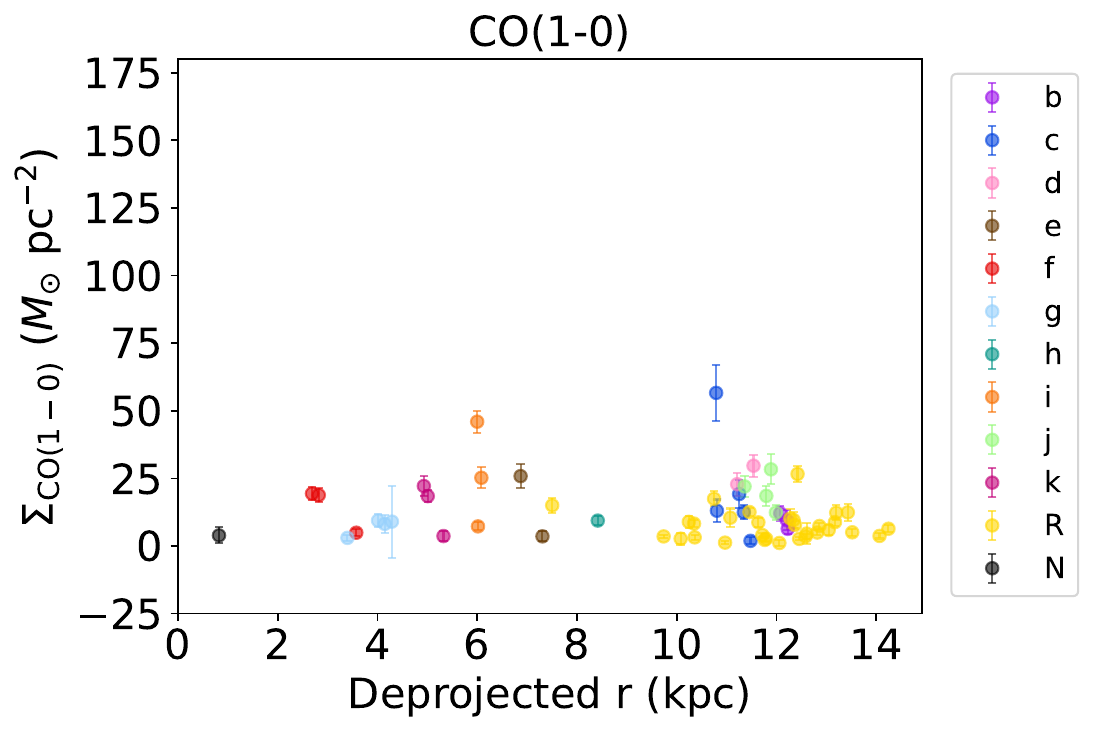}
\end{minipage}
\begin{minipage}{0.32\linewidth}
\centering\includegraphics[width=\textwidth]{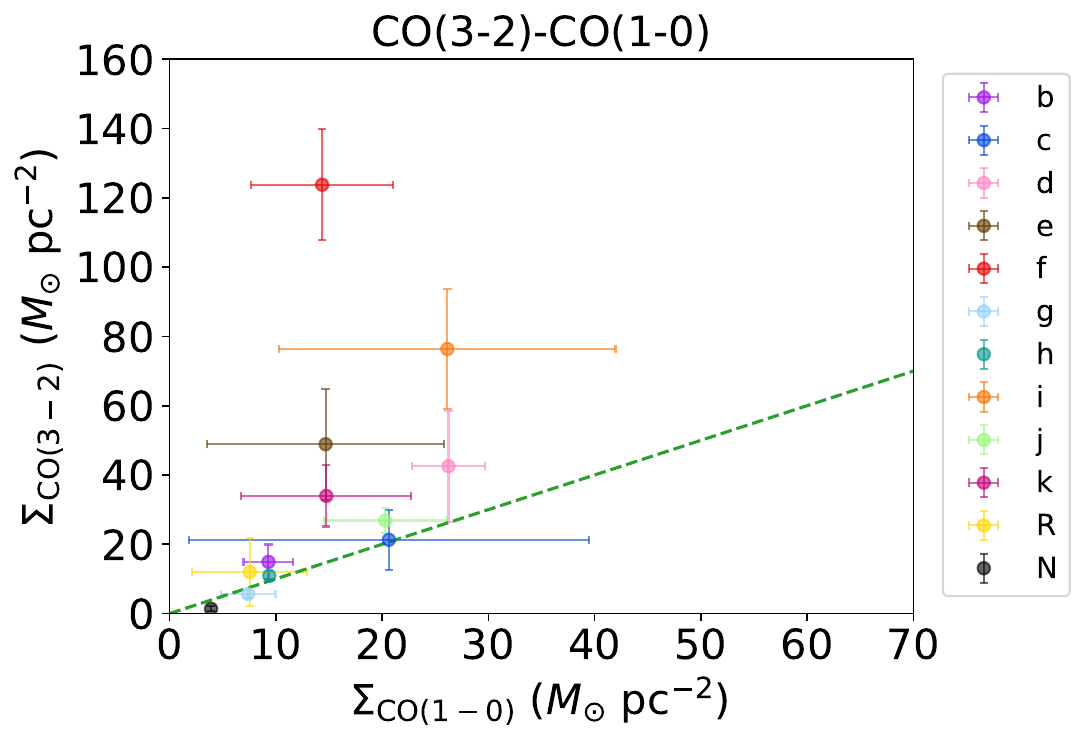}
\end{minipage}
\caption{Radial distribution of mean surface density of identified structures and average comparison. \label{fig-Sig}}
\end{figure*}

The mean surface density of the clouds estimated with Equation \eqref{eq-7} is shown in Figure \ref{fig-Sig}. \citet{2006A&A...453..459N} pointed out that the CO-to-H$_2$ conversion factor in M31 is unlikely to vary much. Therefore, we assume a constant conversion factor $\alpha_{\rm CO}=$ 4.35 $M_{\odot}$ pc$^{-2}$ (K km s$^{-1}$)$^{-1}$ \citep{2013ARA&A..51..207B}. This value is derived from the Milky Way disc and is widely used as a canonical conversion factor \citep{2013ARA&A..51..207B}. The adoption of a constant value can conveniently calculate and compare properties, but the influence of varying conversion factors should also be kept in mind and will be discussed in detail in Section \ref{sec:conversion}. 
The mean surface density remains below 200 $M_{\odot}$ pc$^{-2}$, with a majority around 10 $M_{\odot}$ pc$^{-2}$, and does not exceed 100 $M_{\odot}$ pc$^{-2}$ for CO(1-0). Individual structures show a high surface density in the disc, whereas in the nuclear ring, the mean surface density is significantly lower. It should be noted that certain values in the data set exhibit large errors. Since PYCPROPS does not directly provide errors through bootstrapping, we estimate the error using the error propagation formula. However, it is important to note that the estimated error can be significantly influenced by the values themselves. Hence, care should be taken when considering the error values.
Furthermore, it is observed that the mean surface density derived from CO(3-2) is generally higher than that derived from CO(1-0). The physical meanings for the differences between these two transitions will be discussed in Section \ref{sec: CO transition}.

\subsubsection{Mean internal turbulent pressure}

\begin{figure*}
\begin{minipage}{0.32\linewidth}
\centering\includegraphics[width=\textwidth]{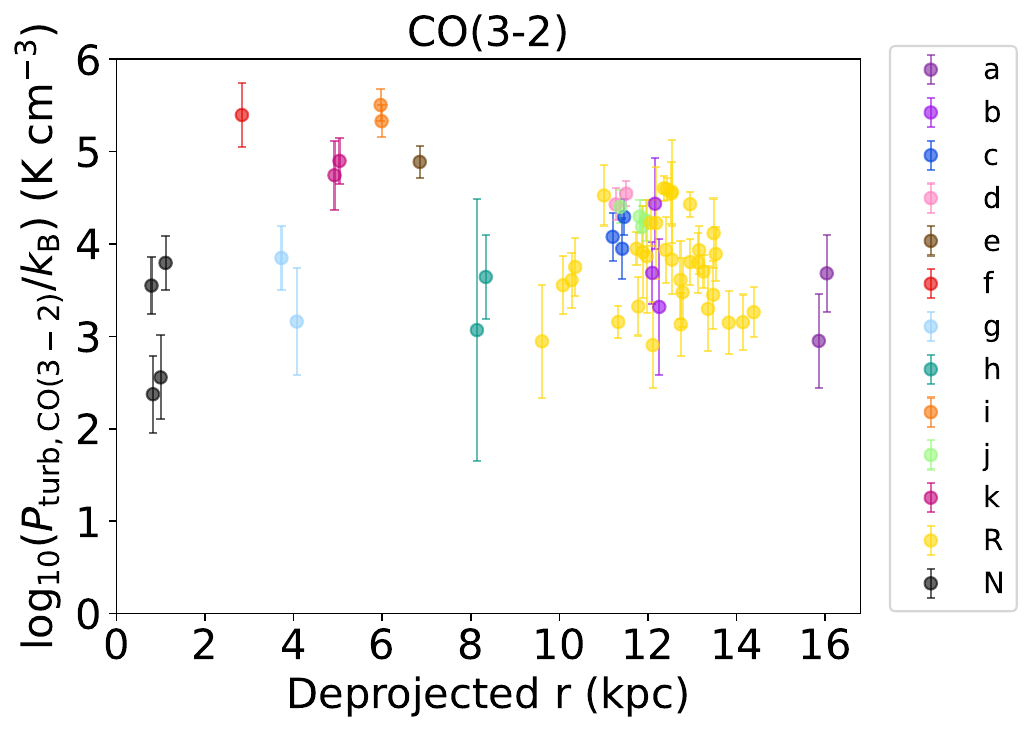}
\end{minipage}
\begin{minipage}{0.32\linewidth}
\centering\includegraphics[width=\textwidth]{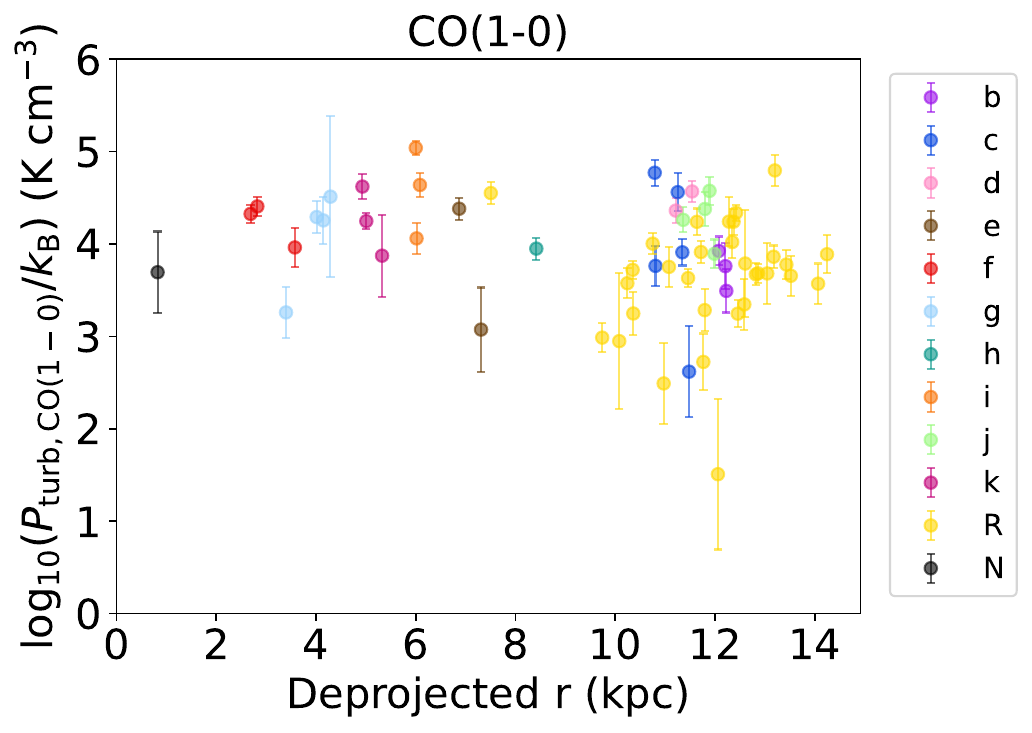}
\end{minipage}
\begin{minipage}{0.32\linewidth}
\centering\includegraphics[width=\textwidth]{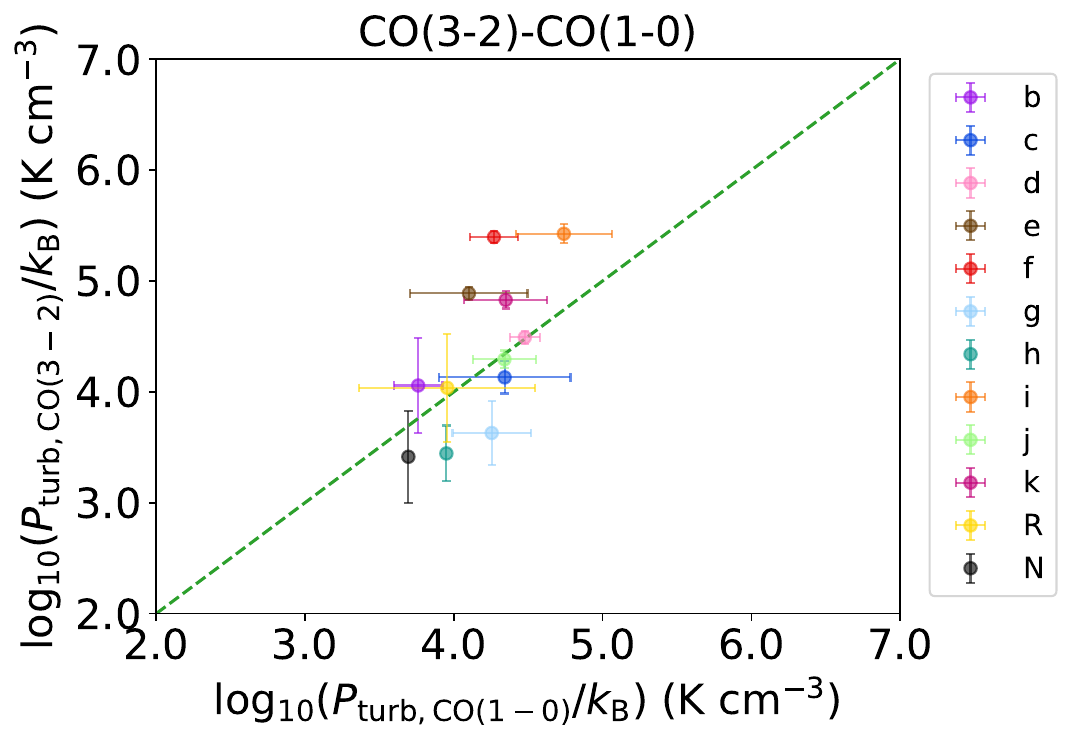}
\end{minipage}
\caption{Radial distribution of mean internal turbulent pressure of identified structures and average comparison. \label{fig-P}}
\end{figure*}

The turbulent pressure reflects the level of gas turbulence. While internal pressure can arise from various sources such as thermal motion and magnetic fields, numerous studies suggest that these contributions are relatively minor compared to turbulence \citep[e.g.][]{1981MNRAS.194..809L,1987ApJ...319..730S}. Therefore, for simplicity, we assume that the turbulent pressure represents the internal pressure, and thus the observed velocity dispersion directly reflects the level of turbulence. 
We also assume spherical clouds with constant density, and isotropic turbulence, using Equation \eqref{eq-9} to calculate the mean turbulent pressure, and the result is presented in Figure \ref{fig-P}.
The turbulent pressure varies greatly with the galactic radius. It ranges from $10^2$ to 10$^6$ K cm$^{-3}$ for CO(3-2), and from 10$^1$ to 10$^5$ K cm$^{-3}$ for CO(1-0). The error is also estimated from the error propagation formula. 
The scatter of turbulent pressure estimated from CO(3-2) is generally larger than from CO(1-0).

\subsection{Relations between properties}\label{sec:relation}

Many observations have found certain relationships among the properties of molecular clouds, which may reveal the condition of the clouds. \citet{1981MNRAS.194..809L} found that there are correlations among the size, velocity dispersion, and mass of molecular clouds, i.e., $\sigma\propto L^\alpha$, $\frac{2GM}{L}\propto \sigma^2$, $\rho\propto L^\beta$, which provide some understanding of the physical conditions of clouds. The first relation, also called the "size-linewidth relation", is thought to reflect properties of turbulence, $\sigma\propto L^\alpha$, where the different index $\alpha$ represents different kinds of turbulence. For example, $\alpha=\frac{1}{2}$ for highly compressible turbulence and strong shock dominant, $\alpha=\frac{1}{3}$ for incompressible turbulence \citep{1981MNRAS.194..809L,1991RSPSA.434....9K,2016ApJ...822...11P}. Larson obtained $\alpha=0.38$, which is often explained as subsonic turbulence in incompressible flows. The second relation indicates self-gravitational equilibrium. The third relation indicates a constant column density for all molecular clouds, with $\beta\sim -1$ obtained by Larson. The "Larson's Laws" are ceaselessly reexamined. \citet{1987ApJ...319..730S} found $\alpha\sim0.5$ using data from 273 clouds, and many other studies also support the result, suggesting that turbulence in ISM is often compressible and supersonic. This is also supported by the fractal dimension found in the Small Magellanic Cloud (SMC) and Large Magellanic Cloud \citep[LMC;][]{2018ApJ...860..172S, 2022MNRAS.512.1196M}, indicating supersonic turbulence in star-forming clouds. \citet{2009ApJ...699.1092H} observed $^{13}$CO emission lines of a range of molecular clouds and found that the column density of the clouds is varied so that the velocity dispersion also depends on the column density, $\sigma=(\frac{\pi G}{5}R\Sigma)^{0.5}$, which reflects the clouds in virial equilibrium. Other extragalactic studies also found the same dependence \citep[e.g.][]{2016ApJ...831...16L}. The slope of the relation is still in debate. \citet{2021MNRAS.500.5268I} simulated the evolution of molecular clouds in the galactic disc and found that $\alpha$ varies from 0.3 to 1.2. 

In this section, we will explore the relations between cloud properties and compare them with those of other studies.

\subsubsection{Size-linewidth relation}

\begin{figure*}
\begin{minipage}{0.48\linewidth}
\centering\includegraphics[width=\textwidth]{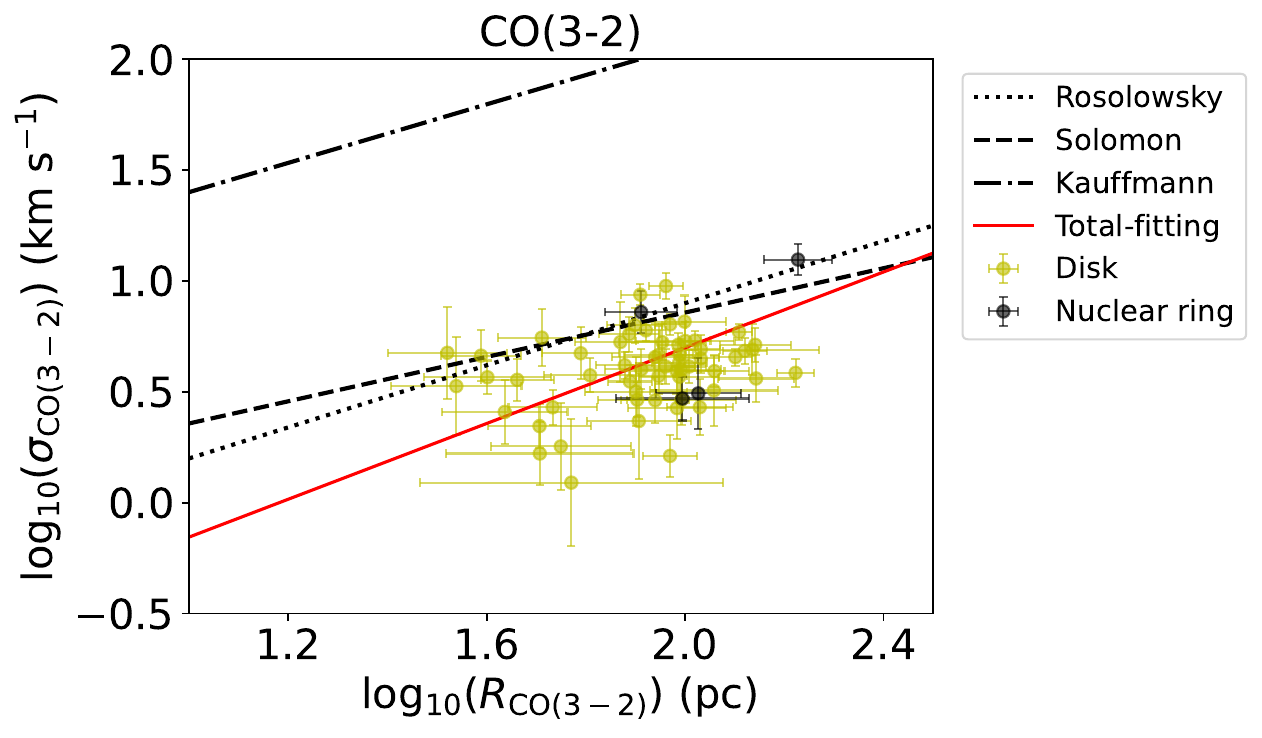}
\end{minipage}
\begin{minipage}{0.48\linewidth}
\centering\includegraphics[width=\textwidth]{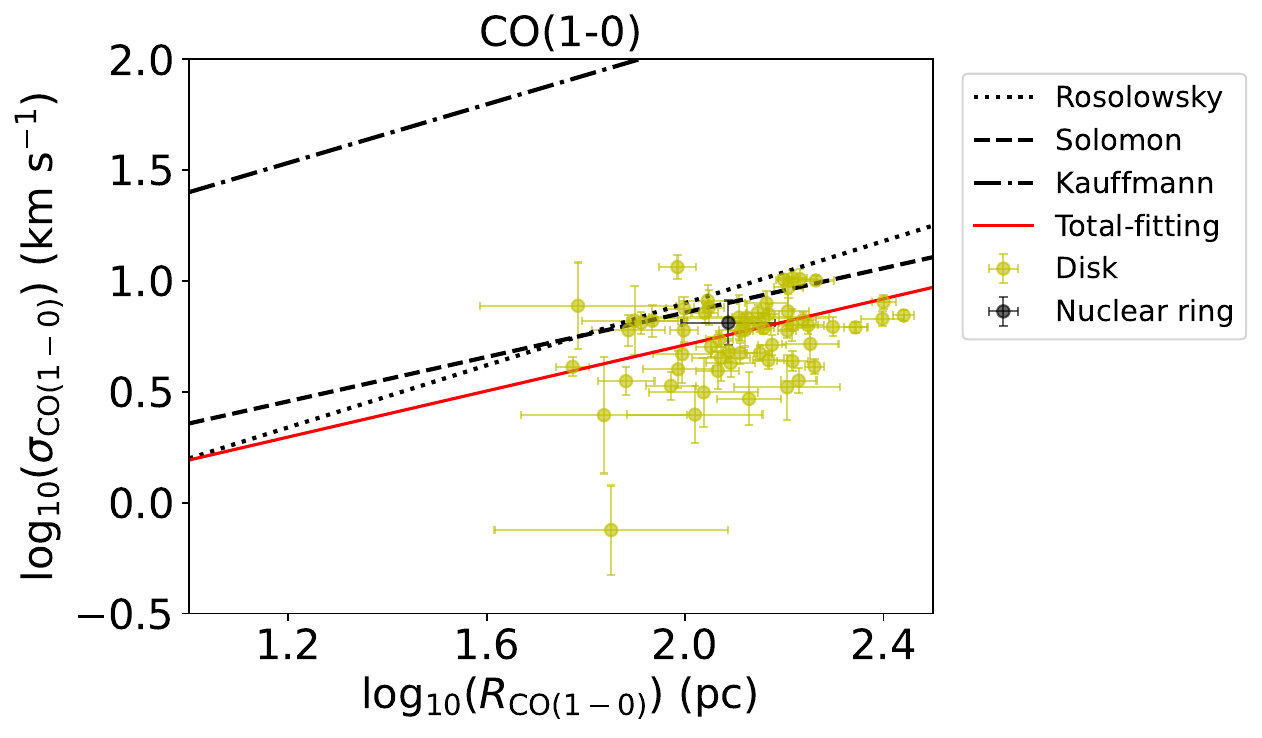}
\end{minipage}
\caption{\textit{Left}: The size-linewidth relationship from CO(3-2). The yellow and black points represent data in the disc and nuclear ring, respectively. Dotted, dashed, and dash-dotted black lines are relations derived from a spiral arm in M31 \citep{2007ApJ...654..240R}, $\sigma_v=0.3R^{0.7}$, from the Milky Way disc \citep{1987ApJ...319..730S}, $\sigma_v=0.72R^{0.5}$, and from the central molecular zone of the Milky Way \citep{2017A&A...603A..89K}, $\sigma_v=5.5R^{0.66}$. The solid red line is the best-fit line for all data, including both the disc and nuclear ring. The parameters are listed in Table \ref{tab-res}. \textit{Right}: The size-linewidth relation is derived from CO(1-0), with symbols the same as the left panel. 
\label{fig-sr}}
\end{figure*}

\citet{1981MNRAS.194..809L} pointed out a relation between size and velocity dispersion, $\sigma_v=\rm C{\it R}^{\rm a}$, and different studies attained diverse $\rm C$ and $\rm a$. Figure \ref{fig-sr} shows our data compared to relations from other studies. Data from CO(3-2) is highly similar to CO(1-0). Both deviate far from the central molecular zone of the Milky Way and, in principle, coincide with the Milky Way disc but are a little lower. There is also little difference between the nuclear ring and the disc. We utilized the least-squares fitting method implemented in the kmpfit module in the Python package {\sc kapteyn} to fit the data, which accounts for uncertainties in both quantities. The results are listed in Table \ref{tab-res}. The CO(1-0) data in the nuclear ring cannot be fitted since only a single data point exists. The limited number of data points available in this region is possibly attributed to the deficiency of molecular gas in the nuclear region \citep{2019MNRAS.484..964L}. We calculate the Spearman's coefficient and the $p$ value, which are 0.24, 0.06 for CO(3-2), and 0.28, 0.03 for CO(1-0), suggesting a weak positive correlation. The slope of the fitted size-linewidth relation of CO(1-0) is 0.52, consistent with the canonical value \citep{1987ApJ...319..730S}, while that of CO(3-2) is steeper. The disc exhibits a slope of 0.61, while including the nuclear region results in an even steeper slope of 0.85. We caution that this high value is largely influenced by a single cloud with the highest $\sigma_{\rm CO(3-2)}$. Nevertheless, even excluding this point results in a slope of 0.64, which remains steeper than the canonical value 0.5. A steep slope could indicate that the turbulence is highly compressible. 
Some studies also got steeper relations, e.g. 0.6 in a variety of extragalactic
systems \citep{2008ApJ...686..948B} as well as the entire Galactic plane \citep{2017ApJ...834...57M}, 0.8 in the LMC \citep{2011ApJS..197...16W} and a nearby dwarf galaxy NGC 404 \citep{2022MNRAS.517..632L}, 1.2 in the nearby barred spiral galaxy NGC 5806 \citep{2023MNRAS.522.4078C}. On the other hand, a high-resolution survey based on JCMT CO(3-2) observations toward the Milky Way first quadrant reported a shallower slope of 0.3 \citep{2019MNRAS.483.4291C}, indicative of subsonic turbulence. 

\subsubsection{Size-linewidth-surface density relation and virial parameter}

\begin{figure*}
\begin{minipage}{0.48\linewidth}
\centering\includegraphics[width=\textwidth]{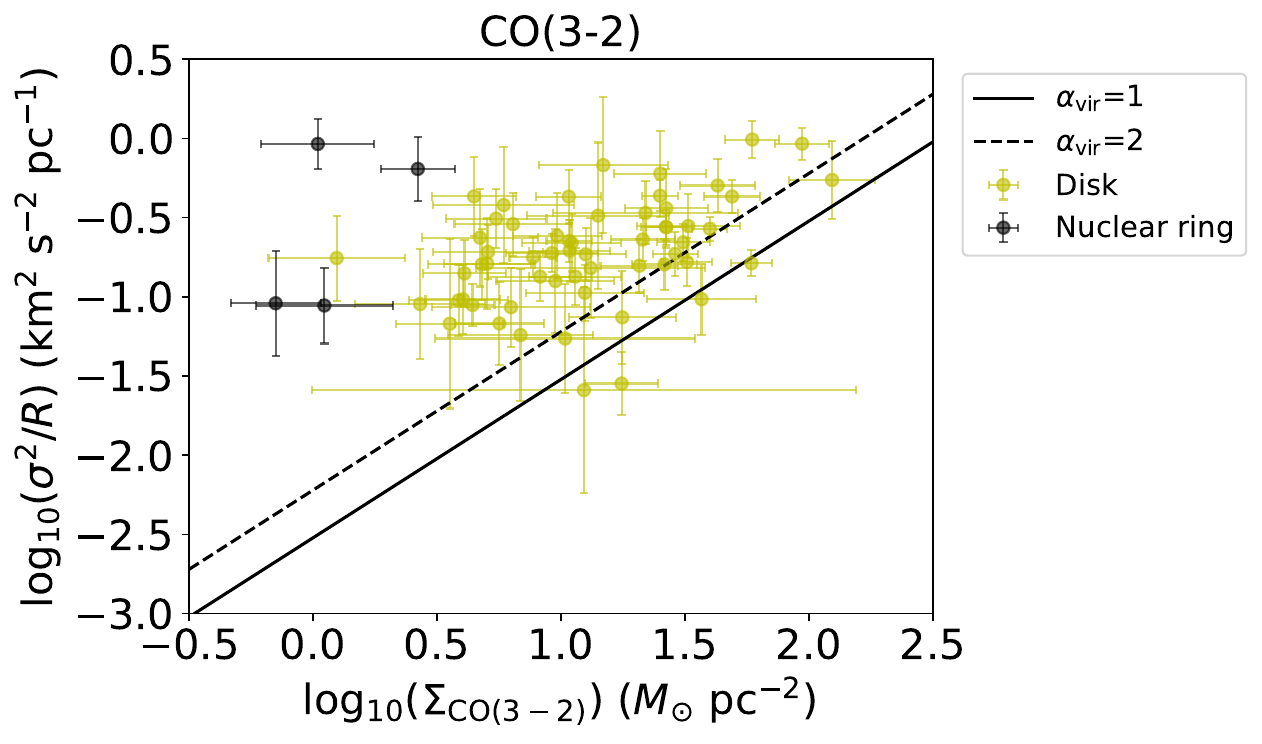}
\end{minipage}
\begin{minipage}{0.48\linewidth}
\centering\includegraphics[width=\textwidth]{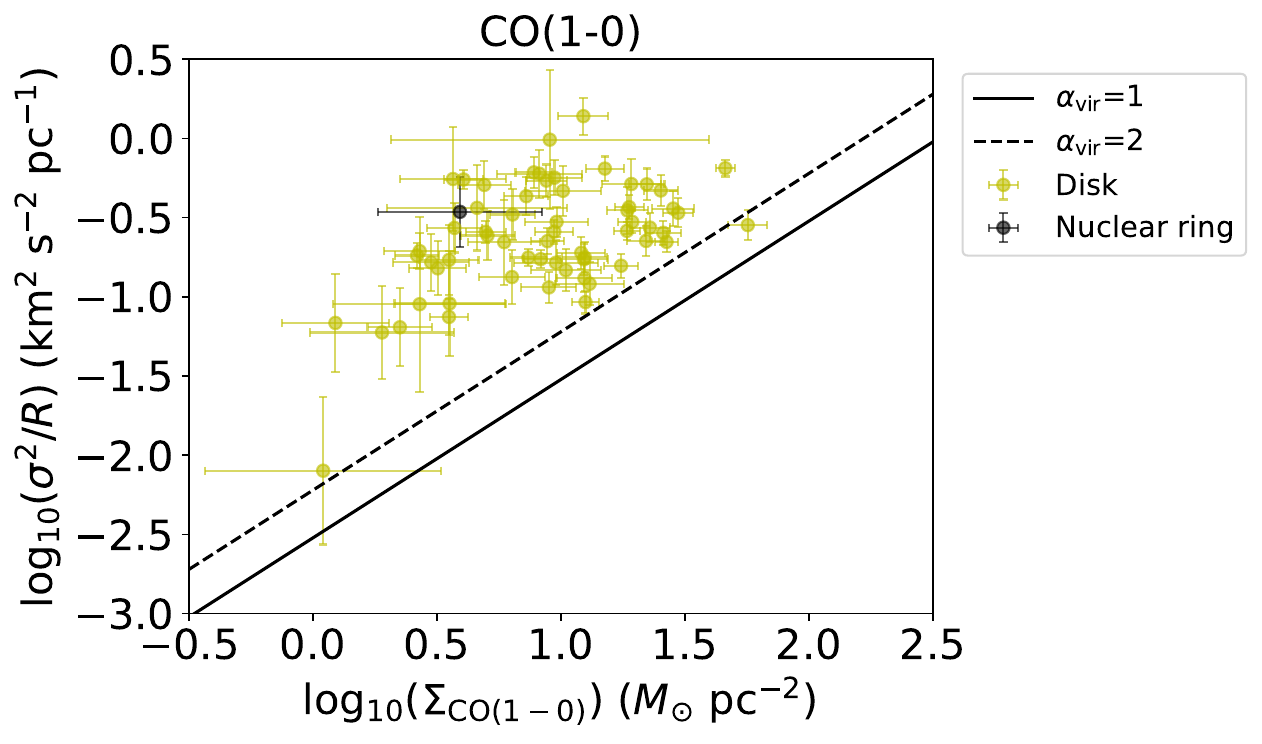}
\end{minipage}
\caption{\textit{Left}: The size-linewidth-surface density relation from CO(3-2). Solid and dashed lines represent $\alpha_{\rm vir}=1$, virial equilibrium, $\alpha_{\rm vir}=2$, marginally bound. \textit{Right}: The size-linewidth-surface density relation from CO(1-0). \label{fig-sss}}
\end{figure*}

More universally, the incorporation of surface density into the size-linewidth relation \citep{2009ApJ...699.1092H}, as depicted in Figure \ref{fig-sss}, allows for a concise description of Larson's scaling relations. This size-linewidth-surface density relation provides a more accurate representation of the clouds' physical condition.
Additionally, the slope of the relation is a direct reflection of the virial parameter $\alpha_{\rm vir}$ (Equation \eqref{eq-10}). 
$\alpha_{\rm vir}=1$ represents virial equilibrium without surface pressure or magnetic support, while $\alpha_{\rm vir}=2$ represents marginally bound clouds \citep{2018ApJ...860..172S}, as shown by solid and dashed lines in Figure \ref{fig-sss}. Most of the CO(3-2) data points are located above the two lines, albeit with significant scatter. Notably, the four data points corresponding to the nuclear ring deviate greatly from the disc points toward the upper left corner. 
On the other hand, CO(1-0) data are relatively concentrated and collectively located above the two lines. 

To quantify the dynamical state of different environments, we fit the virial parameter $\alpha_{\rm vir}=\frac{5\sigma_v^2R}{fGM_{\rm gas}}$ of the disc and the nuclear ring respectively using the {\sc kapteyn} package. The results are listed in Table \ref{tab-res}. Here, we adopt $f=\frac{10}{9}$, corresponding to a density profile $\rho \propto R^{-1}$. All fitted $\alpha_{\rm vir}$ are greater than $1$. For CO(3-2), $\alpha_{\rm vir}$ is 3.59 in the disc, indicating that the clouds are slightly unbound. As for the nuclear ring, $\alpha_{\rm vir}\gg1$, suggesting that the gas is unbound and highly turbulent. It is noteworthy that the error for this value is quite large, which may be partly due to the small number of points. On the contrary, the $\alpha_{\rm vir}$ estimated from CO(1-0) is approximately twice the CO(3-2) estimates, indicating an unbound state of the molecular clouds traced by CO(1-0).

\begin{figure*}
\begin{minipage}{0.32\linewidth}
\centering\includegraphics[width=\textwidth]{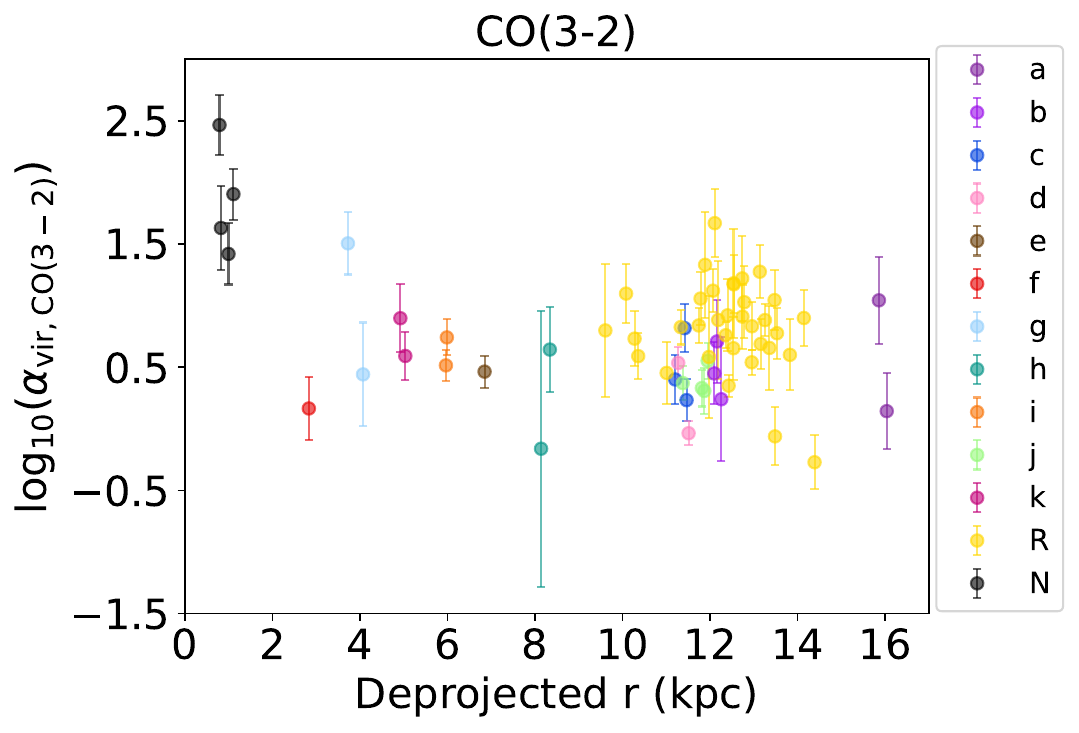}
\end{minipage}
\begin{minipage}{0.32\linewidth}
\centering\includegraphics[width=\textwidth]{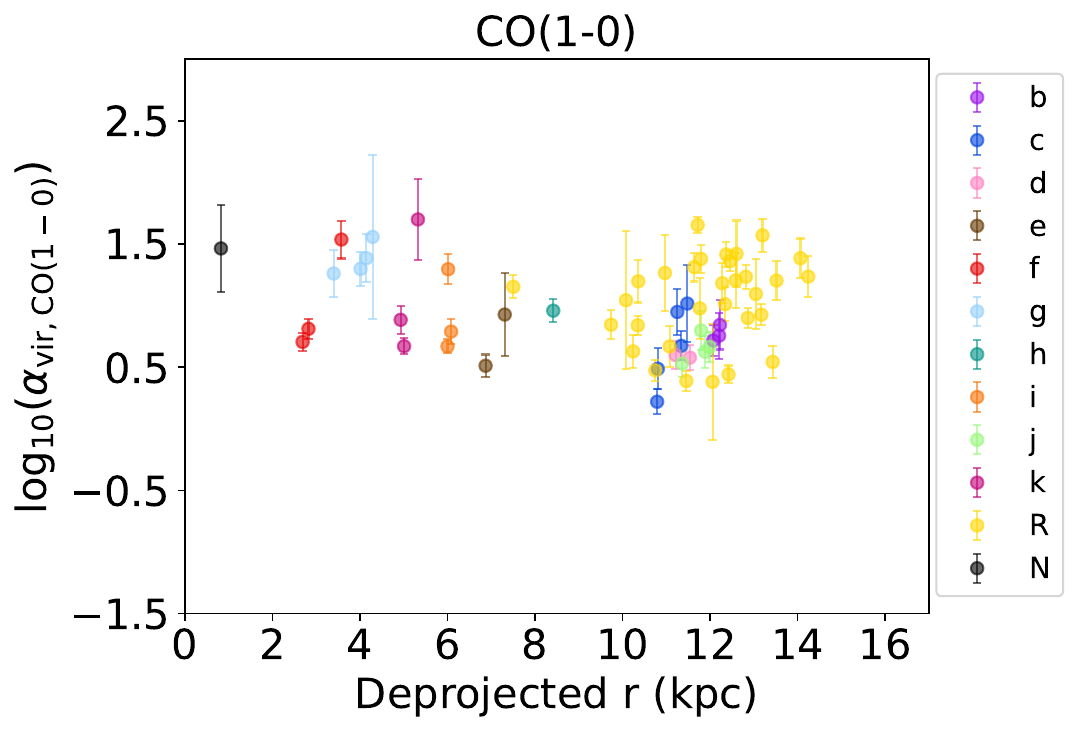}
\end{minipage}
\begin{minipage}{0.32\linewidth}
\centering\includegraphics[width=\textwidth]{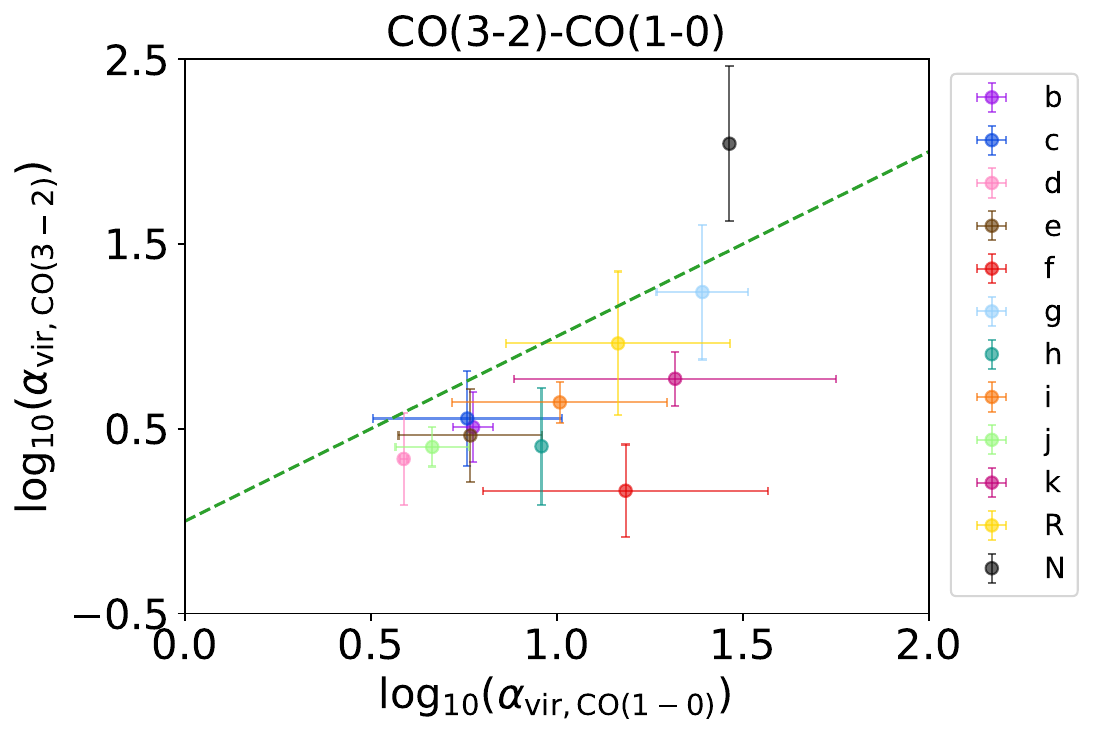}
\end{minipage}
\caption{Radial distribution of virial parameter of identified structures and average comparison. \label{fig-al}}
\end{figure*}

We also calculated the virial parameters of individual clouds, and the resultant radial distribution is shown in Figure \ref{fig-al}. The virial parameter of the nuclear ring clouds derived from CO(3-2) data is collectively higher than that in the disc, while for CO(1-0), there is no clear trend and the data points exhibit larger scatter.

\begin{figure*}
\begin{minipage}{0.48\linewidth}
\centering\includegraphics[width=\textwidth]{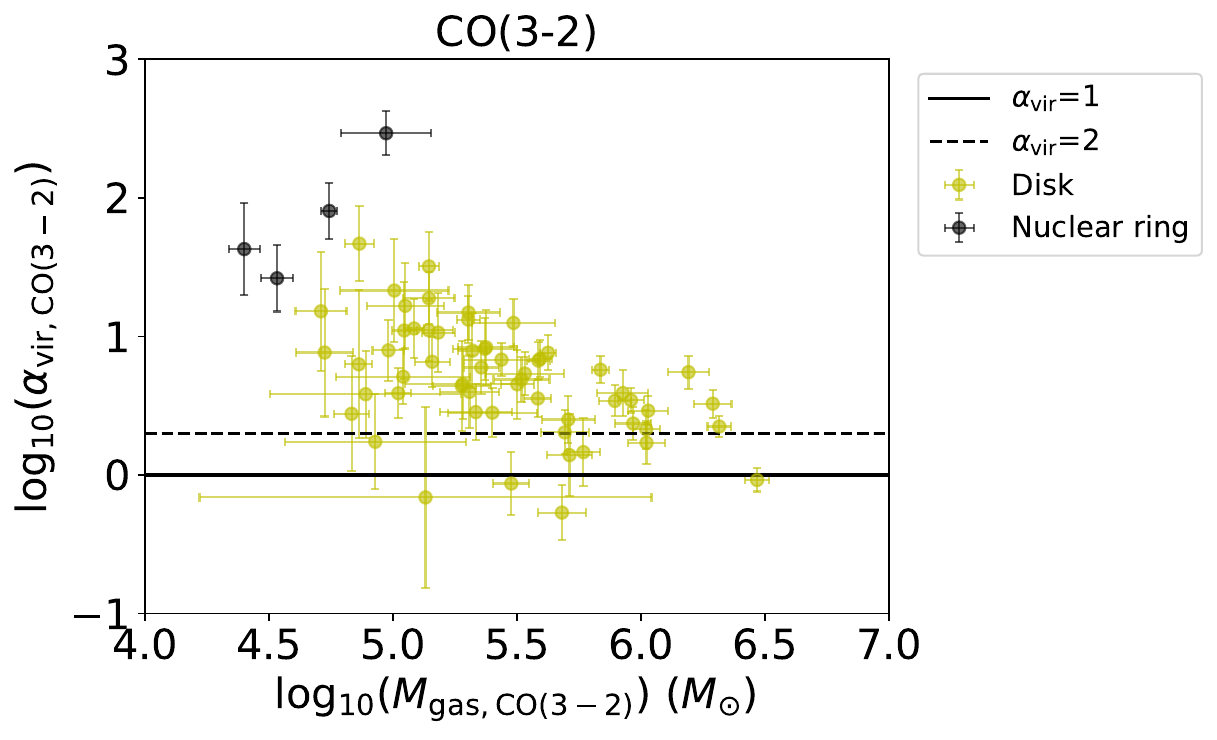}
\end{minipage}
\begin{minipage}{0.48\linewidth}
\centering\includegraphics[width=\textwidth]{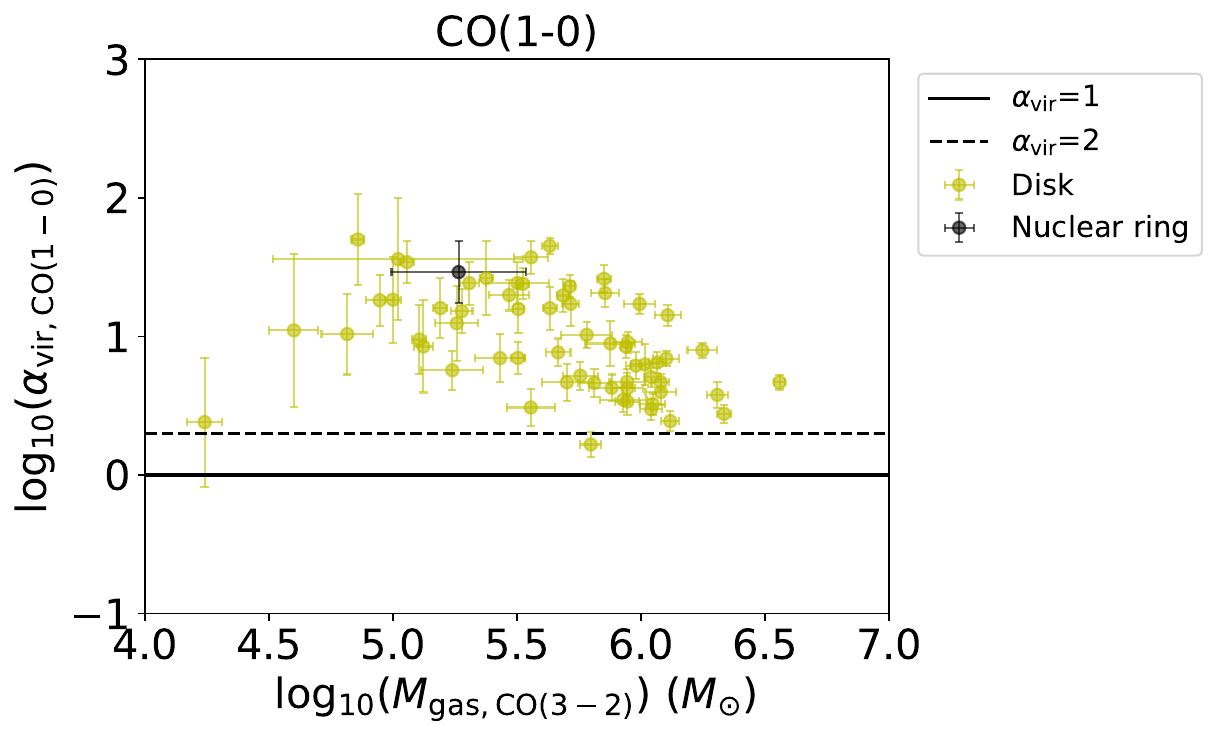}
\end{minipage}
\caption{\textit{Left}: Virial parameter vs. gas mass  from CO(3-2). Solid and dashed lines represent $\alpha_{\rm vir}=1$, virial equilibrium, $\alpha_{\rm vir}=2$, marginally bound. \textit{Right}: Virial parameter vs. gas mass of CO(1-0). \label{fig-vs}}
\end{figure*}

Figure \ref{fig-vs} displays the distribution of the virial parameter as a function of the cloud mass. The majority of the data points lie above the line representing $\alpha\rm_{vir}=2$. Notably, the points from the nuclear ring of CO(3-2) are situated in the upper left corner. Overall, the CO(1-0) data show little difference compared to CO(3-2). Additionally, there is a clear trend of decreasing $\alpha\rm_{vir}$ with increasing gas mass. The Spearman's coefficient and $p$ value are $-$0.59, $10^{-7}$ for CO(3-2) and $-$0.53, $10^{-6}$ for CO(1-0). A similar anticorrelation has been observed in the Milky Way \citep{2016A&A...585A.117Z, 2018MNRAS.475.2215V}, M33 \citep{2023ApJ...953..164M}, as well as in NGC404 \citep{2022MNRAS.517..632L}, suggesting that more massive clouds tend to be more gravitationally bound.

\subsubsection{Size-linewidth-surface density relation with external pressure}

\begin{figure*}
\begin{minipage}{0.48\linewidth}
\centering\includegraphics[width=\textwidth]{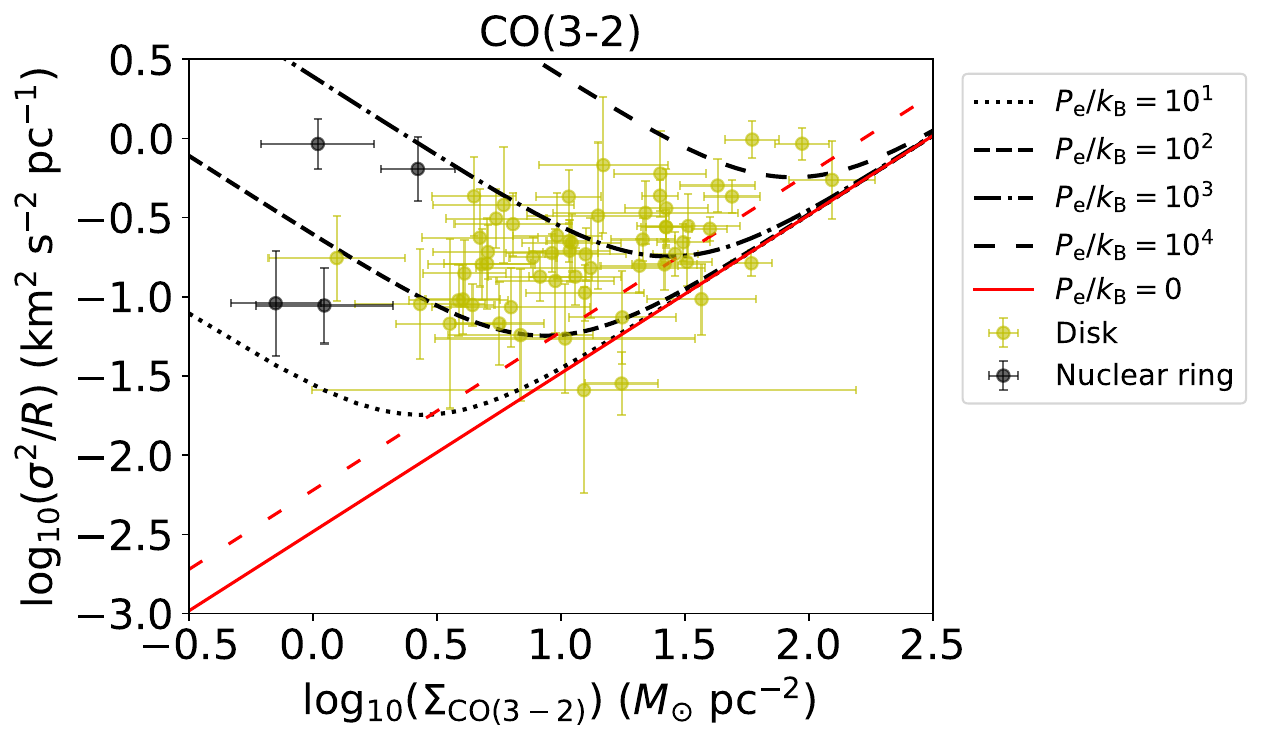}
\end{minipage}
\begin{minipage}{0.48\linewidth}
\centering\includegraphics[width=\textwidth]{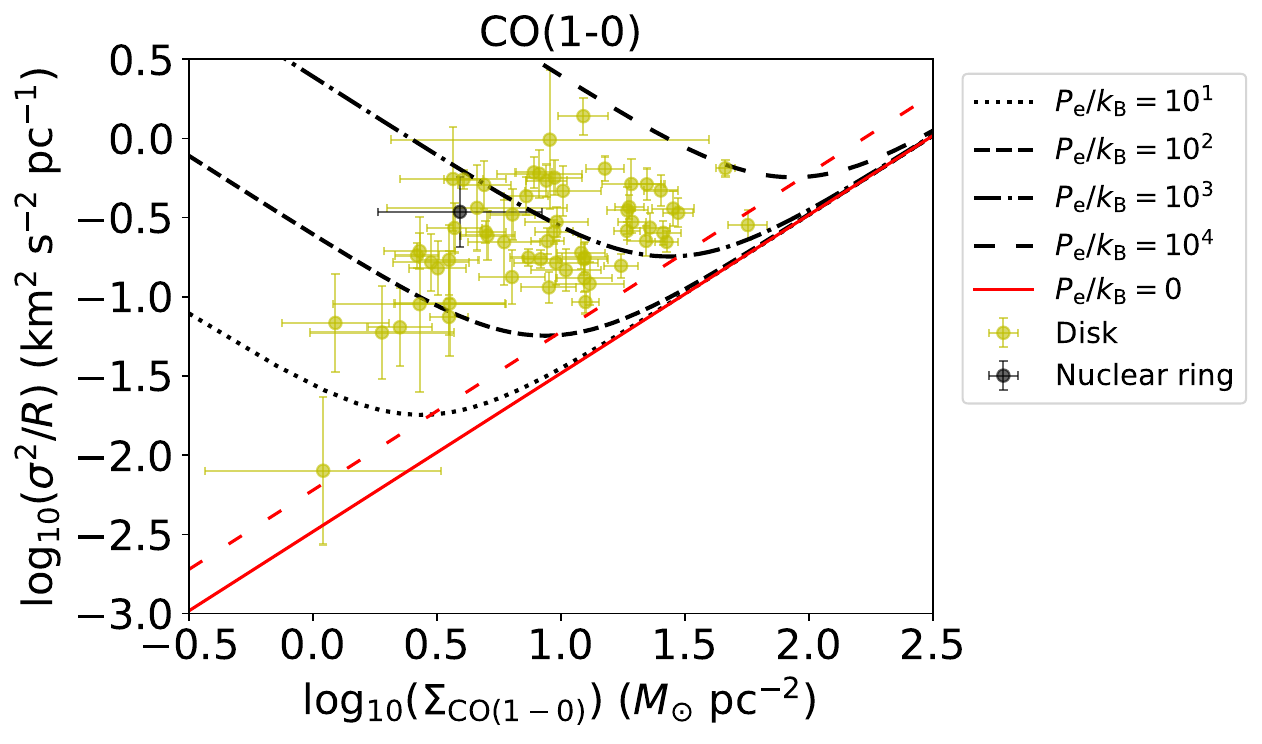}
\end{minipage}
\caption{\textit{Left}: Size-linewidth-surface density relation for CO(3-2) incorporates external pressure. The straight line represents the condition without external pressure $P_{\rm e}$, corresponding to virial parameter $\alpha_{\rm vir}$=1. And the red dashed line corresponding to $\alpha_{\rm vir}$=2. Furthermore, four V-shaped lines correspond to $P_{\rm e}/k_{\rm B}$=10$^1$, 10$^2$, 10$^3$, 10$^4$ K cm$^{-3}$, respectively. \textit{Right}: The size-linewidth-surface density relation considering the external pressure of CO(1-0). \label{fig-ssp}}
\end{figure*}

An alternative explanation for the high $\alpha_{\rm vir}$ is external pressure, which helps to maintain equilibrium. \citet{2011MNRAS.416..710F} introduced a size-linewidth-surface density relation that incorporates external pressure, as expressed in eq. \eqref{eq-13}, which allows us to estimate the external pressure. The variations of the specific form factors have a minimal impact on the results, so we adopt a form factor $\Gamma=0.73$, consistent with \citet{2011MNRAS.416..710F}. Figure \ref{fig-ssp} illustrates the size-linewidth-surface density relation that incorporates external pressure for both CO(3-2) and CO(1-0). Data points span a range from 10$^1$ to 10$^4$ K cm$^{-3}$. Comparatively, the CO(1-0) data points are all positioned above the straight line representing zero pressure, while some CO(3-2) data lie below. 
The external pressures fitted are listed in Table \ref{tab-res}. In particular, the nuclear ring exhibits pressure higher than that of the disc, albeit with considerable uncertainty. Furthermore, the CO(3-2) pressure values are lower than those derived from CO(1-0).

\subsubsection{CO-to-H$_2$ conversion factor}
\label{sec:conversion}

\begin{figure*}
\begin{minipage}{0.48\linewidth}
\centering\includegraphics[width=\textwidth]{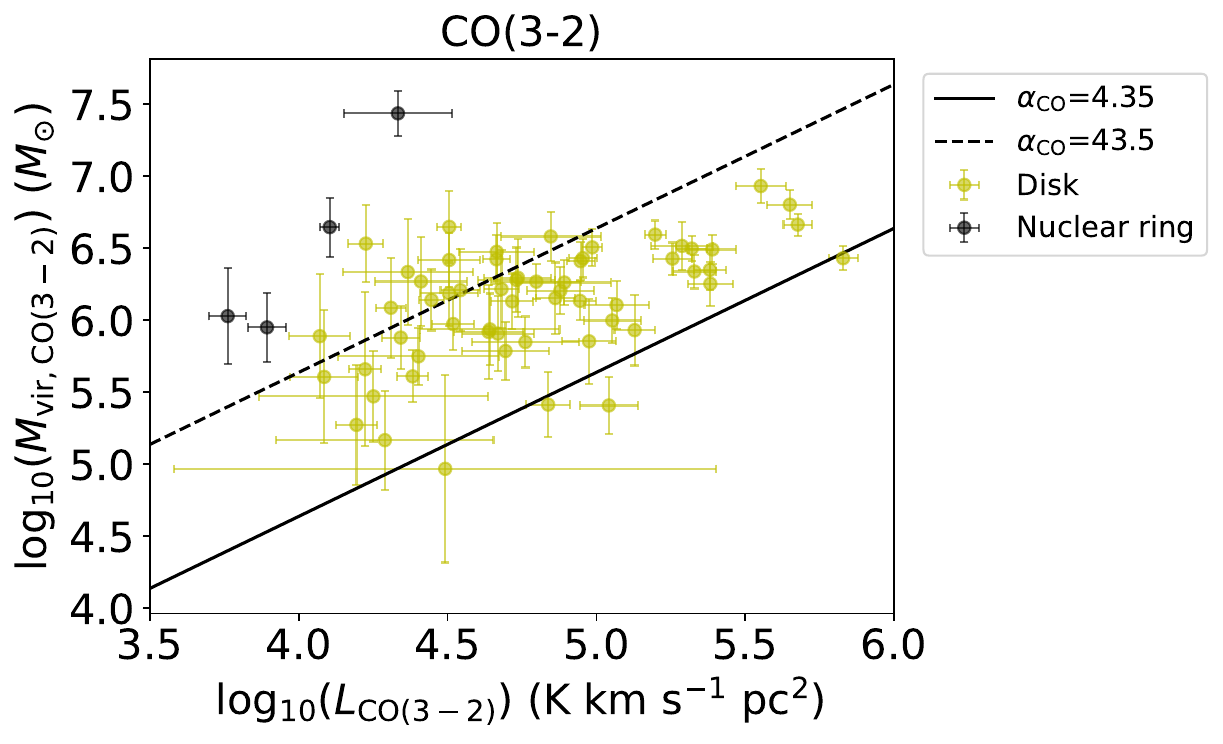}
\end{minipage}
\begin{minipage}{0.48\linewidth}
\centering\includegraphics[width=\textwidth]{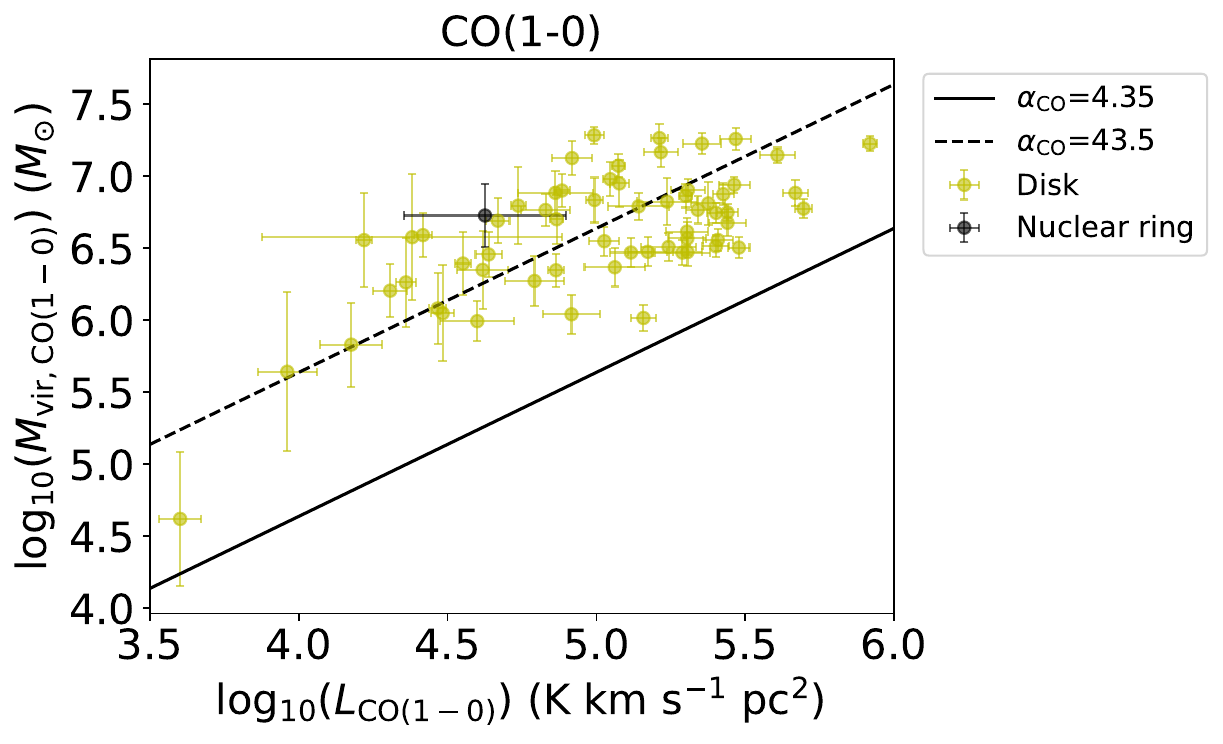}
\end{minipage}
\caption{\textit{Left}: Relation between virial mass and luminosity from CO(3-2). Solid and dashed lines represent the conversion factor $\alpha\rm_{vir}=4.35, 43.5~\it M_{\odot}\rm ~pc^{-2}~(K~ km~ s^{-1})^{-1}$. \textit{Right}: Relation between the viral mass and the luminosity of CO(1-0). \label{fig-conv}}
\end{figure*}

The CO-to-H$_2$ conversion factor $\alpha\rm_{CO}$ is a crucial parameter for estimating cloud mass. In the previous analysis, we assumed a constant conversion factor. However, it varies with environments and is influenced by factors such as optical depth, kinetic temperature, and metallicity \citep{2013ARA&A..51..207B, 2023ApJ...950..119T, 2024A&A...681A..14R}. To further explore its effects, we now employ the method proposed by \citet{2013ARA&A..51..207B} to estimate the conversion factor under the virilization assumption using eq. \ref{eq-12}. The luminosity of CO(3-2) is converted to the luminosity of CO(1-0) using the $R_{31}$ of each field of \citet{2020MNRAS.492..195L}. The viral mass and luminosity relation for both CO(3-2) and CO(1-0) are shown in Figure \ref{fig-conv}, from which we can estimate the conversion factor. The estimated conversion factors differ significantly from the typical value observed in the Milky Way disc (4.35 $M_{\odot}$ pc$^{-2}$ (K km s$^{-1}$)$^{-1}$), as depicted by the solid line in Figure \ref{fig-conv}. A dashed line representing a 10 times higher $\alpha\rm_{CO} = 43.5$ $M_{\odot}$ pc$^{-2}$ (K km s$^{-1}$)$^{-1}$ is also shown. The data points for both CO(3-2) and CO(1-0) in the disc generally fall between the two lines, indicating a conversion factor higher than the standard value of 4.35 $M_{\odot}$ pc$^{-2}$ (K km s$^{-1}$)$^{-1}$. Interestingly, while the nuclear ring shows little difference from the disc for CO(1-0), there is a significant distinction for CO(3-2). This disparity can be mainly attributed to the higher velocity dispersion and radius of the CO(3-2) cloud in the nuclear ring compared to the disc. The results of the conversion factor obtained using {\sc kapteyn} are presented in Table \ref{tab-res}. In fact, the fitted $\alpha\rm_{CO}$ is 2-4 times higher than the standard value for CO(3-2) and CO(1-0) in the disc, while the $\alpha\rm_{CO}$ from CO(3-2) in the nuclear ring is much higher than in the disc.

\begin{table*}
\centering
\caption{Best-fitting parameters of the correlations}\label{tab-res}
 \begin{threeparttable}
\begin{tabular}{ccccccccc}
\hline
 & Transition &   disc &  Nuclear ring & Total \\
\hline
Size-linewidth relation & CO(3-2) & 
$\sigma_v=(0.29\pm0.22)R^{0.61\pm0.17}$ & 
$\sigma_v=(0.00\pm0.00)R^{2.22\pm1.68}$$^{(a)}$ &
$\sigma_v=(0.10\pm0.08)R^{0.85\pm0.18}$ \\
$\sigma_v=\rm{C}R^{\rm{a}}$ & CO(1-0) & 
$\sigma_v=(0.47\pm0.29)R^{0.52\pm0.12}$ & 
one point & 
$\sigma_v=(0.49\pm0.28)R^{0.52\pm0.12}$ \\
\hline
Virial parameter & CO(3-2) & 
$3.59\pm0.35$  & $83.84\pm44.80$ & $3.62\pm0.37$\\
$\alpha_{\rm vir}$ & CO(1-0) & 
$5.60\pm0.55$ & one point & $5.62\pm0.54$\\
\hline
External pressure & CO(3-2) & 
$1.85\pm0.40$  & $4.10\pm3.21$ & $1.95\pm0.40$\\
$P_{\rm e}$ (10$^3$ K cm$^{-3}$) & CO(1-0) & 
$6.12\pm0.82$ & one point & $6.11\pm0.81$\\
\hline
CO-to-H$_2$ conversion factor & CO(3-2) &
 $11.76\pm1.23$  & $175.79\pm92.67$ & $11.81\pm1.27$ \\
$\alpha_{\rm CO}$ ($M_{\odot}\rm ~pc^{-2}(K~ km ~s^{-1})^{-1}$) & CO(1-0) & $20.42\pm1.88$ & one point & $20.44\pm1.88$ \\
\hline
\end{tabular}
\begin{tablenotes}
\footnotesize
\item \emph{Notes.} Best-fitting results of the relations discussed in Section \ref{sec:relation} using {\sc kapteyn} from two transition lines, and from disc, nuclear ring, and all data respectively. \\
$^{(a)}$The value of the fitted C for the nuclear ring, obtained from the CO(3-2) data, is quite small, approximately $10^{-4}$ km s$^{-1}$, and the error is $\sim 10^{-3}$ km s$^{-1}$.  
\end{tablenotes} \end{threeparttable}
\end{table*}

\subsection{CO transition comparison}\label{sec: CO transition}

The choice of CO transitions has a non-negligible influence on cloud property measurements. Notable differences exist between CO(3-2) and CO(1-0). For example, CO(1-0) data are more dispersed, and the identified structures differ in number and properties. The velocity dispersion and the equivalent radius of CO(3-2) are generally smaller than those of CO(1-0), with smaller scatters, while the surface density and turbulent pressure are higher. We believe that these differences are mainly due to three factors.

First, data quality plays a critical role. The initial spatial and spectral resolution of the CO(3-2) data is better than that of CO(1-0). For example, in the nuclear ring, CO(1-0) identifies only a single structure, potentially missing others with significant velocity dispersion. Even after smoothing the resolution to match between the transitions, CO(3-2) may still detect additional small structures that do not fully merge. Consequently, these disparities persist in the overall comparison.

Second, the algorithm and selection criteria also contribute to the differences. As shown above, the structures identified in CO(3-2) and CO(1-0) are not identical. Due to the selection process, even when both transitions detect structures at the same location, their properties may differ, or some structures may be excluded because of insufficient spatial and spectral resolution. Consequently, we did not rely on strictly corresponding structures. Nonetheless, the structures identified in each transition are robust, and considering them representative of their respective regions still provides valuable insights.

Finally, the different CO transitions reflect distinct physical states, which we believe to be the most fundamental factor. CO(3-2) has a higher transition energy and critical density compared to CO(1-0), and it often traces warmer, denser gas \citep{2009ApJ...693.1736W}. In contrast, CO(1-0) is easier to excite and is typically used to map the overall extent of the gas. Consequently, the equivalent radius from CO(1-0) is expected to be larger, and its surface density is lower than that of CO(3-2). Since the noise level differs between the two transitions in terms of H$_2$ mass, with CO(1-0) data being deeper, we tested identifying clouds using a higher CO(1-0) threshold to match the H$_2$ mass sensitivity of CO(3-2). The conclusion remains unchanged even with this adjustment, suggesting an intrinsic difference between the two transitions. As suggested by \citet{2020MNRAS.492..195L}, the intensity of CO(3-2) may be lower than CO(1-0), especially in quiescent environments where CO(1-0) might be the dominant emission, with CO(3-2) contributing only a small fraction. In such cases, weaker structures traced by CO(3-2) might be missed, even if they exist. In contrast, in regions where CO(3-2) dominates, the situation can be reversed, leading to a mismatch in the identified structures.

We consider the primary cause of the observed differences to be the distinct transition states, with data quality and selection factors being secondary. In the future, if instruments with similar observational capabilities can sufficiently mitigate the effects of data quality, the combination of multiple transitions will enable more precise constraints on the properties of molecular clouds.

\section{Discussion}\label{sec:discussion}
\subsection{Property distributions and comparison}
\label{sec:compare}

Turbulence is one of the main factors impacting the properties of clouds, which can be quantified by the velocity dispersion. The velocity dispersion of atomic \citep{2013ApJ...773...88S} and molecular hydrogen \citep{2013AJ....146..150C, 2021ApJ...909...12G} is $\sim$10 km s$^{-1}$ in the Local Group galaxies, typical for self-virialized clouds. In contrast, clouds in regions with high ambient pressure \citep[e.g., the Galactic centre;][]{2001ApJ...562..348O} tend to show higher velocity dispersion. Other mechanisms, e.g., magnetic fields and external pressure, might also act to support or confine the clouds, influencing the dynamical state of molecular clouds, changing the internal structure, and thereby affecting star formation. The presence of larger structures in a galaxy, such as spirals, bars, and rings, is often associated with specific conditions such as high pressure \citep{2001ApJ...562..348O}, abundant molecules \citep{2012MNRAS.424.3050W}, and low gas density \citep{2001ApJ...551..852H}.

The large velocity dispersion may reflect violent turbulence. From the CO(3-2) data, we found that velocity dispersion in the nuclear ring is on average larger than in the disc. It is well established that gas in galactic centres behaves differently from that in discs \citep{2001ApJ...562..348O,2019MNRAS.484..964L}. Numerous studies have found that the velocity dispersion is higher in galactic bars or bulges \citep[e.g.][]{2020ApJ...892..148S, 2021MNRAS.502.1218R}. \citet{2019A&A...625A.148D} measured the velocity dispersion of molecular gas in M31's circumnuclear region, with an average value of $6.8\pm 1.3$ km s$^{-1}$, which aligns with our results of $\sim \rm 6~km~s^{-1}$ in the nuclear ring. 
Studies also indicate larger dispersion in the spiral arms \citep[e.g.][]{2018ApJ...860..172S} than interarm regions, which is also the case in our study since the interarm region (Region h) generally has a lower $\sigma$. Some studies suggest that the observed velocity dispersion may include significant contributions from galactic motion \citep[e.g.][]{2015ApJ...803...16U, 2023MNRAS.522.4078C}, potentially leading to overestimations. We estimated the velocity gradients for each cloud by performing least-squares fitting of the central velocity and position on a pixel-by-pixel basis. The resulting gradients are small, less than 0.3 km s$^{-1}$ pc$^{-1}$, consistent with the findings of \citet{2007ApJ...654..240R}. Consequently, we did not apply any corrections. 

The equivalent radius in the nuclear ring is generally larger than in the disc for the CO(3-2) data, suggesting that molecular gas in the nuclear ring may be more extended, whereas gas in the disc tends to clump more easily. Despite this, the mean surface density in the nuclear ring is relatively low, even though it has a relatively large velocity dispersion and equivalent radius. This is expected since M31 is a galaxy with low molecular gas density and star formation rate \citep{2018ApJ...860..172S}. Our estimated surface density is less than 200 $M_{\odot}$ pc$^{-2}$, consistent with the findings of \citet{2018ApJ...860..172S}, and supporting previous observations that the centre of M31 lacks molecular gas \citep{2017A&A...607L...7M, 2019A&A...625A.148D}.
Additionally, studies have found that the velocity dispersion may be larger in regions with low gas density compared to that estimated from virialized gas \citep[e.g.][]{2001ApJ...551..852H}, which is also observed in the centre of M31.


The turbulent pressure in the nuclear ring is quite low compared to that in the disc. This is likely due to the ring's lower surface density, despite the higher mean velocity dispersion. 
It is noteworthy that our estimation of the turbulent pressure may be subject to uncertainties due to the limitations of our study. 
Firstly, we use a constant conversion factor, which should actually vary with the environment. Studies have found that the conversion factor can be lower due to high metallicity \citep{2013ARA&A..51..207B}, and higher if the [C\,{\sc ii}]/CO ratio is high \citep{2024A&A...681A..14R}. \citet{2020ApJ...905..138L} found that the [C\,{\sc ii}]/CO ratio in M31's circumnuclear region is higher than in the disc, suggesting that the conversion factor may be higher. Therefore, we could be underestimating the gas mass and, consequently, the turbulent pressure in the nuclear ring. For more accurate measurements, we should use varying conversion factors correlated with metallicity or [C\,{\sc ii}] intensity \citep{2013ARA&A..51..207B, 2020ApJ...892..148S, 2024A&A...681A..14R}. Secondly, the 13 selected regions in our study cover only a limited portion of the galaxy's radius and may not fully represent the overall environmental conditions. More observations are needed to better understand the trends across the galaxy.

\subsection{Size-linewidth relation comparison}

The size-linewidth relation is a reflection of the underlying turbulence within molecular clouds. Our data for both CO(3-2) and CO(1-0) follow the classical distribution $\sigma \propto R^{0.5}$ found in the Milky Way's disc \citep[][Figure \ref{fig-sr}]{1987ApJ...319..730S}, which suggests highly compressible supersonic turbulence or virial equilibrium. This is consistent with the relation found in a spiral arm of M31 \citep{2007ApJ...654..240R}. In comparison, the size-linewidth relation in Milky Way's centre lies well above \citep{2017A&A...603A..89K}, which is mainly attributed to a much higher velocity dispersion \citep{2001ApJ...562..348O}. The generally lower velocity dispersions observed in M31 suggest a relatively quiescent environment throughout the galaxy. 

In particular, our data fall slightly below the classical relation, as illustrated in Figure \ref{fig-sr}. If we expect our estimated properties to align with \citet{1987ApJ...319..730S}, we might consider that we are underestimating the velocity dispersion or overestimating the equivalent radius. However, underestimating velocity dispersion seems unlikely, as most factors—whether related to physical state and environment \citep{2018ApJ...860..172S} or observational effects \citep{2016AJ....151...34C}—tend to increase it. Overestimating the equivalent radius, on the other hand, is more plausible. The minimum pixel number ($N = 16$) we set in the algorithm may be too large, causing some structures to merge and inflating the radius estimates. Additionally, as discussed by \citet{2019A&A...625A.148D} regarding velocity resolution, our low angular resolution might lead to the omission of smaller structures. It is also possible that the observed clouds genuinely have different properties. Given the quiescent environment of the M31 disc, GMCs are expected to exhibit lower turbulence due to less energy injection from star formation activities. Similarly, recent ALMA-ACA 7m Array CO(2-1) mapping of M33 revealed that the velocity dispersion of M33 GMC is intrinsically lower \citep{2023ApJ...953..164M}. This could be attributed to the presence of low surface density clouds supported by lower velocity dispersion, as observed in 12 other external galaxies \citep{2008ApJ...686..948B}. Overall, these results show minimal differences between the two transitions. 

\subsection{Size-linewidth-surface density relation and virial parameter analysis}
\label{sec:virial}

The size-linewidth-surface density relation is a size-linewidth relation considering surface density, which provides a better constraint on molecular cloud state. 
The $\alpha_{\rm vir}$ derived from this relation is 3.59 for CO(3-2), as listed in Table \ref{tab-res}, indicating a state close to marginally bound. In comparison, \citet{2018ApJ...860..172S} estimated $\alpha_{\rm vir}$ $\sim 10$ in the M31 disc, though their calculation uses a fixed equivalent radius based on the beam size. The velocity dispersion could be overestimated due to factors such as beam smearing and overlapping components, which are commonly found in high-density regions or highly inclined galaxies. 
Additionally, other factors, such as stellar winds, interstellar shocks and/or magnetic fields, can significantly contribute to the velocity dispersion, especially in low-density environments where the cloud's self-gravity is weaker, leading to an elevated $\alpha_{\rm vir}$. 
Furthermore, the average of the surface density over a large area can reduce the estimated $\Sigma$, which further contributes to an overestimation of $\alpha_{\rm vir}$.
In fact, some identified GMCs may consist of multiple components, leading to an overestimation of their size. 

As the cloud size approaches the thickness of the disc, the spherical approximation may not be valid, and a spheroidal geometry should instead be adopted. If we approximate the cloud size using a 3D mean radius of a spheroidal cloud following \citet{2021MNRAS.502.1218R}, the estimated $\alpha_{\rm vir}$ would generally decrease by 30–40\%. Since we did not correct for the galaxy inclination or large-scale motion, it is reasonable to assume that our results also contain such deviations, suggesting that the true $\alpha_{\rm vir}$ might be lower and closer to virial equilibrium. As \citet{2022AJ....164...43S} pointed out, if the clouds are anisotropic and thus inclination correction is adopted, $\alpha_{\rm vir}$ would decrease by 40\%.
Furthermore, the potential for an axisymmetric ellipsoid would deviate from a spherical potential by a geometric factor $\beta = (\rm arcsin~\it e\rm)/\it e$. Here, $e$ is the eccentricity defined by the axis ratio of the cloud, which is generally smaller than observed due to the projection of a prolate cloud, as demonstrated by \citet{2013ApJ...768L...5L}. Including this correction in the virial mass ($M_{\rm vir} = \frac{5\sigma_v^2R}{f\beta \rm G}$) would result in a $\lesssim$20\% decrease in the $\alpha_{\rm vir}$.

Finally, the presence of undetected "CO-dark" clouds could result in an underestimation of $M_{\rm gas}$ and contribute to the overestimation of $\alpha_{\rm vir}$ (Equation \ref{eq-10}). In the centre of M31, the strong [C\,{\sc ii}] emissions observed by \textit{Herschel} suggest a strong radiation field and low surface density \citep{2020ApJ...905..138L}, indicating the probable presence of a significant amount of CO-dark gas in this region. It is also found that there are dust-traced clouds with no CO emission at a scale of 30 pc in the CARMA survey, suggesting the existence of CO-dark gas \citep{2022MNRAS.511.5287A}.

The $\alpha_{\rm vir}$ estimated from CO(1-0) is higher, $\sim 6$. As mentioned above, CO(1-0) has a lower surface density and larger equivalent radius compared to CO(3-2). Furthermore, CO(1-0) traces a larger portion of molecular gas, which may blend more components and exacerbate the beam-smearing effect, leading to higher velocity estimates than CO(3-2). Taking into account these factors, the $\alpha_{\rm vir}$ estimated from CO(1-0) is expected to be higher than that from CO(3-2). 

The nuclear ring shows a significant difference from the disc, with $\alpha_{\rm vir}\sim100$. 
Similarly, \citet{2019A&A...625A.148D} reported an average $\alpha_{\rm vir}$ of 140 in the M31 circumnuclear region, indicating high values as well. They suggested that these molecular clouds are not in virial equilibrium but instead are unbound, temporary aggregates of smaller virialized structures. \citet{2017ApJ...834...57M} also noted that true virial equilibrium is rare, with molecular clouds being dynamic and short-lived. Achieving virial equilibrium in the nuclear region is particularly challenging. The strong gravitational potential and dynamically complex environment induce large-scale motions, broadening the linewidth. Additionally, \citet{2020MNRAS.492..195L} found that the gas temperature in M31's circumnuclear region is significantly higher than that in the disc, likely due to heating by the old stellar population, which can increase thermal pressure. Meanwhile, the low-density conditions make the gas more susceptible to intense radiation \citep{2020ApJ...905..138L}, further enhancing turbulence. Together, these factors contribute to the elevated $\alpha_{\rm vir}$ observed in the nuclear ring.

The interpretation of $\alpha_{\rm vir}$ should be considered with caution. Traditional $\alpha_{\rm vir}$ only considers self-gravity and often assumes isolated, spherical, uniform-density clouds. However, molecular clouds are dynamic, compressible, irregular, and influenced by galactic environments, such as tidal forces. To more accurately reflect the state of molecular clouds, future analyses should incorporate corrections for additional factors, including external gravity \citep{2021MNRAS.505.4048L,2022MNRAS.515.2822R}.

\subsection{External pressure explanation}

One possible explanation for the high $\alpha_{\rm vir}$ observed in the M31 nuclear region is external pressure. When external pressure is taken into account, we find that the pressure required to maintain virial equilibrium ranges from $10^1$ to $10^4$ K cm$^{-3}$ (Figure \ref{fig-ssp}). Notably, our estimated internal turbulent pressure ranges from $10^1$ to $10^6$ K cm$^{-3}$ (Figure \ref{fig-P}), which is higher than the external pressure required to maintain balance. This is in line with the findings of \citet{2020ApJ...892..148S}, suggesting that the turbulent pressure exceeds the external dynamic equilibrium pressure.
In comparison, \citet{2011MNRAS.416..710F} estimated that the external pressure required for Galactic GMCs lies between $10^4$ and $10^7$ K cm$^{-3}$ based on $^{13}$CO(1-0) observations \citep{2009ApJ...699.1092H}. 
Despite this difference, our results are consistent with the typical theoretical pressure for the neutral ISM of $\sim \rm 5\times 10^3~K ~cm^{-3}$ \citep{1989ApJ...338..178E}, as well as the observed pressure in individual Galactic GMCs of $\sim \rm 10^5~K ~cm^{-3}$ \citep[e.g.][]{2008ApJ...672..410L}. 
Therefore, the molecular clouds in M31 could be confined by external pressure from the surrounding interstellar medium, particularly the atomic gas, which balances turbulent pressure alongside self-gravity. This idea is aligned with the fact that M31 is rich in atomic gas \citep{2009ApJ...695..937B}. However, atomic gas is relatively scarce in the centre of M31 \citep{2009ApJ...695..937B}, casting doubt on this explanation. Nonetheless, preliminary estimates of external pressure in the nuclear ring region, based on warm ionized gas (Li et al. 2025, submitted), suggest values of $10^{5-6}$ K cm$^{-3}$, indicating that confinement by warm gas may be a viable alternative. 
It is important to note that this explanation mainly considers self-gravity, turbulent pressure, and interstellar medium pressure. Other factors, such as magnetic fields, external gravitational potential, or stellar feedback, may also play important roles and should be considered in interpreting the dynamical state of GMCs. A comprehensive analysis of the dynamical equilibrium of M31 GMCs is beyond the scope of this paper and will be addressed in future work.

\subsection{Virial-based CO-to-H$_2$ conversion factor estimate}

We estimate the CO-to-H$_2$ conversion factor using a virial-based method \citep{2013ARA&A..51..207B}, which assumes that the molecular clouds are in virial equilibrium. In the M31 disc, we find a large conversion factor from the CO(3-2) observations, $\alpha_{\rm CO} \sim 10~M_{\odot}\rm ~pc^{-2}(K~km~s^{-1})^{-1}$ (Table \ref{tab-res}), which is much higher than the canonical value of 4.35 \citep{2013ARA&A..51..207B}. The wide dispersion of the data points indicates substantial variation in the factor. Estimates from CO(1-0) are even higher than those from CO(3-2), likely due to larger velocity dispersion and equivalent radius, which results in a higher virial mass. Additionally, the estimated conversion factor in the nuclear ring is considerably higher than in the disc. 

The high conversion factor seems unlikely, as there are few reported cases where the virial-based method yields values exceeding a dozen. Across M31, a conversion factor close to the canonical value has been derived using dust column density from \textit{Herschel} observations \citep{2012ApJ...756...40S}, suggesting that the conversion factor in M31 is consistent with typical expectations. Resolution plays a role in these estimates, with lower resolution generally leading to higher conversion factors \citep{2013ARA&A..51..207B}. Since we smoothed the data to a lower resolution, this may have contributed to the higher factors. Smoothing dilutes CO intensity, reducing luminosity and thus elevating $\alpha\rm_{CO}$. 
Additionally, the choice of parameters results in larger cloud sizes, and the virial-based method is sensitive to how clouds are defined. Some of the identified clouds may actually be composites of multiple physical clouds, which are not in virial equilibrium, leading to overestimated conversion factors. 
To investigate the impact of resolution, we attempted to identify clouds using CO(3-2) data at its original resolution. The resultant $\alpha_{\text{CO(3-2)}}$ in the disc is smaller ($\sim$5.5 $M_{\odot}\rm ~pc^{-2}(K~km~s^{-1})^{-1}$) and closer to the canonical value, confirming that the conversion factor is indeed sensitive to resolution. Applying geometric and projection corrections would further reduce the virial mass \citep{2013ApJ...768L...5L} and thus the virial-based conversion factor, as discussed in Section \ref{sec:virial}. However, due to the low S/N of the original data, the identified clouds are biased toward denser, more gravitationally bound regions. Nevertheless, the nuclear region still shows a substantially higher conversion factor than the disc, indicating that the environments in these two regions are quite different.

Furthermore, the presence of "CO-dark" clouds in regions with intense radiation fields or low metallicity can increase the conversion factor. Although the metallicity in M31's centre is high \citep{2014ApJ...780..172D}, our results indicate a higher $\alpha\rm_{CO}$ in the nuclear ring. This may be attributed to the intense radiation field from old stellar populations as manifested by the high [C\,{\sc ii}]/CO(3-2) line ratio in \textit{Herschel} observations \citep{2020ApJ...905..138L}. This strong FUV radiation field could dissociate CO molecules and produce more CO-dark gas traced by [C\,{\sc ii}] emission, leading to an elevated $\alpha\rm_{CO}$ factor. 

Since the virial-based estimation of $\alpha_{\rm CO}$ depends on the virial parameter, and an exceptionally high $\alpha_{\rm CO}$ appears unlikely, a more plausible explanation could be that the clouds are not truly in virial equilibrium. As discussed in Section \ref{sec:virial}, the clouds in M31 exhibit a large virial parameter and are not gravitationally bound. If this is indeed the case, the conversion factor could be much lower and closer to the canonical value of 4.35. 

\section{Summary}
\label{sec:sum}

We utilized CO(3-2) data from the JCMT and CO(1-0) data from the IRAM 30m telescope to estimate properties of molecular clouds in the nuclear ring and selected regions in the disc of M31, exploring the relationships between these properties. Our main findings are as follows:

\begin{itemize}
\item The velocity dispersion and size of molecular clouds in the nuclear ring are generally larger than those in the disc, while the mean surface density and turbulent pressure are lower. This suggests that the nuclear ring may have a unique environment, with the lower turbulent pressure likely attributed to the reduced surface density. Additionally, the assumption of a constant CO-to-H$_2$ conversion factor could influence the observed differences between the nuclear ring and the disc. 
\item The estimated velocity dispersion and equivalent radius of CO(3-2) are smaller than those of CO(1-0), while the surface density and turbulent pressure are higher. The choice of CO transition has a significant impact on the results. Apart from data quality and algorithm selection bias, the fundamental cause is the different physical conditions reflected by different transitions.
\item The derived $\alpha_{\rm vir}$ values from the size-linewidth-surface density relation are greater than 1, suggesting that the clouds are not in virial equilibrium, with even higher values in the nuclear ring. This indicates that the clouds in this region may be short-lived and dynamic structures. Overestimating velocity dispersion and underestimating surface density can impact the accuracy of $\alpha_{\rm vir}$ estimates. Caution is needed when interpreting high traditional $\alpha_{\rm vir}$ values, as molecular clouds are also influenced by factors such as external gravitational potential.
\end{itemize}

In the future, improved observations will allow for more precise estimates of cloud properties, enabling us to account for the influence of various factors and gain a deeper understanding of the relationships between them and the conditions of molecular clouds.

\section*{Acknowledgments}

This work was supported by MOST 2022YFA1605300, the National Natural Science Foundation of China
(grant 11988101 and 12225302) and the National Key Research and Development Program of
China (No. 2022YFA1605000 and NO.2022YFF0503402). Z.N.L. acknowledges support by the China National Postdoctoral Program for Innovation Talents (grant BX20220301) and the East Asian Core Observatories Association Fellowship. J.W. thank the support from NSFC 12041302 and National Key Research and Development Program of China (No.2023YFA1608004). R.d.G. was supported in part by the Australian Research Council Centre of Excellence for All Sky Astrophysics in 3 Dimensions (ASTRO 3D), through project number CE170100013. B.L. acknowledges support by the National Research Foundation of Korea (NRF), grant Nos. 2022R1A2C100298213. D.L. is a New Cornerstone investigator.

\section*{Data Availability}

The raw CO(1-0) data of the disc used for this work are available on the IRAM data archive\footnote{https://iram-institute.org/science-portal/proposals/lp/early-science/lp009-andromeda/}
The CO(3-2) data in the disc are available on the HASHTAG archive\footnote{https://hashtag.astro.cf.ac.uk/DR1.html}. The CO(1-0) and CO(3-2) data in the nuclear ring used
for this work are available upon request to the corresponding author.

\bibliographystyle{mn2e_new}
\bibliography{ref} 


\appendix
\section{Clump catalog}

The following tables list the main properties of identified clumps within each field from CO(3-2) and CO(1-0) data, respectively, using the PYCPROPS algorithm. The ID letters represent the field names, and the numbers mark each cloud. Position, radius, line-of-sight velocity, velocity dispersion, and luminosity are also given.

\setcounter{table}{0}
\renewcommand\thetable{A\arabic{table}}
\begin{table*}
\centering
\caption{Clump properties from CO(3-2)
\label{tab-clumps}}
\begin{threeparttable}
\begin{tabular}{ccccccccc}
\hline
ID &  RA & DEC &  $v$ &  $R$ & $\sigma_{v}$ & $L_{\rm CO}$ & $M\rm_{mol}$ \\
& (J2000) & (J2000) & $\rm (km~s^{-1})$ & (pc) & $\rm (km~s^{-1})$& $(10^4~\rm K~km~s^{-1}~pc^2)$ & $(10^5~  M_{\odot}$)\\
(1) & (2) & (3) & (4) & (5) & (6) & (7) & (8)\\
\hline
a1&11.641&42.195&$-$45.2&$96.3\pm21.8$&$2.7\pm0.9$&$2.4\pm0.5$&$5.1\pm1.1$\\
a2&11.621&42.190&$-$44.2&$114.3\pm34.0$&$3.2\pm1.2$&$0.5\pm0.1$&$1.1\pm0.1$\\
b1&11.397&41.972&$-$49.4&$80.1\pm14.5$&$2.9\pm0.5$&$1.4\pm0.6$&$2.5\pm1.0$\\
b2&11.387&41.986&$-$53.0&$50.9\pm22.2$&$1.7\pm0.6$&$0.5\pm0.4$&$0.8\pm0.7$\\
b3&11.378&41.981&$-$56.1&$40.0\pm11.8$&$3.7\pm0.6$&$0.6\pm0.4$&$1.1\pm0.7$\\
c1&11.156&41.885&$-$109.2&$64.4\pm13.8$&$3.8\pm0.7$&$1.0\pm0.2$&$1.4\pm0.2$\\
c2&11.140&41.879&$-$95.7&$102.0\pm8.1$&$4.1\pm0.7$&$7.0\pm1.2$&$10.5\pm1.8$\\
c3&11.164&41.882&$-$98.8&$88.4\pm11.7$&$3.7\pm0.7$&$3.4\pm0.8$&$5.1\pm1.3$\\
d1&11.262&41.922&$-$77.1&$96.9\pm9.0$&$5.2\pm0.6$&$4.1\pm0.6$&$7.8\pm1.2$\\
d2&11.237&41.924&$-$81.5&$126.3\pm9.6$&$4.5\pm0.4$&$15.5\pm1.7$&$29.3\pm3.3$\\
e1&11.108&41.629&$-$55.6&$83.4\pm7.6$&$6.0\pm0.7$&$5.2\pm1.0$&$10.7\pm2.0$\\
f1&10.781&41.411&$-$70.4&$38.8\pm7.2$&$4.6\pm1.2$&$1.2\pm0.2$&$5.8\pm0.9$\\
g1&10.598&41.109&$-$514.3&$99.8\pm19.0$&$6.6\pm1.8$&$0.5\pm0.0$&$1.4\pm0.1$\\
g2&10.588&41.093&$-$507.1&$56.2\pm18.2$&$1.8\pm0.8$&$0.2\pm0.0$&$0.7\pm0.1$\\
h1&11.013&41.705&$-$74.9&$59.0\pm41.6$&$1.2\pm0.8$&$0.3\pm0.7$&$1.4\pm2.8$\\
h2&11.013&41.711&$-$85.0&$79.6\pm18.9$&$3.2\pm1.1$&$0.4\pm0.1$&$1.9\pm0.5$\\
i1&11.053&41.586&$-$89.4&$81.2\pm7.4$&$8.7\pm0.9$&$6.3\pm1.1$&$19.4\pm3.4$\\
i2&11.054&41.595&$-$92.4&$91.5\pm7.4$&$9.5\pm1.2$&$5.0\pm1.0$&$15.5\pm3.0$\\
j1&11.372&41.751&$-$80.2&$75.6\pm10.2$&$4.2\pm0.6$&$2.1\pm0.3$&$3.8\pm0.5$\\
j2&11.352&41.753&$-$87.9&$97.5\pm8.8$&$4.6\pm0.6$&$5.1\pm0.9$&$9.3\pm1.6$\\
j3&11.361&41.735&$-$97.2&$77.4\pm7.8$&$3.5\pm0.6$&$2.7\pm0.6$&$4.9\pm1.1$\\
j4&11.368&41.753&$-$93.7&$107.5\pm9.0$&$4.5\pm0.7$&$5.8\pm0.7$&$10.5\pm1.3$\\
k1&10.953&41.557&$-$115.3&$51.4\pm10.5$&$5.6\pm1.6$&$0.7\pm0.1$&$2.1\pm0.3$\\
k2&10.969&41.564&$-$113.9&$79.2\pm10.1$&$6.3\pm1.1$&$2.7\pm0.6$&$8.4\pm2.0$\\
R1&11.247&41.474&$-$201.3&$93.3\pm11.8$&$1.6\pm0.4$&$3.5\pm0.8$&$4.8\pm1.1$\\
R2&11.180&41.468&$-$213.5&$136.7\pm42.3$&$4.9\pm1.3$&$0.5\pm0.1$&$0.7\pm0.1$\\
R3&11.156&41.499&$-$170.6&$54.1\pm11.1$&$2.7\pm0.5$&$0.8\pm0.1$&$1.0\pm0.1$\\
R4&11.141&41.484&$-$176.3&$114.6\pm18.4$&$3.9\pm0.6$&$2.5\pm0.9$&$3.4\pm1.2$\\
R5&11.160&41.442&$-$204.8&$34.6\pm10.5$&$3.4\pm1.7$&$0.4\pm0.1$&$0.5\pm0.1$\\
R6&11.159&41.451&$-$214.0&$74.1\pm15.6$&$5.3\pm2.2$&$0.7\pm0.4$&$1.0\pm0.5$\\
R7&11.186&41.443&$-$181.5&$90.0\pm18.1$&$5.3\pm1.0$&$1.0\pm0.2$&$1.4\pm0.3$\\
R8&11.116&41.459&$-$186.7&$138.5\pm22.9$&$5.2\pm0.9$&$2.2\pm0.9$&$3.1\pm1.2$\\
R9&11.095&41.448&$-$191.7&$80.6\pm19.5$&$2.3\pm1.4$&$0.5\pm0.1$&$0.7\pm0.1$\\
R10&11.187&41.489&$-$177.7&$97.4\pm18.1$&$3.7\pm0.8$&$0.9\pm0.1$&$1.2\pm0.1$\\
R11&11.194&41.445&$-$213.2&$98.4\pm24.7$&$2.9\pm0.8$&$1.4\pm0.8$&$1.9\pm1.0$\\
R12&11.177&41.445&$-$214.1&$87.0\pm17.5$&$4.5\pm1.5$&$0.8\pm0.3$&$1.1\pm0.4$\\
R13&11.200&41.437&$-$197.0&$107.0\pm16.3$&$2.7\pm0.8$&$1.5\pm0.4$&$2.0\pm0.6$\\
R14&11.194&41.420&$-$202.1&$87.0\pm13.7$&$2.9\pm0.7$&$0.7\pm0.1$&$1.0\pm0.1$\\
R15&11.184&41.429&$-$208.2&$96.2\pm21.3$&$3.9\pm0.9$&$1.0\pm0.2$&$1.4\pm0.3$\\
R16&11.210&41.482&$-$181.7&$98.0\pm18.3$&$4.0\pm1.3$&$1.1\pm0.2$&$1.5\pm0.2$\\
R17&11.201&41.484&$-$158.8&$88.5\pm14.3$&$4.6\pm1.4$&$1.7\pm0.5$&$2.4\pm0.6$\\
R18&11.211&41.493&$-$158.9&$61.6\pm11.0$&$4.7\pm1.3$&$2.3\pm0.4$&$3.2\pm0.5$\\
R19&11.187&41.478&$-$168.3&$77.1\pm10.8$&$5.8\pm1.0$&$1.5\pm0.2$&$2.0\pm0.2$\\
R20&11.151&41.459&$-$190.7&$167.0\pm14.4$&$3.8\pm0.6$&$2.8\pm0.3$&$3.9\pm0.4$\\
R21&11.162&41.420&$-$203.0&$104.6\pm7.6$&$5.4\pm0.5$&$6.7\pm0.8$&$9.1\pm1.1$\\
R22&11.143&41.405&$-$205.1&$139.0\pm24.5$&$3.6\pm0.9$&$1.7\pm0.5$&$2.3\pm0.7$\\
R23&11.118&41.430&$-$204.6&$45.9\pm7.8$&$3.6\pm0.8$&$1.6\pm0.5$&$2.2\pm0.7$\\
R24&11.135&41.427&$-$205.0&$107.6\pm9.8$&$4.9\pm0.7$&$2.9\pm0.3$&$3.9\pm0.4$\\
R25&11.206&41.472&$-$172.5&$97.3\pm12.3$&$4.3\pm0.5$&$2.0\pm0.7$&$2.7\pm1.0$\\
R26&11.207&41.463&$-$181.9&$131.9\pm13.5$&$4.8\pm0.7$&$3.1\pm0.2$&$4.2\pm0.3$\\
R27&11.233&41.488&$-$167.1&$50.9\pm12.1$&$2.2\pm0.5$&$2.2\pm0.4$&$3.0\pm0.5$\\
R28&11.226&41.477&$-$174.1&$81.6\pm13.8$&$4.0\pm0.8$&$1.7\pm0.3$&$2.3\pm0.4$\\
R29&11.173&41.452&$-$188.2&$93.3\pm6.8$&$6.4\pm0.7$&$5.0\pm0.4$&$6.9\pm0.6$\\
R30&11.185&41.465&$-$187.2&$128.5\pm7.7$&$5.9\pm0.5$&$15.2\pm1.7$&$20.7\pm2.3$\\
R31&11.169&41.441&$-$189.2&$100.1\pm22.9$&$5.4\pm1.1$&$1.5\pm0.4$&$2.0\pm0.6$\\
R32&11.179&41.434&$-$191.6&$91.3\pm12.4$&$4.1\pm0.7$&$2.4\pm0.6$&$3.3\pm0.9$\\
R33&11.156&41.444&$-$227.5&$43.3\pm12.6$&$2.6\pm0.8$&$0.6\pm0.5$&$0.8\pm0.7$\\
R34&11.154&41.424&$-$227.7&$33.2\pm9.2$&$4.7\pm2.3$&$0.4\pm0.1$&$0.5\pm0.1$\\
\hline
\end{tabular}
 \end{threeparttable}
\end{table*}

\begin{table*}
\centering
\contcaption{Clump properties from CO(3-2)}
\begin{threeparttable}
\begin{tabular}{ccccccccc}
\hline
ID &  RA & DEC &  $v$ &  $R$ & $\sigma_{v}$ & $L_{\rm CO}$ & $M\rm_{mol}$ \\
& (J2000) & (J2000) & $\rm(km~s^{-1})$ & (pc) & $\rm(km~s^{-1})$ &$(10^4~\rm K~km~s^{-1}~pc^2)$ & $(10^5~ M_{\odot}$)\\
(1) & (2) & (3) & (4) & (5) & (6) & (7) & (8)\\
\hline
N1&10.771&41.275&$-$157.6&$81.6\pm13.7$&$7.2\pm1.6$&$0.8\pm0.1$&$0.6\pm0.0$\\
N2&10.748&41.287&$-$163.6&$168.9\pm26.8$&$12.5\pm2.0$&$1.4\pm0.6$&$0.9\pm0.4$\\
N3&10.756&41.260&$-$194.4&$98.8\pm30.6$&$2.9\pm0.7$&$0.5\pm0.1$&$0.3\pm0.0$\\
N4&10.743&41.260&$-$205.8&$106.3\pm20.9$&$3.1\pm1.1$&$0.4\pm0.1$&$0.3\pm0.0$\\
\hline
\end{tabular}
\begin{tablenotes}
\footnotesize
\item[(1)] The ID of identified clouds within each field.
\item[(2)] Right ascension of identified clouds centre.
\item[(3)] Declination of identified clouds centre.
\item[(4)] Velocity of identified clouds centre.
\item[(5)] Equivalent radius of identified clouds.
\item[(6)] Velocity dispersion of identified clouds.
\item[(7)] 
Total luminosity of identified clouds.
\item[(8)] Luminosity mass of identified clouds.
\end{tablenotes}
 \end{threeparttable}
\end{table*}

\begin{table*}
\centering
\caption{Clump properties from CO(1-0)
\label{tab-clumps2}}
\begin{threeparttable}
\begin{tabular}{ccccccccc}
\hline
ID &  RA & DEC &  $v$ &  $R$ & $\sigma_{v}$ & $L_{\rm CO}$ &$M\rm_{mol}$ \\
& (J2000) & (J2000) & $\rm(km~s^{-1})$ & (pc) & $\rm(km~s^{-1})$ &$(10^4~\rm K~km~s^{-1}~pc^2)$ & $(10^5~ M_{\odot}$)\\
\hline
b1&11.396&41.971&$-$51.3&$122.5\pm9.4$&$4.8\pm0.5$&$13.1\pm2.1$&$5.7\pm0.9$\\
b2&11.394&41.983&$-$48.0&$116.4\pm11.6$&$3.9\pm0.8$&$6.2\pm1.4$&$2.7\pm0.6$\\
b3&11.383&41.984&$-$53.6&$76.0\pm10.0$&$3.5\pm0.5$&$4.0\pm1.1$&$1.7\pm0.5$\\
c1&11.142&41.876&$-$95.3&$149.8\pm9.7$&$5.2\pm0.5$&$20.2\pm3.3$&$8.8\pm1.4$\\
c2&11.173&41.871&$-$92.0&$59.3\pm4.6$&$4.1\pm0.4$&$14.4\pm1.4$&$6.3\pm0.6$\\
c3&11.163&41.883&$-$98.9&$111.6\pm7.9$&$7.6\pm1.4$&$17.3\pm4.0$&$7.5\pm1.7$\\
c4&11.134&41.876&$-$116.2&$104.7\pm32.8$&$2.5\pm0.7$&$1.5\pm0.4$&$0.7\pm0.2$\\
c5&11.156&41.865&$-$118.1&$93.6\pm10.7$&$3.4\pm0.5$&$8.3\pm1.8$&$3.6\pm0.8$\\
d1&11.261&41.919&$-$78.2&$129.3\pm7.6$&$5.9\pm0.6$&$27.6\pm3.9$&$12.0\pm1.7$\\
d2&11.237&41.925&$-$80.2&$147.5\pm7.3$&$7.1\pm0.7$&$46.6\pm4.5$&$20.3\pm2.0$\\
e1&11.129&41.619&$-$53.5&$109.0\pm27.4$&$3.1\pm1.1$&$3.0\pm0.3$&$1.3\pm0.1$\\
e2&11.109&41.630&$-$55.6&$117.1\pm7.3$&$5.4\pm0.4$&$25.6\pm2.9$&$11.1\pm1.3$\\
f1&10.748&41.409&$-$100.3&$85.9\pm10.1$&$6.6\pm1.1$&$2.6\pm0.2$&$1.1\pm0.1$\\
f2&10.779&41.405&$-$70.1&$134.4\pm7.4$&$6.3\pm0.5$&$25.3\pm1.7$&$11.0\pm0.8$\\
f3&10.771&41.405&$-$86.9&$140.1\pm7.5$&$7.2\pm0.6$&$26.7\pm2.1$&$11.6\pm0.9$\\
g1&10.612&41.104&$-$479.8&$60.8\pm27.7$&$7.7\pm3.5$&$2.4\pm2.8$&$1.0\pm1.2$\\
g2&10.589&41.092&$-$508.5&$111.1\pm14.9$&$8.2\pm1.3$&$7.3\pm2.1$&$3.2\pm0.9$\\
g3&10.604&41.106&$-$508.0&$99.7\pm8.5$&$7.5\pm0.9$&$6.7\pm1.3$&$2.9\pm0.5$\\
g4&10.591&41.113&$-$535.7&$96.8\pm15.9$&$4.0\pm0.8$&$2.0\pm0.3$&$0.9\pm0.1$\\
h1&11.013&41.714&$-$86.4&$173.0\pm10.3$&$6.7\pm0.5$&$20.3\pm2.7$&$8.8\pm1.2$\\
i1&11.055&41.600&$-$109.9&$145.9\pm14.3$&$7.9\pm1.0$&$11.1\pm0.6$&$4.8\pm0.3$\\
i2&11.059&41.576&$-$81.4&$109.6\pm6.8$&$7.2\pm0.8$&$21.9\pm2.1$&$9.5\pm0.9$\\
i3&11.055&41.589&$-$92.2&$158.2\pm5.7$&$10.1\pm0.6$&$83.1\pm4.3$&$36.1\pm1.9$\\
j1&11.373&41.751&$-$81.3&$129.2\pm10.3$&$4.7\pm0.5$&$14.9\pm2.3$&$6.5\pm1.0$\\
j2&11.352&41.754&$-$88.1&$112.8\pm7.8$&$5.0\pm0.5$&$20.2\pm2.3$&$8.8\pm1.0$\\
j3&11.367&41.752&$-$94.8&$133.4\pm9.4$&$6.8\pm1.2$&$23.8\pm3.1$&$10.3\pm1.4$\\
j4&11.362&41.736&$-$97.1&$99.4\pm6.4$&$6.0\pm0.8$&$20.2\pm3.0$&$8.8\pm1.3$\\
k1&10.966&41.579&$-$96.1&$79.1\pm19.3$&$6.6\pm2.4$&$1.7\pm0.1$&$0.7\pm0.0$\\
k2&10.952&41.558&$-$116.1&$81.5\pm4.8$&$6.5\pm0.7$&$10.6\pm1.2$&$4.6\pm0.5$\\
k3&10.971&41.562&$-$116.2&$143.9\pm6.8$&$6.1\pm0.4$&$27.6\pm2.2$&$12.0\pm0.9$\\
R1&11.055&41.474&$-$121.4&$164.7\pm10.7$&$10.3\pm0.8$&$29.4\pm3.5$&$12.8\pm1.5$\\
R2&11.226&41.520&$-$153.0&$76.9\pm13.0$&$6.0\pm1.0$&$4.3\pm0.4$&$1.9\pm0.2$\\
R3&11.187&41.553&$-$148.1&$68.6\pm26.4$&$2.5\pm1.5$&$0.9\pm0.2$&$0.4\pm0.1$\\
R4&11.198&41.531&$-$157.4&$160.6\pm39.6$&$3.3\pm1.1$&$2.3\pm0.2$&$1.0\pm0.1$\\
R5&11.130&41.424&$-$210.0&$161.6\pm11.1$&$9.4\pm1.0$&$16.5\pm2.1$&$7.2\pm0.9$\\
\hline
\end{tabular}
 \end{threeparttable}
\end{table*}

\begin{table*}
\centering
\contcaption{Clump properties from CO(1-0)}
\begin{threeparttable}
\begin{tabular}{ccccccccc}
\hline
ID &  RA & DEC &  $v$ &  $R$ & $\sigma_{v}$ & $L_{\rm CO}$ & $M\rm_{mol}$ \\
& (J2000) & (J2000) & $\rm(km~s^{-1})$ & (pc) & $\rm(km~s^{-1})$ &$(10^4~\rm K~km~s^{-1}~pc^2)$ & $(10^5~ M_{\odot}$)\\
\hline
R6&11.191&41.420&$-$201.0&$131.8\pm15.9$&$6.0\pm1.0$&$4.7\pm0.4$&$2.0\pm0.2$\\
R7&11.186&41.445&$-$213.5&$98.7\pm13.2$&$4.7\pm1.4$&$4.1\pm0.8$&$1.8\pm0.4$\\
R8&11.151&41.446&$-$208.0&$183.5\pm15.2$&$10.0\pm0.5$&$9.8\pm0.7$&$4.3\pm0.3$\\
R9&11.176&41.455&$-$199.5&$170.5\pm13.4$&$10.2\pm1.1$&$16.3\pm1.1$&$7.1\pm0.5$\\
R10&11.194&41.439&$-$190.8&$164.4\pm20.0$&$6.3\pm1.1$&$9.9\pm0.6$&$4.3\pm0.3$\\
R11&11.183&41.483&$-$174.5&$198.6\pm32.3$&$6.2\pm0.6$&$7.7\pm0.5$&$3.3\pm0.2$\\
R12&11.180&41.433&$-$200.7&$96.5\pm8.3$&$11.5\pm1.5$&$8.3\pm1.3$&$3.6\pm0.6$\\
R13&11.213&41.493&$-$185.0&$118.2\pm15.8$&$4.5\pm1.1$&$3.6\pm0.2$&$1.5\pm0.1$\\
R14&11.206&41.484&$-$161.1&$128.5\pm53.7$&$6.8\pm1.6$&$5.5\pm0.3$&$2.4\pm0.1$\\
R15&11.127&41.396&$-$231.6&$250.1\pm24.8$&$6.8\pm0.5$&$11.9\pm0.6$&$5.2\pm0.3$\\
R16&11.087&41.435&$-$198.4&$169.4\pm14.1$&$3.6\pm0.4$&$7.3\pm0.5$&$3.2\pm0.2$\\
R17&11.160&41.422&$-$201.2&$251.3\pm14.0$&$8.0\pm0.6$&$22.6\pm3.3$&$9.8\pm1.5$\\
R18&11.138&41.395&$-$210.2&$275.9\pm13.6$&$7.0\pm0.4$&$40.6\pm5.6$&$17.7\pm2.4$\\
R19&11.166&41.513&$-$165.7&$178.9\pm23.4$&$5.2\pm1.0$&$7.4\pm0.4$&$3.2\pm0.2$\\
R20&11.141&41.485&$-$175.7&$164.7\pm11.7$&$4.3\pm0.5$&$17.5\pm3.7$&$7.6\pm1.6$\\
R21&11.121&41.456&$-$187.8&$220.6\pm11.4$&$6.2\pm0.4$&$29.0\pm3.5$&$12.6\pm1.5$\\
R22&11.212&41.500&$-$161.2&$141.5\pm12.5$&$6.5\pm0.6$&$13.9\pm3.3$&$6.0\pm1.4$\\
R23&11.234&41.490&$-$166.9&$147.3\pm10.6$&$4.4\pm0.4$&$19.5\pm4.2$&$8.5\pm1.8$\\
R24&11.208&41.467&$-$180.5&$177.0\pm12.7$&$6.3\pm0.5$&$19.9\pm1.0$&$8.7\pm0.4$\\
R25&11.184&41.464&$-$187.1&$160.7\pm6.7$&$6.0\pm0.4$&$49.7\pm3.3$&$21.6\pm1.4$\\
R26&11.158&41.463&$-$188.2&$182.3\pm8.3$&$4.1\pm0.3$&$30.1\pm2.5$&$13.1\pm1.1$\\
R27&11.103&41.421&$-$209.7&$141.4\pm8.1$&$4.7\pm0.4$&$25.2\pm2.6$&$11.0\pm1.1$\\
R28&11.102&41.409&$-$221.8&$123.6\pm14.0$&$4.3\pm0.6$&$11.5\pm2.7$&$5.0\pm1.2$\\
R29&11.208&41.516&$-$195.4&$134.6\pm19.8$&$2.9\pm0.8$&$2.9\pm0.2$&$1.3\pm0.1$\\
R30&11.246&41.478&$-$199.7&$161.1\pm15.6$&$7.3\pm1.3$&$11.9\pm0.9$&$5.2\pm0.4$\\
R31&11.196&41.490&$-$206.4&$71.0\pm38.4$&$0.8\pm0.3$&$0.4\pm0.1$&$0.2\pm0.0$\\
N1&10.751&41.283&$-$163.1&$122.3\pm26.6$&$6.5\pm1.5$&$4.2\pm2.6$&$1.8\pm1.1$\\
\hline
\end{tabular}
 \end{threeparttable}
\end{table*}






\bsp	
\label{lastpage}
\end{document}